\documentclass[final,leqno,onefignum,onetabnum]{siamltexmm}
\usepackage{color}
\usepackage{graphicx}
\usepackage{amssymb,amsmath}
\graphicspath{{Figures/}}

\newcommand{\ThetaH}{\Theta_{\rm H}}
\newcommand{\RR}{\mathbb{R}}
\newcommand{\CC}{\mathbb{C}}
\newcommand{\abs}[1]{{\left\lvert#1\right\rvert}}
\newcommand{\xX}{\vec{x}_{\rm X}}
\newcommand{\xY}{\vec{x}_{\rm Y}}
\newcommand{\cC}{\mathcal{C}}
\newcommand{\cF}{\mathcal{F}}
\newcommand{\cS}{\mathcal{S}}
\newcommand{\cH}{\mathcal{H}}
\newtheorem{rem}{Remark}[section]
\newcommand{\rc}{\rho_{\rm c}}
\newcommand{\tmf}{\mathfrak{t}}

\title{Dynamics of vortex dipoles in anisotropic Bose-Einstein condensates}

\author{%
Roy H. Goodman%
{\thanks{%
Department of Mathematical Sciences,
New Jersey Institute of Technology,
University Heights, Newark, NJ 07102, USA
(\email{goodman@njit.edu})
}},
P.G. Kevrekidis%
{\thanks{%
Department of Mathematics and Statistics,
University of Massachusetts,
Amherst, MA 01003-4515, USA and
Center for Nonlinear Studies and Theoretical Division, Los Alamos
National Laboratory, Los Alamos, NM 87544
(\email{kevrekid@math.umass.edu})
}},
and
R. Carretero-Gonz{\'a}lez%
{\thanks{%
Nonlinear Dynamical System Group,
({\texttt{URL}: http://nlds.sdsu.edu}),
Computational Science Research Center,
({\texttt{URL}: http://www.csrc.sdsu.edu}),
and Department of Mathematics and Statistics,
San Diego State University,
San Diego, CA 92182-7720, USA
(\email{rcarretero@mail.sdsu.edu})
}}
}

\begin{document}
\maketitle
\newcommand{\slugmaster}{%
\slugger{siads}{xxxx}{xx}{x}{x--x}}%

\begin{abstract}
We study the motion of a vortex dipole in a Bose-Einstein condensate confined
to an anisotropic trap. We focus on a system of ordinary 
differential equations describing the vortices' motion, which is in turn a reduced model 
of the Gross-Pitaevskii equation describing the condensate's motion.
Using a sequence of canonical changes of variables, we reduce the
dimension and simplify the equations of motion. We uncover two interesting
regimes. Near a family of periodic orbits known as guiding centers, we find
that the dynamics is essentially that of a pendulum coupled to
a linear oscillator, leading to stochastic reversals in the overall
direction of rotation of the dipole. Near the separatrix orbit in the
isotropic system, we find other families of  periodic, quasi-periodic,
and chaotic trajectories. In a neighborhood of the guiding
center orbits, we derive an explicit iterated map that simplifies the
problem further. Numerical calculations are used to illustrate the
phenomena discovered through the analysis. Using the results from the
reduced system we are able to construct complex periodic orbits in
the original, partial differential equation, mean-field model for
Bose-Einstein condensates, which corroborates the phenomenology
observed in the reduced dynamical equations.
\end{abstract}



\pagestyle{myheadings}
\thispagestyle{plain}
\markboth{R.H. Goodman, P.G. Kevrekidis, and R. Carretero-Gonz\'alez}%
{Dynamics of vortex dipoles in anisotropic Bose-Einstein condensates}

\section{Introduction}

An atomic Bose-Einstein condensate (BEC) is a state of matter occurring only at extremely low temperatures. It is a gas composed of atoms (typically
alkali ones that behave as bosons), which, near absolute zero, lose their individual identity and share a single macroscopic wave function.
The wavefunction of the cloud of BEC particles obeys the Gross-Pitaevskii (GP)
equation, namely, the cubic nonlinear Schr\"odinger equation with an
added external potential used for confining the atoms.
BECs are inherently three-dimensional, although strong confinement
in one or two directions can effectively render the BEC two- or even
one-dimensional.
In non-dimensional units (see, e.g., Ref.~\cite{emergent} for a discussion
of the relevant adimensionalization), a cloud of repulsive BEC particles
confined in a quasi-two-dimensional trap is
described by the GP equation
\begin{equation}
i u_t = -\frac{1}{2}\Delta u + \abs{u}^2 u + V(x,y) u,
\label{GP}
\end{equation}
where $u(x,y,t)$ is the BEC wavefunction---whose observable in
the experiments is its associated density $|u|^2$---and the
potential $V(x,y)$ contains the effects of externally applied
magnetic and optical fields which are used to trap the condensate
at a particular location in space.
In most studies, $V(x,y)$ is taken to be isotropic about the origin;
yet, in a few more recent studies, the \emph{critical} potential
role of anisotropy in
$V(x,y)$ has been investigated.
The current study uses methods from Hamiltonian perturbation theory to gain a deeper understanding of the effects this anisotropy can have on the dynamics.

We note that this anisotropy is entirely straightforward to implement
in ongoing experiments (see relevant discussion below), by means of
the detailed available control over the magnetic traps that typically
are used to induce the parabolic confinement~\cite{becbook1,becbook2}.

Like an ordinary inviscid fluid, the Bose-Einstein condensate flow may possess highly localized vortex lines, or, if confined in a quasi-two-dimensional geometry, localized  vortices.
These may be idealized as point vortices, in which the vorticity is non-zero only at the vortex's center. Following a closed curve that encircles exactly one such vortex, one finds the phase of the wavefunction to have increased, or decreased, by $2\pi$ (or multiples thereof); this quantization of the circulation
is, arguably, the most fundamental difference between the superfluid
BEC vortices and the ones of an ordinary inviscid fluid. In order for a smooth solution of equation~\eqref{GP} to support a phase-singularity, $u$ must vanish at
the point vortex core. Notice in addition, the freedom in the
clockwise or counterclockwise nature of the phase rotation
(mirrored in the respectively, negative or positive charge of the vortex).

The first experimental observation of vortices in atomic BECs~\cite{Matthews99}
by means of a phase-imprinting method between two hyperfine
spin states of a $^{87}$Rb BEC~\cite{Williams99}
paved the way for a systematic
investigation of their dynamical properties.
Stirring the BECs~\cite{Madison00} above a certain critical angular speed~\cite{Madison01,Recati01,Sinha01} led to the production of few vortices~\cite{Madison01} and even of
robust vortex lattices~\cite{Raman}. Other vortex-generation
techniques were also
used in experiments, including the breakup of the BEC superfluidity by
dragging obstacles through the condensate~\cite{kett99}, as well as
nonlinear interference between condensate fragments~\cite{BPAPRL}.
In addition, apart from
unit-charged vortices,
higher-charged vortex structures were produced~\cite{S2Ket} and their dynamical
(in)stability was examined.

\subsection{Recent precedents}
Although much of this earlier work was focused on single vortices
(or large clusters of vortices constituting vortex lattices),
recently, a lot of attention has been paid to the problem of the motion that arises when few vortices (i.e., small clusters) interact.
These studies were partially seeded by the use of the so-called Kibble-Zurek
mechanism in order to quench a gas of atoms from well-above to well-below
the BEC transition in the work of Ref.~\cite{BPA}.
The result of this was that phase gradients would
not have sufficient time to ``heal'', as would happen by adiabatically
crossing the transition, but would rather often freeze, resulting
in the formation of vortices and even multi-vortices.
Subsequently,
a technique was devised
that enabled for the first time the systematic
dynamical visualization of such ``nucleated'' vortices~\cite{dshall}
and even of vortex pairs, i.e.\ dipoles consisting of two oppositely
charged vortices.
The technique involved pulsing a microwave beam through the BEC that
would expel a small fraction of it that could be imaged.
This, in turn, spearheaded further studies~\cite{dshall1,Torres:2011fp}
which
developed particle models that predicted the dipole dynamics
(equilibria, near-equilibrium epicyclic precessions and far from
equilibrium quasi-periodic motions) observed in these experiments.
A nearly concurrent development
produced such vortex dipoles (one or multiple such),
by the superfluid analogue of dragging a cylinder through a fluid~\cite{BPA3}.
The role of the cylinder here was played by a laser beam. More recently,
use of the above visualization  scheme~\cite{dshall}
together with rotation has allowed to ``dial in'' and observe the dynamics of
vortex clusters of, controllably, any number of vortices between 1
and 11~\cite{Navarro:2013uv}.
This led to the observation
that such configurations may suffer symmetry breaking events.
As a result,
instead of the commonly expected anti-diametric pair,
equilateral triangle, or square
configurations, it is possible to observe
symmetry broken configurations featuring asymmetric pairs,
isosceles triangles, and rhombi or general/asymmetric
quadrilaterals~\cite{zampetaki}. To further add to these developments,
yet another experimental group~\cite{tripole} produced 3-vortex configurations but of
alternating charge in the form of a tripole (i.e., a positive-negative-positive
or its opposite).

Naturally, this considerable volume of experimental developments has
triggered a number of corresponding theoretical efforts.
Again, as for ordinary fluids, it is possible to derive a finite-dimensional system of ordinary differential equations that describes the motion of these vortices, in which each vortex induces a velocity field, and each vortex moves under the velocity field induced by the other vortices.
Numerical studies have shown that the motion of vortices evolving under the GP equation~\eqref{GP} is mimicked by the motion arising in the ODE dynamics~\cite{komineas,Navarro:2012vn,Navarro:2013uv,Stockhofe:2013,Stockhofe:2011df}. Analysis of the ODE system, displayed below in equation~\eqref{eq:xy}, has allowed for the prediction of many different dynamical phenomena in the GP equation, but also
detailed numerical computations have been systematically used
to
unravel the evolution of these few vortex
clusters~\cite{mikko2,mikko1}.

For example, Torres et al.~\cite{Torres:2011fp} examined the \emph{vortex dipole} consisting of two counter-rotating vortices of equal and opposite vorticity, showing the existence of fixed points, in which the two vortices sit stably at a fixed distance on opposite sides of the magnetic trap's center. They additionally found periodic orbits in which the two vortices travel at constant angular velocity around a circle centered at the trap's center, keeping an angle of 180 degrees between them. Both of these types of orbits, known as guiding centers, are shown to be neutrally stable. Finally, they found families of quasi-periodic orbits, or more specifically, relative periodic orbits. These solutions appear periodic when viewed in an appropriate rotating reference frame, so that in the laboratory reference frame, these orbits display whirls upon whirls (i.e., epitrochoidal motion).

Other studies~\cite{Navarro:2012vn,Navarro:2013uv} have looked at small systems of two to four co-rotating vortices. The same types of solutions are found
such as  
relative stationary,  and relative periodic ones. However, the stationary and periodic solutions are shown to lose stability in Hamiltonian pitchfork bifurcations that result in the creation of new solutions in which the symmetry of the solutions is broken, even in the absence of anisotropy. The case of three
vortices but with opposite charges (i.e., two positive and one
negative or vice-versa) has also been a focal
point of recent interest~\cite{Koukouloyannis:2013,Kyriakopoulos:2013},
especially due to its potential for chaotic dynamics.

An additional direction that has been receiving a fair amount of
interest is that of imposing  asymmetries on the potential $V$. For instance, McEndoo and Busch~\cite{McEndoo:2009tp} use a variational method to study the existence and stability of steady arrangements of small numbers of vortices in an anisotropic trap. They find that above a critical level of anisotropy the ground state arrangement undergoes a bifurcation, from a lattice  in the isotropic case to a linear arrangement along the major axis of the anisotropic trap.
Subsequently, the same authors explored the vortex dynamics in such
anisotropic traps~\cite{mcendoo2}.
Stockhofe et al.~\cite{Stockhofe:2013,Stockhofe:2011df} showed that of the one-parameter family of stationary arrangements of a vortex dipole, only two survive the imposition of anisotropy: the arrangement with both vortices along the minor axis of the trap is always unstable, while the arrangement along the major axis is stable for small anisotropy but destabilizes when the anisotropy is increased beyond some threshold. This conclusion is then generalized to larger clusters of vortices. It is  the dynamics that arises in the presence of small anisotropy that will be the systematic focus of the present study.

\subsection{Organization of article}
The remainder of the article is organized as follows. Section~\ref{sec:formulation} introduces the differential equations and their Hamiltonian form. A change of variables is made that reduces the equations by one degree of freedom in the case of zero anisotropy.
This isotropic reduced equation is studied in detail in Section~\ref{sec:unperturbed}. This is very similar to work in Refs.~\cite{Navarro:2012vn,Navarro:2013uv}, but the reduction allows us to understand the solutions more thoroughly.
In Section~\ref{sec:near_resonance} we derive a further reduced ordinary differential system that explains the effects of anisotropy very clearly. We then sketch the derivation of a separatrix map for this system, simplifying the dynamics further and allowing us to explain the various families of periodic orbits that arise due to the anisotropy.
Section~\ref{sec:numerics} contains the results of numerical simulations including studies of the ODE's of vortex motion, and a bifurcation diagram based on the separatrix map.
In Section~\ref{sec:numericalGP} we use the results obtained for the reduced
ODE to construct complex periodic solutions for the original
partial differential equation, the GP equation.
Here, despite the apparent complexity of the orbits, good correspondence
is found between the ODE and the PDE findings.
Finally, we discuss the impact of this work and future directions in Section~\ref{sec:discussion}.

\section{Mathematical Formulation}
\label{sec:formulation}

Although our principal focus will be on the study of the ODEs,
as indicated above the results will be corroborated by full numerical
simulations of the PDE of the GP type from which these ODEs are
derived (see Section~\ref{sec:numericalGP}). 
This derivation can be obtained in a wide variety of ways.
In the mathematical literature, it can be obtained
by reverting to the semi-classical
limit of the GP equation (i.e., the limit of large density, using as
a small parameter the inverse of the density) and applying
a variational approach~\cite{peli} or by means of moment
methods~\cite{jerrard,spirn} (by suitably decomposing the field), or
through techniques based
on the Fredholm alternative~\cite{Chang:2002}. All of these
techniques derive effective equations for the vortices
as particles which incorporate two crucial features: the
rotation of the vortices in the (here, anisotropic) trap
and the pairwise velocity-field induced interactions between the
vortices. On the physical side, there have also been numerous
derivations of such equations in both isotropic setting
(see, e.g., Ref.~\cite{castin}, for relevant reviews~\cite{fetter,fetter1} and for a
recent extension summarizing earlier literature Ref.~\cite{fetter2})
and even in the anisotropic~\cite{McEndoo:2009tp} setting, based chiefly
on applying variational methods to a suitable
vortex-bearing ansatz.

\subsection{Equations of motion and fixed points}
A cluster of $N$ vortices in an anisotropic Bose-Einstein condensate confined in a magnetic trap satisfies the ordinary differential equations
\begin{equation}
\begin{split}
\dot{x}_k & =- s_k Q \omega_y^2 y_k + \frac{B}{2} \sum_{j\neq k} s_j \frac{y_j-y_k}{\rho_{jk}^2 },\\
\dot{y}_k & = \phantom{-}s_k Q \omega_x^2 x_k - \frac{B}{2} \sum_{j\neq k} s_j \frac{x_j-x_k}{\rho_{jk}^2 } \text{ for } k=1,\ldots,N,
\end{split}
\label{eq:xy}
\end{equation}
where $(x_k,y_k)$ gives the Cartesian coordinates of the $k$th vortex,  $s_k = \pm 1$ is its charge, and $\rho_{jk}^2  = {(x_j-x_k)}^2+{(y_j-y_k)}^2$. By rescaling the independent and dependent variables, we may set $Q\omega_x^2 = 1$ and $Q\omega_y^2 = 1+\epsilon$~\cite{fetter,Stockhofe:2011}.
We here consider the vortex dipole case: $N=2$ with $s_1=1$, $s_2=-1$. For the remainder of the paper, we fix the parameter $B=0.22$ describing the ratio of time-scales of rotation due to the vortex interactions and due to precession induced by the applied magnetic trap.%
\footnote{This number is obtained as the ratio of two frequencies $\omega_{\rm vort} \approx 0.005$ and $\omega_{\rm pr} = 0.023$, associated, respectively,
with inter-vortex interaction and individual vortex precession in Ref.~\cite{Navarro:2013uv}.}  We work in the regime of weak anisotropy, namely $\epsilon\ll 1$.
\begin{rem}
\label{rem:simplification}
The equations, as presented, contain the additional assumption that the vortices remain close to the trap's center. The full equations, in the case of isotropic traps, contain a modification to the precession frequency, i.e.,
an increase in the precession frequency as the outer rim of
the condensate is approached; see, e.g., Refs.~\cite{Kolokolnikov,Navarro:2013uv,zampetaki}
for some of the implications of this modification. While we have not
worked out the modified equations in the anisotropic case, we are here considering the leading-order effects of adding anisotropy, and thus have reason to believe the simpler equations can provide insight into the general behavior;
see also the relevant comparison between our ODE and PDE results below, which
a posteriori justify the present considerations.
\end{rem}
It is shown in Ref.~\cite{Torres:2011fp} that when $\epsilon=0$, system~\eqref{eq:xy} has a one-parameter family of fixed points (i.e., a \emph{resonance}) of the form
\begin{equation}
(x_1,x_2,y_1,y_2)
= \frac{\sqrt{B}}{2} \left(\cos{\theta},-\cos{\theta},\sin{\theta},-\sin{\theta}\right), \,
0\le \theta < 2\pi,
\label{eq:fixedcircle}
\end{equation}
in which the two vortices lie on opposite sides of, and equidistant to, the origin.
When $\epsilon \neq 0$, only four of the fixed points on the resonance survive. Letting $\vec{x}=(x_1,x_2,y_1,y_2)$, these fixed points are
\begin{equation}
\xY =\pm\sqrt{\frac{B}{4(1+\epsilon)}}
\begin{pmatrix}
0 \\ 0 \\ 1 \\ -1
\end{pmatrix}
\text{~and~}
\xX=\pm\frac{\sqrt{B}}{2}
\begin{pmatrix}
1\\-1\\0\\0
\end{pmatrix}.
\label{fixedpoints}
\end{equation}
These solutions consist of orbits aligned on the $y$-axis ($\xY$) and the $x$-axis ($\xX$).
The configurations $-\xX$ and $-\xY$ are also solutions, obtained by reversing the locations of the two vortices.
When $0<\epsilon<1$, the arrangement $\xX$ is stable and $\xY$ is unstable, while when $\epsilon<0$, the opposite is true. We will assume throughout, without
loss of generality, that $\epsilon>0$.

We will be concerned with the dynamics that accompany this symmetry-breaking perturbation. The primary tools will come from Hamiltonian mechanics.

\subsection{The Hamiltonian Formulation}
Stockhofe et al.~\cite{Stockhofe:2011} showed that this system is Hamiltonian with canonical position and momentum variables $q_k=x_k$ and $p_k= -s_k y_k$, respectively, and Hamiltonian
$$
H(q,p) = \sum_{k=1}^N \left[ q_k^2 + (1+\epsilon)p_k^2 + \frac{B s_k}{4} \sum_{j=k+1}^N s_j \ln \rho_{jk} \right],
$$
where $\rho_{jk}^2  = {(q_j - q_k)}^2 + {(s_j p_j - s_k p_k)}^2$.
For the vortex dipole this reduces to
\begin{equation*}
H = H_0 + \epsilon H_1 =
\frac{1}{2}   \left(p_1^2+q_1^2+p_2^2+q_2^2\right)
-\frac{B}{4}  \log{\left({\left(p_1+p_2\right)}^2+{\left(q_1-q_2\right)}^2\right)}
+\frac{\epsilon}{2} \left(p_1^2+p_2^2\right).
\end{equation*}
We introduce two successive canonical changes of variables that clarify the dynamics while preserving the Hamiltonian structure. First, we define action-angle coordinates
$$q_j = \sqrt{2 J_j} \cos{\phi_j} \text{ and } p_j = \sqrt{2 J_j} \sin{\phi_j}, j=1,2,$$
which transforms the Hamiltonian to
\begin{align*}
H_0 &= J_1 + J_2 -\frac{B}{4}  \log {\left(J_1+J_2-2 \sqrt{J_1} \sqrt{J_2} \cos{\left(\phi_1+\phi_2\right)}\right)}, \\
H_1 &= J_1 \sin ^2\left(\phi_1\right) +J_2 \sin ^2\left(\phi_2\right).
\end{align*}
Since $H_0$ depends on the angle variables only through the combination $(\phi_1+\phi_2)$, we define the additional canonical change of variables
\begin{equation}
\theta_1 = -\phi_1 + \phi_2, \,
\theta_2 = \phi_1+\phi_2, \, 
\rho_1 = \frac{-J_1 + J_2}{2}, \,
\rho_2 = \frac{J_1+J_2}{2},
\label{eq:theta_phi_rho_J}
\end{equation}
which yields the Hamiltonian $H = H_0 + \epsilon H_1$, with
\begin{equation}
\label{Hrhotheta}
\begin{split}
H_0 & = 2 \rho_2 -\frac{B}{4}  \log{\left(\rho_2-\sqrt{\rho_2^2-\rho_1^2} \cos \theta_2\right)}, \\
H_1 & = \rho_1
   \sin{\theta_1} \sin{\theta_2} + \rho_2 \left(1 -\cos{\theta_1}\cos{\theta_2}\right). \end{split}
\end{equation}
It is this latter form that we will use.

\begin{rem}
The proper limits on the new angle variables are $-2\pi \le \theta_1 < 2\pi$ and $0\le\theta_2<4 \pi$. Nonetheless the reduced Hamiltonian is $2\pi$-periodic in these angles. The natural limits become important when inverting the change of variables~\eqref{eq:theta_phi_rho_J} to obtain the vortex paths.
\end{rem}

\section{Analysis of the unperturbed equation}
\label{sec:unperturbed}
We first set $\epsilon=0$ and consider the dynamics due to the unperturbed Hamiltonian $H_0$. 
In this limit, the variable $\theta_1$ is cyclic: it does not appear in $H_0$ as a result of the rotation invariance of the underlying system, an invariance that is broken when $\epsilon \neq 0$. 
Thus, the angular momentum $\rho_1$ is an additional conserved quantity, in involution with $H_0$ so that $H_0$ is a completely integrable Hamiltonian.
This conservation law is generic for isotropic traps.
The interesting phenomena described in later sections result mainly from the breaking of this isotropy.

The reduced system obeys the ODEs
\begin{subequations}
\label{thetarhodot}
\begin{align}
\dot{\theta}_2 &= 2 -
\frac
{\left(\sqrt{\rho_2^2 -\rho_1^2} - \rho_2 \cos{\theta_2} \right)B}
{4\left( \rho_2 \sqrt{\rho_2^2 -\rho_1^2}- \left( \rho_2^2 -\rho_1^2 \right) \cos{\theta_2} \right)},
\label{theta2dot}
\\
\dot{\rho_2} & = \frac{B \sqrt{\rho_2^2 -\rho_1^2} \sin{\theta_2}}
{4\rho_2 - 4 \sqrt{\rho_2^2 -\rho_1^2}\cos{\theta_2} }.
\label{rho2dot}
\\
\intertext{The additional angle $\theta_1$ satisfies the evolution equation}
\dot{\theta}_1&= \frac{-B}{4}
\frac{\rho_1 \cos{\theta_2}}{\rho_2\sqrt{\rho_2^2-\rho_1^2} -(\rho_2^2-\rho_1^2) \cos{\theta_2}}.
\notag
\end{align}
\end{subequations}
The system~\eqref{thetarhodot} has a single equilibrium
\begin{equation}
(\theta_2^*,\rho_2^*) = \left(\pi,\sqrt{\rho_1^2 + \frac{B^2}{64}}\right).
\label{fixedpoint}
\end{equation}
In the laboratory frame, the overall angular velocity is
$$
\Omega_1 \equiv \dot{\theta}_1 = \frac{16 \rho_1}{B+\sqrt{B^2 + 64 \rho_1^2}}.
$$
Translating this back to the $(x,y)$ coordinates gives
$$
\begin{pmatrix} x_1 \\ x_2 \\ y_1 \\ y_2 \end{pmatrix}
=
\begin{pmatrix}
\phantom{-}\frac{1}{2} \sqrt{\sqrt{B^2+64 \rho_1^2}-8 \rho_1} \cos {\left(\frac{\Omega_1}{2} (t-t_0)\right)}\\
-\frac{1}{2} \sqrt{\sqrt{B^2+64 \rho_1^2}+8 \rho_1} \cos {\left(\frac{\Omega_1}{2} (t-t_0)\right)}\\
\phantom{-}\frac{1}{2} \sqrt{\sqrt{B^2+64 \rho_1^2}-8 \rho_1} \sin {\left(\frac{\Omega_1}{2} (t-t_0)\right)}\\
-\frac{1}{2} \sqrt{\sqrt{B^2+64 \rho_1^2}+8 \rho_1} \sin {\left(\frac{\Omega_1}{2} (t-t_0)\right)}
\end{pmatrix}.
$$
Here, the two vortices trace circular orbits, with both vortices collinear with and on opposite sides of the origin.
The circle of fixed points~\eqref{eq:fixedcircle} is obtained by setting $\rho_1=0$. This family of periodic orbits forms a two-dimensional tube in $\RR^4$. When $\rho_1 \neq 0$, the direction of rotation depends on its sign. The circle with $\rho_1=0$, on which there is no motion, is called a resonance and is structurally unstable to symmetry-breaking perturbations.

\begin{figure}[htbp]
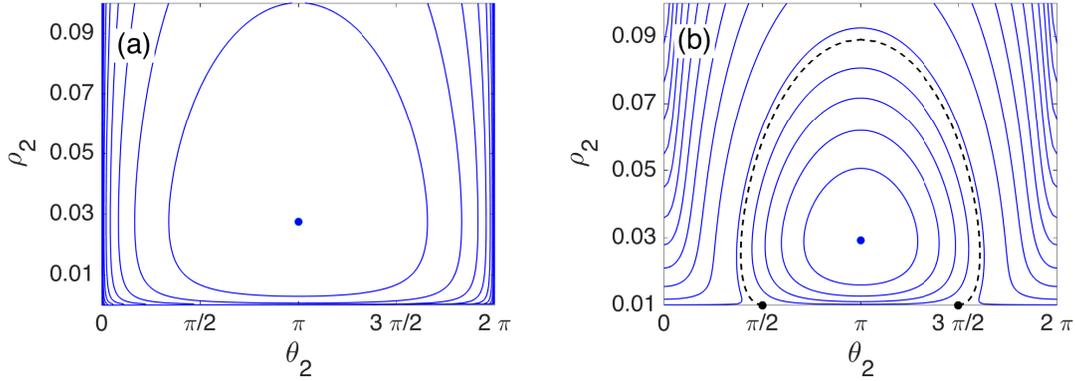
 
   \centering
   \includegraphics[width=.47\textwidth]{phaseplane_rho_0}
   \includegraphics[width=.47\textwidth]{phaseplane_rho_0p01}
   \caption{
(color online)
Phase plane of reduced system~\eqref{thetarhodot} with $B=0.22$ with (a) $\rho_1=0$ and (b) $\rho_1=0.01$.}
\label{fig:phaseplanes}
\end{figure}

Figure~\ref{fig:phaseplanes} shows the typical shape of the phase plane in the case $\rho_1=0$ (left) and $\rho_1 \neq 0$ (right). The case $\rho_1=0$ is simpler. All the orbits, which are traversed counterclockwise, form a nested set of closed curves encircling the fixed point at $(\theta_2,\rho_2) = (\pi, B/8)$ (see [blue] dot). This fixed point is known as a ``guiding center'' in the BEC literature~\cite{Torres:2011fp}. The accessible phase space is $0<\theta_2<2\pi$ and $\rho_2>\abs{\rho_1}\ge0$, and system~\eqref{thetarhodot} shows the vector field to be singular at the boundaries. The singularity at the left and right boundary is due to the impossibility of collisions between the vortices. The singularity along the bottom edge is simply that of polar coordinates.

When $\abs{\rho_1}>0$, the left and right edges of the domain are no longer singular. The bottom edge remains singular, with two apparent hyperbolic fixed points at $\rho_2=\abs{\rho_1}$ and $\theta_2 = \pi/2$ or $3\pi/2$. These are not in fact fixed points or even relative fixed points in the $(x,y)$ or $(p,q)$ coordinate systems. At these points $\dot{\rho}_2=0$, but $\dot{\theta}_2$ is singular. When the solution reaches this point, one of the two vortices reaches the minimum of the trap, and its angular direction in the reduced coordinates, though not in the lab coordinates, jumps discontinuously.

The singularity may be removed by putting the right hand side of $\dot \rho_2$ and $\dot \theta_2$ in equation~\eqref{thetarhodot} over common denominators. The system obtained by considering only the numerators of these expressions is non-singular and has the same trajectories as system~\eqref{thetarhodot}. In the desingularized system, these fixed points are hyperbolic. They are connected by three heteroclinic orbits, two along the line $\rho_2 = \abs{\rho_1}$ (recalling the periodicity in the $\theta_2$ direction) and a third extending into the region $\rho_2 > \abs{\rho_1}$. The solutions near the elliptic fixed point oscillate counterclockwise, and those  outside  the separatrix move to the left. These points and their invariant manifolds play the same role as separatrices as they would if they were actual fixed points.

\begin{figure}[htbp] 
   \centering
   \includegraphics[width=7cm]{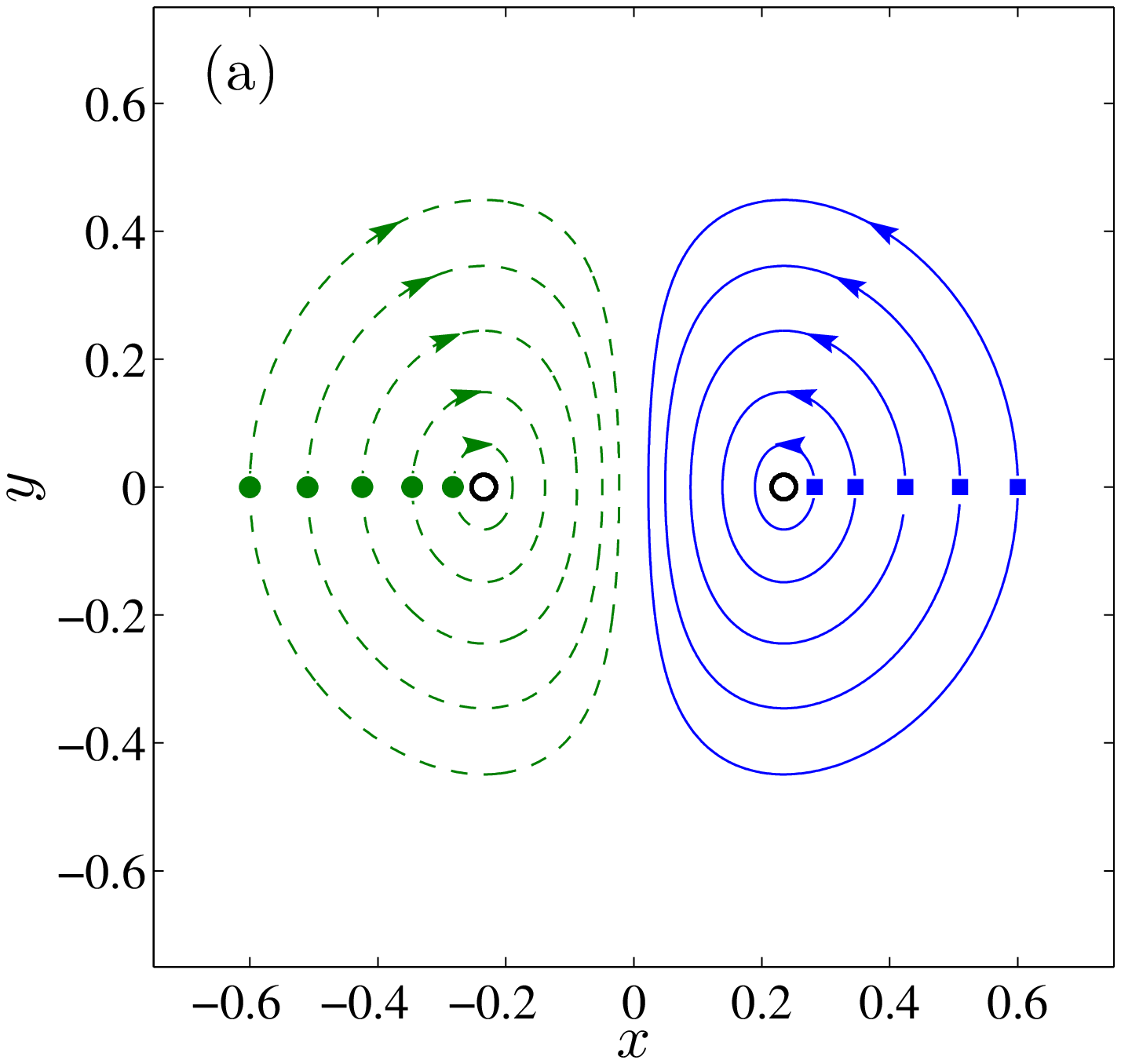}
   \includegraphics[width=7cm]{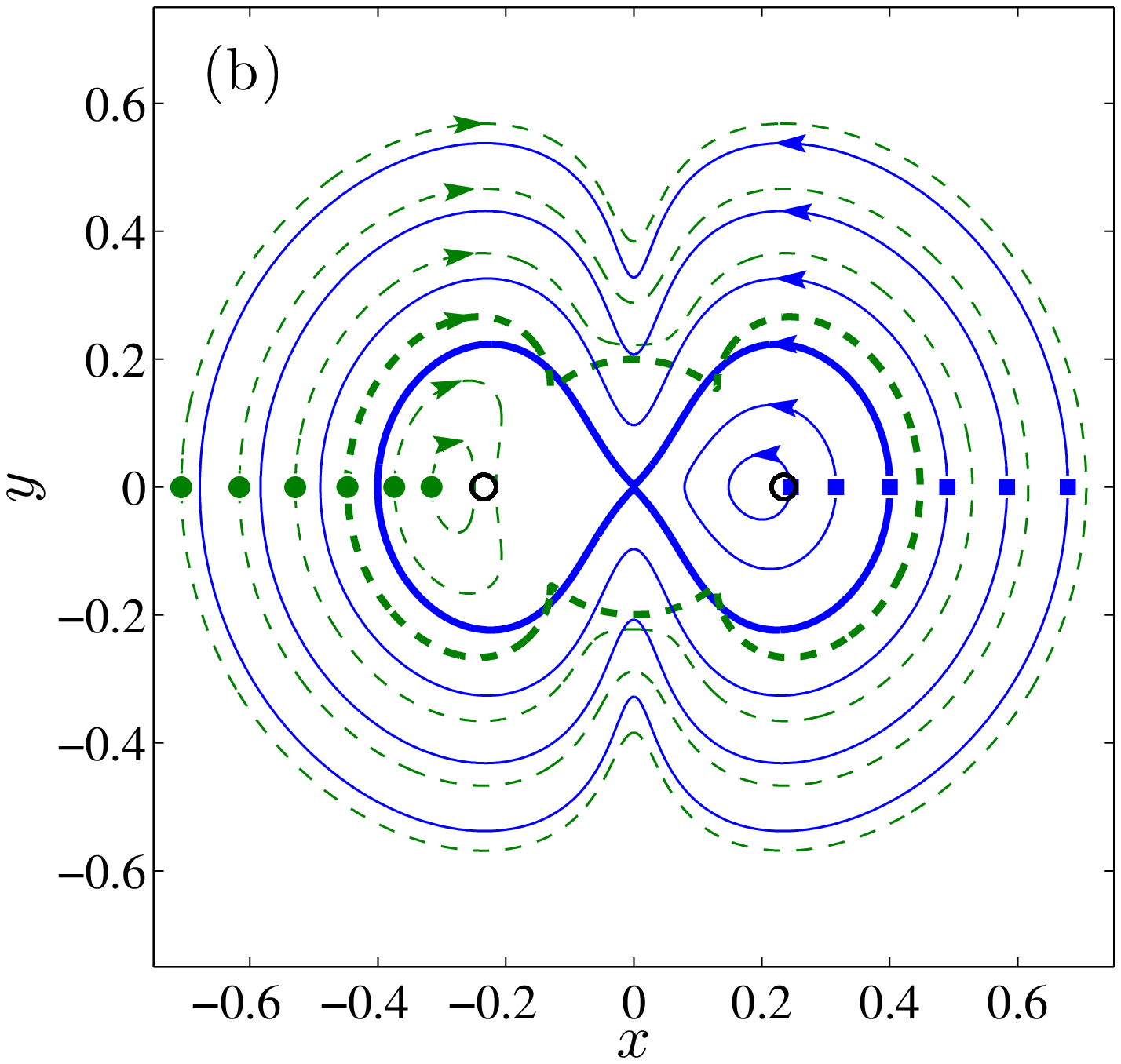}
   \caption{
(color online)
Typical orbits for two interacting vortices in an isotropic
($\epsilon=0$) BEC. The solid (blue) line corresponds to the
orbit of one vortex and the dashed (green) line to the orbit of
the other vortex. Matching orbits for each pair corresponds to
the closest initial conditions (filled squares and circles) pairs
from the origin.
(a) Closed orbits of two vortices when $B=0.22$ and $\rho_1=0$.
At $t=0$, the solutions start on the $x$-axis at maximal distance apart
(see filled squares and circles). They rotate about their centers in opposing
directions (see arrows indicating the direction of the orbits),
nearly collide and then move off again.
(b) ``Closed orbits'' of the system with $\rho_1=0.01$, which are
closed when viewed in an appropriate rotating reference frame.
The orbits depicted with the thick lines corresponds to the
the separatrix in Figure~\ref{fig:phaseplanes}b.
The solutions outside the separatrix cross the line $x=0$ without
colliding.
The equilibrium positions (\ref{fixedpoints}) are depicted
by the black empty circles.}
\label{fig:particle_paths_reduced}
\end{figure}

Each of the orbits in the phase planes of these systems is periodic in the $(\theta_2,\rho_2)$ coordinate system, and quasi-periodic in the full $(x,y)$ coordinates.  The motion in $(x,y)$ coordinates is complicated, but, because $(\theta_2,\rho_2)$ evolves independently of $\theta_1$ we may remove the $\theta_1$ dependence from the solution entirely. We show such solutions in Figure~\ref{fig:particle_paths_reduced}. In the case $\rho_1=0$ all the solutions are in fact periodic. The fixed point of system~\eqref{thetarhodot} corresponds to the two vortices sitting along a line through the origin at equal distance from the origin. In the periodic orbit, the two vortices trace closed paths around these fixed points, the right vortex moving counterclockwise and the left orbit clockwise, satisfying $x_1=-x_2$ and $y_1 = y_2$. The farther the orbits start from the fixed point, the closer they come to colliding.

When $\rho_1 \neq 0$, the motion is more complicated. The vortices can pass close to each other without colliding. There are two families of orbits with very different dynamics, corresponding to solutions exterior to or interior to the separatrix. These orbits are depicted in Figure~\ref{fig:particle_paths_reduced}b together
with the separatrix (thick orbits). On the separatrix orbit one of the vortices crosses the origin, at which point its tangent direction changes discontinuously, as a result of the logarithmic singularity of $H_0$.

Most of the above results are obtained by Torres et al.~\cite{Torres:2011fp}, although the reduced phase space provides a more global picture of the dynamics. 
The features not noticed in that study are the separatrix orbit and the 
corresponding orbits outside of this separatrix which encircle the origin
and both fixed points (see Figure~\ref{fig:particle_paths_reduced}b). 
These authors use the more realistic model for the trap as discussed in Remark~\ref{rem:simplification}. It is straightforward to check that these features persist in the latter case.

\section{Analysis of the perturbed system in a neighborhood of the resonance}
\label{sec:near_resonance}
\subsection{Change of variables and leading-order asymptotics}
We can expect that when $\epsilon >0$, there will be interesting dynamics near the resonance, which is given as the fixed point of the reduced system in equation~\eqref{fixedpoint}. When $\epsilon>0$, $\rho_1$ varies slowly, and so does $\rho_2^*$, which had been a fixed point.  We introduce a further change of variables to fix this point at the origin  by defining new variables $(\Theta_2,R_2) = (\theta_2-\pi,\rho_2 - \rho_2^*)$. This is extended to the full four-dimensional system using the generating function~\cite{Goldstein:2001} of type three%
\footnote{Recall that the angle variable $\theta_j$ is a position and the action variable~$\rho_j$ is a momentum.}
$$
F_3(\rho,\Theta) = - \Theta_2 \left( \rho_2 - \sqrt{\frac{B^2}{64} + \rho_1^2} \right) -\Theta_1 \rho_1 - \pi \rho_2.
$$
This gives the implicit change of variables from $(\theta,\rho)$ to $(\Theta,R)$:
$$
\theta_j = -\frac{\partial F_3}{\partial \rho_j}, \,\; R_j = - \frac{\partial F_3}{\partial \Theta_j}.
$$
In terms of the new variables
$$
 \Theta_1 =  \theta_1 + \frac{(\theta_2-\pi)\rho_1}{\sqrt{\frac{B^2}{64} + \rho_1^2}}, \,
 \Theta_2 = \theta_2 - \pi, \,
  R_1  = \rho_1,~\text{and} \,
  R_2  = \rho_2 -\sqrt{\frac{B^2}{64} + \rho_1^2},
$$
the Hamiltonian takes the form
\begin{equation*}
\begin{split}
H  =  &
\frac{B}{4} \left(  \beta + 8 R_2
-\log {\left(\sqrt{16 R_2   \left(\beta+4 R_2\right)+B^2} \cos{\Theta_2}+\beta+8 R_2\right)}\right) \\
   & +\epsilon   \left(
   \frac{1}{8} \left(\beta+8 R_2\right) \left(\cos{\Theta _2}
   \cos{\left(\Theta _1-\frac{8 \Theta _2 R_1}{\beta}\right)}+1\right)-
   R_1 \sin{\Theta _2}
   \sin{\left(\Theta _1-\frac{8 \Theta _2 R_1}{\beta}\right)}\right).
\end{split}
\end{equation*}
where $\beta= \sqrt{B^2 + 64 R_1^2}$.
Upon initial inspection, this form does not appear to be an improvement over the previous form of $H$. Its utility becomes apparent when we assume the trajectory remains in a neighborhood of the resonance, i.e., when
$R_1$, $R_2$, and $\Theta_2$ are small. A maximal balance is achieved by the assumptions
\begin{equation}
R_1 = O(\sqrt{\epsilon}), \, 
R_2 = O(\sqrt{\epsilon}),\, \text{and} \,
\Theta_2 = O(\sqrt{\epsilon}).
\label{scalingAssumption}
\end{equation}
These make the leading order Hamiltonian $O(\epsilon)$ and the perturbation $O(\epsilon^{3/2})$.
The leading order Hamiltonian then yields
\begin{equation*}
H_{\rm approx} =  \frac{4 }{B} R_1^2 + \frac{8}{B} R_2^2 + \frac{B}{16} \Theta_2^2 +
\epsilon \left(\frac{B}{8} \cos{\Theta_1} +R_2 + R_2 \cos{\Theta_1} \right).
\end{equation*}
An additional canonical change of variables ($R_2 \to R_2  - \epsilon B/16$) removes one of 
the $O(\epsilon)$ terms while adding a new term at the ignored higher order, yielding
\begin{equation*}
H_{\rm approx} =  \frac{4 }{B} R_1^2 + \frac{8}{B} R_2^2 + \frac{B}{16} \Theta_2^2 +
\epsilon \left(\frac{B}{8} \cos{\Theta_1} +R_2 \cos{\Theta_1} \right).
\end{equation*}
Using the above change of variables 
we may rewrite the evolution equations as
\begin{subequations}
\label{eq:Theta_r}
\begin{align}
\ddot{\Theta}_1 - \epsilon \sin{\Theta_1} - \frac{8\epsilon}{B} R_2 \sin{\Theta_1} & = 0 ,\label{eq:Theta1}\\
\ddot{R}_2 + 2 R_2 + \frac{B \epsilon}{8} \cos{\Theta_1} &=0.\label{eq:r2}
\end{align}
\end{subequations}
This is quite a well-known system, and similar systems have been studied many times before and using different methods.  The model system consisting of a pendulum coupled to a harmonic oscillator goes back to Poincar\'e's seminal work on the geometric approach to mechanics~\cite{Poincare:1890} as described by Holmes~\cite{Holmes:1990kz}. A similar system was derived by Lorenz as a model of the ``atmospheric slow manifold''~\cite{Lorenz:1986}, and subsequently  analyzed by Camassa et al.~\cite{Camassa:1998tc}, who rigorously found some of the same features to be described below, but who stopped short of writing down an explicit iterated map like the one identified below.  A simple approach in which  phase-plane analysis and matched asymptotics expansions are used to derive a discrete-time iterated map is developed in a series of papers by Goodman and Haberman~\cite{Goodman:2008,Goodman:2005vv,Goodman:2007wj}. In the final paper~\cite[Section~7B]{Goodman:2008}, this system was reduced, formally, to a singular iterated map from the plane to itself. We briefly reiterate that calculation in the the present context. We compare the analytical predictions of this approach with appropriate simulations of system~\eqref{Hrhotheta} in Section~\ref{sec:bifurcationdiagram}.

The appearance of a normal-form system~\eqref{eq:Theta_r} is a generic feature in the following sense.  In the limit $\epsilon=0$, the $(\Theta_2,R_2)$ equation corresponds to simple harmonic motion, and trajectories of this system are nested ellipses surrounding the origin. The other angle $\Theta_1$ varies at a constant rate, and we may define a Poincar\'e map each time $\Theta_1 = 0 \mod{2\pi}$. The trajectories of this map lie on the same family of ellipses. For sufficiently small nonzero values of  $\epsilon$, KAM theory ensures that most of these ellipses persist, but those whose period is a rational multiple of the frequency $T_2=2 \pi / \sqrt{2} $ of the $(\Theta_2,R_2)$-motion will break up into a sequence of fixed points of alternating elliptic and hyperbolic type. In a band containing these fixed points, the nonlinearity looks precisely like that of a pendulum~\cite{Lichtenberg:1992}.

\subsection{Derivation of an iterated-map approximation}
\label{sec:iterated_map}
Clearly, taking $\epsilon \to 0$ naively in equation~\eqref{eq:Theta_r} yields an uncoupled system, so the limit must be taken carefully. To do this,  scale time by $\tau = \sqrt{\epsilon} t$. In terms of this variable, dropping subscripts on $R_2$ and $\Theta_1$, and letting $(\cdot)'$ denote $\tau$-derivatives, equation~\eqref{eq:Theta_r} is rescaled to
\begin{subequations}
\label{eq:Theta_r_scaled}
\begin{align}
\Theta'' -  \sin{\Theta} - \frac{8}{B} R \sin{\Theta} & = 0 ,\label{scaled_a}
\\
R''+ \lambda^2  R + \frac{B }{8} \cos{\Theta} &=0,\label{scaled_b}
\end{align}
\end{subequations}
with $\lambda^2 = 2/\epsilon$. Note that the coupling is now formally $O(1)$, but, because $\Theta$ and $R$ evolve on such different timescales, the energy transfer between the two modes is exponentially small. Temporarily setting $R= 0$, the first component  conserves a pendulum-like energy
\begin{equation}
E = \frac{1}{2}{\Theta'}^2 + \cos{\Theta}-1.
\label{eq:E}
\end{equation}
Because equation~\eqref{scaled_a} depends on $R(\tau)$, the energy $E$ also evolves in time under the full dynamics of~\eqref{eq:Theta_r_scaled}.

\begin{figure}[htbp] 
   \centering
   \includegraphics[width=4in]{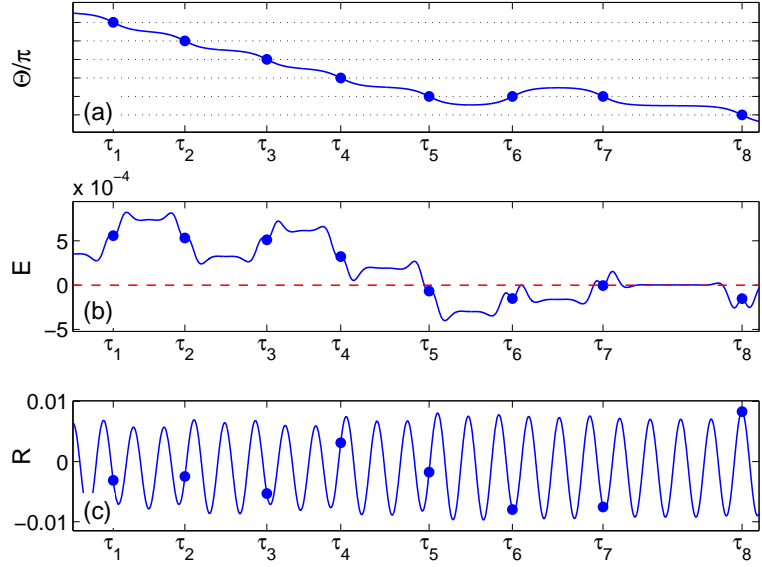}
   \caption{
(color online)
Figure used in constructing the iterated map. (a) $\Theta(\tau)$ with horizontal dashed lines at $\Theta=\pi\mod{2\pi}$, the minima of the potential $V(\Theta) = \cos{\Theta}-1$. This defines a sequence of times $\tau_j$ (see filled [blue] circles in all panels) at which $\Theta(\tau_j)$ reaches these minima. (b) The energy $E(\tau)$ in the pendulum like component.  (c) The evolution of the oscillatory variable $R(\tau)$. }
\label{fig:map_explanation}
\end{figure}

The map is constructed in consultation with Figure~\ref{fig:map_explanation},
which shows a numerical simulation of the equations of motion in
form~\eqref{eq:Theta_r_scaled}.  Panel~(a) shows $\Theta(\tau)$ and identifies a
sequence of ``transition times'' $\tau_j$ at which $\Theta = \pi \mod{2\pi}$. We also define a sequence of ``plateau times''~$\tmf_j$ (not shown) satisfying $\tau_{j-1} < \tmf_j <\tau_j$, defined as the times 
at which the solution makes its closest approach 
to the saddle point at $\Theta=0 \mod{2\pi}$. 
Panel~(b) shows that at the times $\tau_j$, the energy  undergoes a rapid
jump between two plateaus. On the plateaus where $E<0$ the solution librates,
with zero net change of the angle over one period, while on those with $E>0$, the solution
rotates, with $\Theta$ changing monotonically by $\pm 2\pi$. Panel~(c)
shows the evolution of $R(\tau)$. The amplitude and phase of
this oscillatory variable is different at each $\tau_j$ and its
amplitude changes slightly at each transition time. The
interval between transitions is longer on plateaus where $E$ is near zero, so that
$R(\tau)$ oscillates more on these periods.

We define $E_j$ to be the plateau value of energy on the interval $(\tau_{j-1},\tau_j)$ and assume that on this interval
\begin{equation}
R= R_{\rm before}(\tau) \approx -\frac{B}{8\lambda^2} + 
\cC\left(c_j \cos{\lambda(\tau-\tau_j)} + s_j \sin{\lambda(\tau-\tau_j)}\right).
\label{Rbefore}
\end{equation}
Although written in terms of the transition time $\tau_j$, this represents the state of the system near the plateau time $\tmf_j$, where $\cos{\Theta}\approx 1$.
The constant $\cC$ is chosen below to normalize the variables
and eventually simplify the derived map and $c_j$ and $s_j$ are the
corresponding amplitudes for the cosine and sine components.
Thus, we seek a map of the form $(E_{j+1},c_{j+1},s_{j+1}) = F(E_j,c_j,s_j)$. This is done for a very similar system in Ref.~\cite{Goodman:2008}, so we include here only a brief sketch of the derivation, with somewhat less precise language.

The solution is approximated using a matched asymptotic expansion,  alternating between ``outer solutions'' near the saddle points and ``inner solutions'', along which the approximate solution is given by the heteroclinic trajectory~$\ThetaH(\tau) = 4 \tan^{-1}{(e^{\pm \tau})}$.  First, an approximate solution to equation~\eqref{scaled_b} satisfying condition~\eqref{Rbefore} is given by  variation of parameters,
\begin{equation}
\label{eq:Rintegral}
\begin{split}
R(\tau) = R_{\rm before}(\tau) 
&- \frac{B \cos{\lambda(\tau-\tau_j)}}{8 \lambda} \int_{-\infty}^{\tau} \sin{\lambda(s-\tau_j)} 
\cos{\left(\ThetaH(s-\tau_j)\right) ds}\\
&+ \frac{B\sin{\lambda(\tau-\tau_j)}}{8 \lambda}\int_{-\infty}^{\tau} \cos{\lambda(s-\tau_j)} \cos{\left(\ThetaH(s-\tau_j)\right) ds}.
\end{split}
\end{equation}
In the limit $\tau\to\infty$, the first of the above integrals vanishes by symmetry, so as $\tau-
\tau_j \to + \infty$
$$
R(\tau) = R_{\rm after}(\tau) \sim R_{\rm before} +  \frac{B\sin{\lambda(\tau-\tau_j)}}{8 \lambda}\int_{-\infty}^{\infty} \cos{\lambda(s-\tau_j)} \cos{\left(\ThetaH(s-\tau_j)\right) ds}
$$
We define $\cC$ to be the coefficient of $\sin{\lambda(\tau-\tau_j)}$ in the above equation. Evaluating the integral by residues then gives
$$
\cC = \frac{B \pi}{4 \sinh{\frac{\pi \lambda}{2}}}.
$$ 
This then simplifies the above formula to
\begin{equation}
R_{\rm after}(\tau) =  -\frac{B}{8\lambda^2} +
\cC\left(c_j \cos{\lambda(\tau-\tau_j)} + (s_j+1) \sin{\lambda(\tau-\tau_j)}\right).
\label{eq:Rafter}
\end{equation}

Next, the change in energy is calculated using
$$
E_{j+1}- E_{j} 
= \int_I \frac{dE}{d\tau} d\tau
= \int_I \left(\ThetaH''(\tau)-\sin{\ThetaH} \right) \ThetaH'(\tau)\, d\tau
= \int_I \left(\frac{8}{B} R(\tau) \sin{\ThetaH(\tau)}\right) \ThetaH'(\tau)\, d\tau,
$$
where $E$ is given by equation~(\ref{eq:E}) and $I$ is the time interval $[\tmf_j,\tmf_{j+1}]$. The integral is approximated by replacing the interval $I$ with the whole line and using the formula~\eqref{eq:Rintegral} for $R(\tau)$. Two more steps complete the calculation: first an integration by parts to move the $\tau$ derivative from $\ThetaH'\sin{(\ThetaH)}$ to $R$, and a residue integral identical to the one performed above. These lead to the following recurrence relationship for consecutive $E$ values
$$
E_{j+1} = E_j - \frac{32 \lambda^2 \cC^2}{B^2} ( 1 + 2 s_j ).
$$

Formula~\eqref{eq:Rafter} is given in terms of $(\tau-\tau_j)$. To complete the specification of the map, we must rewrite it in terms of $(\tau-\tau_{j+1})$, and thus need to find $(\tau_{j+1}-\tau_j)$.  The matching procedure, described in detail in~\cite[Sec. III]{Goodman:2008}, yields the expected result to leading order: it is the time between two successive approaches to $\Theta=\pi \mod{\pi}$ along the periodic orbit of equation~\eqref{scaled_a} with $R$ set to zero and energy $E_{j+1}$. This is found to be
$$
\tau_{j+1}-\tau_{j} \approx \log{\frac{32}{\abs{E_{j+1}}}}.
$$
With this, we find
\begin{equation*}
\binom{c_{j+1}}{s_{j+1}} =
\begin{pmatrix}
\cos{\psi_{j+1}} & \sin{\psi_{j+1}} \\
-\sin{\psi_{j+1}} & \cos{\psi_{j+1}}
\end{pmatrix}
\binom{c_j}{s_j +1},
\text{ where }
\psi_{j+1} = \lambda \left(\tau_{j+1}-\tau_{j} \right).
\end{equation*}
Close inspection shows that the reduced discrete map conserves an energy
\begin{equation*}
\cH =
\frac{B^2}{32 \cC^2\lambda^2}E_j +  \left(c_j^2 + s_j^2 \right).
\end{equation*}
Using this energy, we may eliminate $E_j$ and reduce the dimension of the map 
from three to two. Defining $Z_j = c_j + i \left(s_j - \tfrac{1}{2} \right)$ and
$\Psi(Z) =2 \lambda \log{\frac{\cC\lambda}{B}}+\lambda \log{\abs{\abs{Z}^2-\cH}}$,
the map takes the particularly simple form $Z_{j+1} = \cF(Z_j)$ with
\begin{equation}
\label{Zmap}
\cF(Z) =
 e^{-i \Psi\left(Z+\frac{i}{2}\right)}
\left(Z+\frac{i}{2}\right) + \frac{i}{2}.
\end{equation}
The constant $-\tfrac{i}{2}$ used in defining $Z_j$ is convenient because the inverse map now takes an almost identical form.
\begin{equation*}
\cF^{-1}(Z) =
 e^{i \Psi\left(Z-\frac{i}{2}\right)}
\left(Z-\frac{i}{2}\right) - \frac{i}{2}.
\end{equation*}

When $\cH<0$, the argument in the logarithm in the above functions is bounded below, but for $\cH>0$, the logarithmic term in $\cF(Z)$ and, respectively $\cF^{-1}(Z)$, are singular on the circles
$$
\Gamma_+ =
\left\{ Z \in \CC   : {\abs{Z +\frac{i}{2}}}^2 = \cH \right\}
\text{~and~}
\Gamma_- =
\left\{ Z \in \CC   : {\abs{Z -\frac{i}{2}}}^2 = \cH \right\}.
$$

\subsection{Interpretation of the maps}
Further, let $D_+$ and $D_-$ be the discs interior to these circles, and $D_+^{\rm c}$ and $D_-^{\rm c}$ be their complements. The phase space of the map can have three different configurations depending on $\cH$.
\begin{itemize}
\item \emph{Configuration 1:} When $\cH<0$, the system has too little energy for the pendulum to rotate and all solutions stay in the librating mode. The discs do not exist in this configuration.
\item \emph{Configuration 2:} For $0<\cH<\frac{1}{4}$, the discs are disjoint.
\item\emph{Configuration 3:} Finally, for $\frac{1}{4}<\cH$, the two discs have nontrivial intersection.
\end{itemize}
In Configuration 1,  generalize the definitions of the disks to 
$D_+ = D_- = \emptyset$ 
and their complements $D_+^{\rm c}=D_+^{\rm c} = \CC$.  With this notation, if $Z_j \in D_-$, then the pendulum is outside the separatrix (rotating) on iterate $j$, whereas it is inside (librating) if $Z_j  \in D_+^{\rm c}$. Similarly, if $Z_j \in D_+$, the pendulum rotates on step $j+1$, and for $Z_j \in D_+^{\rm c}$ it executes a libration at step $j+1$. In configuration 2, the empty intersection of the two disks indicates that at these energy levels, the pendulum mode can  be in the rotational state for at most one iteration at a time before returning to the libration state, while in configuration 3, the solution may stay in the rotational mode arbitrarily long.

\subsection{Fixed points and periodic points}
\label{sec:fixedpts}
As an example of the type of calculation that can be accomplished using this map, we note that fixed points of map~\eqref{Zmap} correspond to periodic orbits of the ODE system. Clearly, solutions to the fixed points of the map~\eqref{Zmap} satisfy $\abs{Z_{j+1}-\frac{i}{2}}=\abs{Z_j + \frac{i}{2}}$, so that a fixed point $Z^*$ satisfies $Z^* = X \in \RR$. Using this value in the map yields an implicit formula for these fixed points that can be solved numerically:
\begin{equation}
X = -\frac{1}{2} \cot{\frac{\Psi\left(\sqrt{X^2+\frac{1}{4}}\right)}{2}}.
\label{eq:fixpt}
\end{equation}
Note that $R(t)$ is assumed small enough that it satisfies a linear ODE, and that the magnitudes of $X(t)$ and $R(t)$ are of the same size up to the scaling constant $\cC$, but that this equation has a countably infinite and unbounded family of solutions. Therefore large solutions fail to satisfy assumption~\eqref{scalingAssumption} and likely do not correspond to actual solutions of the ODE from which the map was derived.

 We also find two families of period-two points. The first family, which we call type-1, consists of pairs of complex points $Z_1$ and $Z_2$ that are complex conjugates of each other: $Z_2 = \bar{Z}_1$. In the second family, type-2 period-two points, $Z_1$ and $Z_2$ are real, with $Z_2=-Z_1$. Again, we can find explicit values of $\cH$ where these families arise in period doubling bifurcations. While we do not write down these families of solutions, we will draw a partial bifurcation diagram below in Section~\ref{sec:bifurcationdiagram}.

\section{Numerical experiments}
\label{sec:numerics}

We turn now to numerical studies, guided by the analysis presented above. We begin with two numerical simulations of the initial value problem, with initial conditions chosen to display behavior in the interesting regimes already identified. We then use Poincar\'e sections to place these simulations within a wider view of the dynamics, showing how chaotic dynamics coexists with periodic and quasi-periodic dynamics when the anisotropy parameter $\epsilon$ is non-zero. Finally, we discuss the bifurcations of periodic orbits based on the iterated map derived in Section~\ref{sec:iterated_map}.

\subsection{Initial-value problem simulations}
Figure~\ref{fig:phaseplanes} shows two distinguished regions in the phase space of the isotropic $\epsilon=0$ system: near the separatrix between librating and rotating periodic orbits and near the guiding center fixed point $(\theta_2^*,\rho_2^*)$.  By performing simulations with initial conditions near these regions, we are able to find interesting dynamics.

\begin{figure}[htbp] 
   \centering
   \includegraphics[width=9.0cm]{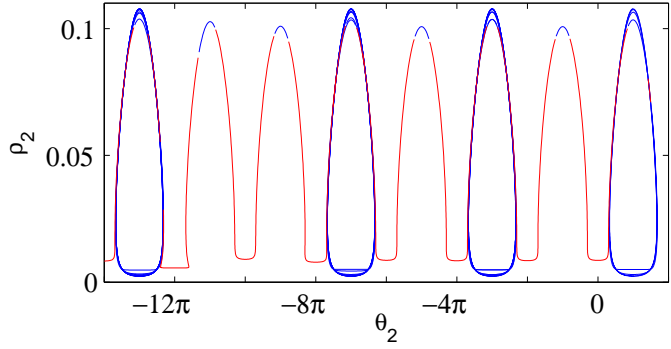}
   \caption{
(color online)
The $(\theta_2,\rho_2)$ component of the motion near the separatrix of Figure~\ref{fig:phaseplanes}. The solution is plotted in red when the instantaneous energy $H_0(\theta_2,\rho_1,\rho_2)$ lies below the separatrix value and in blue when it lies above.}
\label{fig:theta2rho2}
\end{figure}

\paragraph{Near the separatrix} First, we show the dynamics in a neighborhood of the separatrix  that exists when $\rho_1 \neq 0$. We initialize the solution with parameters $B=0.22$ and $\epsilon=0.2$, and with initial conditions close to the point $(\theta_2,\rho_2 )= (3\pi/2,\abs{\rho_1})$ where  the separatrix orbit leaves the bottom edge of the reduced phase diagram. The actual initial condition is $(\theta_1,\theta_2,\rho_1,\rho_2) = (4.7421,  4.7068, -0.0050, 0.0052)$, which lies on the energy level $H=0.3$, for later reference. A portion of the projection of the phase plane onto the coordinates $(\theta_2,\rho_2)$ is shown in Figure~\ref{fig:theta2rho2}. The solution jumps between the interior of the separatrix, and the exterior several times, in a manner we show below to be chaotic.

\begin{figure}[htbp] 
   \centering
   \includegraphics[width=5.0cm]{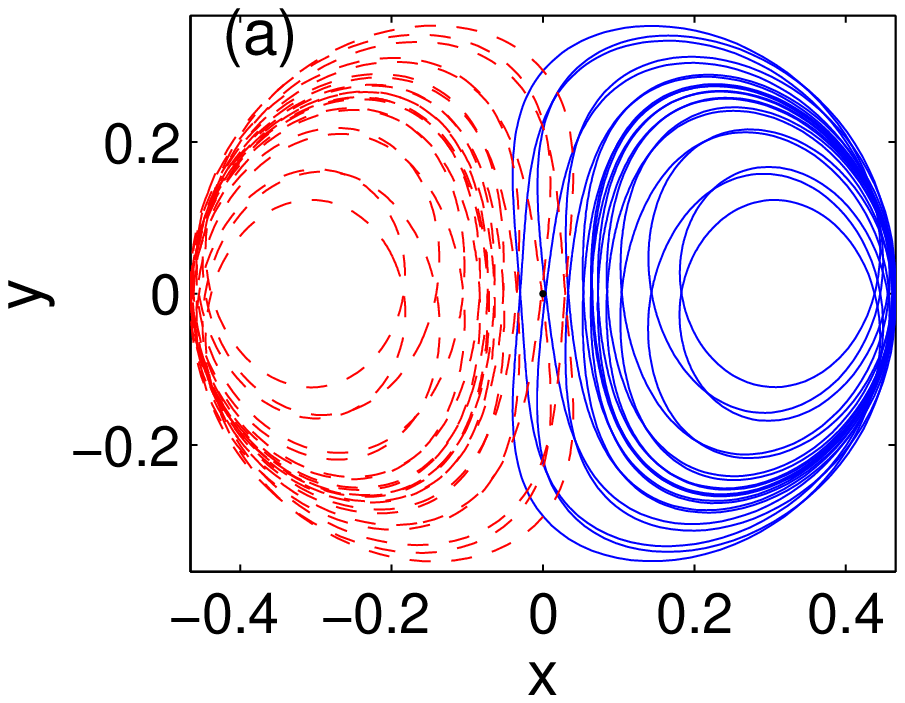}
   \includegraphics[width=5.0cm]{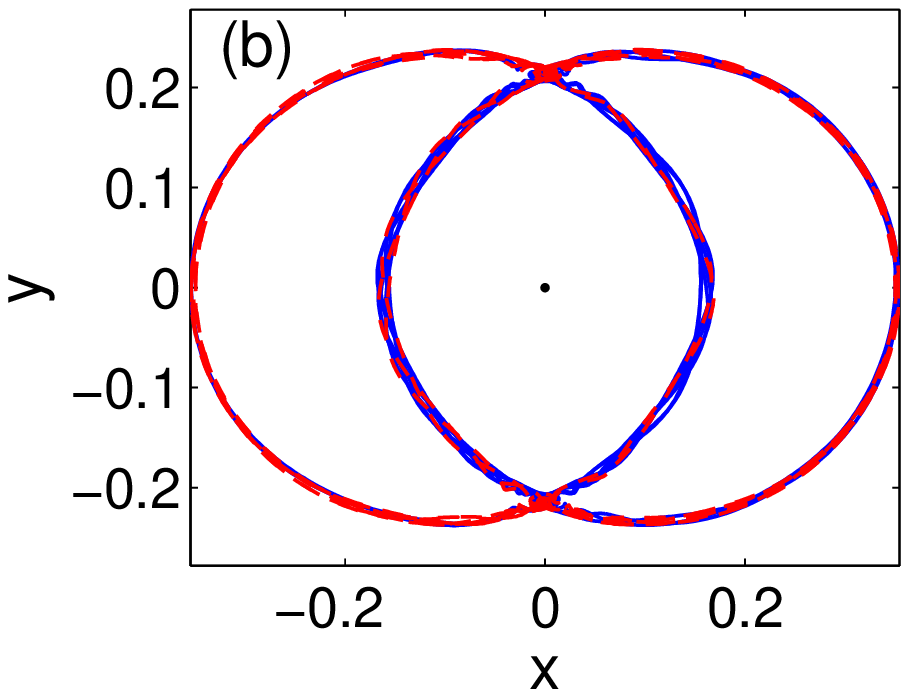}
   \caption{
(color online)
The reconstructed particle trajectories from (a)~the near-separatrix orbit, 
and (b)~the near guiding center.}
\label{fig:reconstruction}
\end{figure}

To explain the dynamics of the vortices, we show the solutions in the position variable $(x_1,x_2,y_2,y_2)$ obtained by undoing the repeated change of coordinates performed in Section~\ref{sec:formulation}. A small portion of this numerical trajectory is shown in Figure~\ref{fig:reconstruction}a. The dynamics is reminiscent of the $\epsilon=0$, $\rho_1=0$ case shown in Figure~\ref{fig:particle_paths_reduced}a. While that figure represents the relative motion of the vortices from a rotating reference frame, the present figure is in the fixed reference frame. In the unperturbed case, the two particles' positions are mirror images of each other across the $y$-axis, while for the perturbed problem, there is some small deviation from this symmetry. The perturbed orbits are chaotic and do not close. Occasionally one vortex or the other encircles the origin. This happens precisely when the position $(\theta_2,\rho_2)$ crosses to the outside of the separatrix shown in Figure~\ref{fig:theta2rho2}.

\begin{figure}[htbp] 
   \centering
   \includegraphics[width=14.0cm]{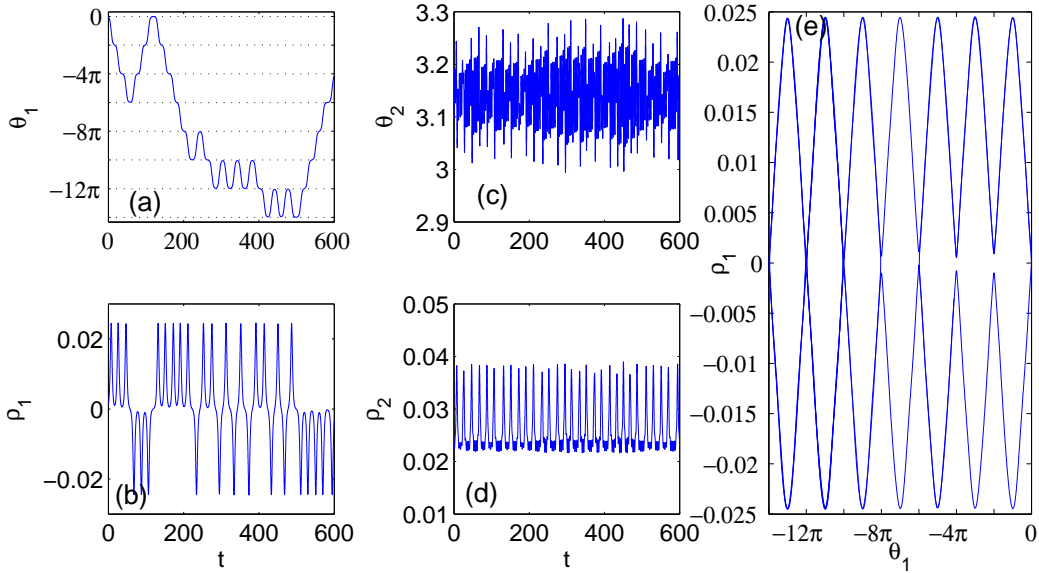}
   \caption{
(color online)
Numerical solution of initial value problem near the guiding center orbit. \textbf{(a)-(d)}, evolution of the components of the solution, \textbf{(e)}~phase portrait projected to $(\theta_1,\rho_1)$ coordinates.}
\label{fig:nearPendulumSeparatrix}
\end{figure}

\paragraph{Near the guiding center}
Figs.~\ref{fig:reconstruction}b and~\ref{fig:nearPendulumSeparatrix} show a simulation of dynamics near the guiding-center fixed point. The parameters $B$ and $\epsilon$ are as in the previous simulation, and the initial conditions chosen are $(\theta_1,\theta_2,\rho_1,\rho_2)=(0.11, \pi, 0, 0.024)$, which lie close to an unstable fixed point of the Poincar\'e map described in the next section. In this figure, note that $\theta_1$ switches chaotically between three behaviors: monotonically increasing, monotonically decreasing, and oscillating. The sign of $\rho_1$ is opposite that of $d \theta_1/dt$. Also note that $\theta_2\approx \pi$ to within $0.15$ radians (about~$9^\circ$), and the variation of $\rho_2$ is much smaller than that of $\rho_1$, indicating that the two vortices remain nearly collinear with, and on opposite sides of, the center of the magnetic trap. The phase plane diagram shows the $(\theta_1,\rho_1)$ dynamics stay close to what appears to be a pendulum separatrix, consistent with the reduced system~\eqref{eq:Theta_r}.

Referring to the reconstruction of the laboratory-frame trajectories in Figure~\ref{fig:reconstruction}b, we describe the dynamics of the two vortices. The vortices remain nearly opposite each other. When they approach the unstable solution $\xY$ on the $y$-axis, they slow down. From the reduced system~\eqref{eq:Theta_r}, the potential energy is high at this point, so the kinetic energy is small. Depending on their kinetic energy, they may either turn around or else may, after a pause, continue in the direction they were headed. Depending on which orientation they are rotating, one vortex will move on the ``inside track'' closer to the origin and the other along the ``outside track.''

\subsection{Poincar\'e sections}

The results of a few initial-value simulations are insufficient to understand the dynamics and can be put into better context by plotting Poincar\'e sections. We define the Poincar\'e section to be the set of intersections of a given trajectory with the hyperplane $\theta_2 = (2n+1)\pi$ that cross with $\theta_2'(t)<0$. This corresponds in the limiting case $\epsilon=0$ to a half-line extending upward from $\rho_2= \rho_2^*$ in Figure~\ref{fig:phaseplanes}. In order to plot multiple trajectories in the same figure, all trajectories must be chosen to lie on the same level set of the full Hamiltonian $H^*=H(\theta,\rho;\epsilon)$. Taking $(\theta_1,\rho_1)$ as parameterizing the section, this implicitly fixes the value of $\rho_2$. For a given value of $H^*$, the set of accessible coordinates $(\theta_1,\rho_1)$ may be significantly smaller than the natural domain $\theta_1 \in S^1$, $\abs{\rho_1}\le \rho_2$.

Reducing the dynamics from system~\eqref{eq:xy} to system~\eqref{Hrhotheta} and then 
to the Poincar\'e section in two dimensions increases the density of information 
in the figure. Complete understanding of the two-vortex dynamics, implied 
by the Poincar\'e section, requires ``undoing'' the corresponding reductions.  
For $H \ge H(\xY)$, the Poincar\'e map has a fixed point $(\theta_1,\rho_1)=(0,0)$ which corresponds to the unstable fixed point $\xY$ when $H= H(\xY)$  and to a periodic orbit surrounding it otherwise, corresponding to the periodic orbits of Figure~\ref{fig:particle_paths_reduced}a with the orientation of the plot rotated by $90^\circ$. These fixed points are always unstable as is $\xY$. For $H \ge H(\xX)$, the Poincar\'e map has a fixed point at $(\pi,0)$ corresponding to the stable equilibrium $\xX$ at $H=H(\xY)$ and the periodic orbits surrounding it otherwise.
As $H$ decreases, the region of the $(\theta_1,\rho_1)$ plane accessible to intersect with physical orbits of system~\eqref{Hrhotheta} shrinks. For $H< H(\xY)$, the fixed point at $(0,0)$ disappears, and for $H<H(\xX)$, the Poincar\'e section is empty.

\begin{figure}[t] 
   \centering
   \includegraphics[width=6.0cm]{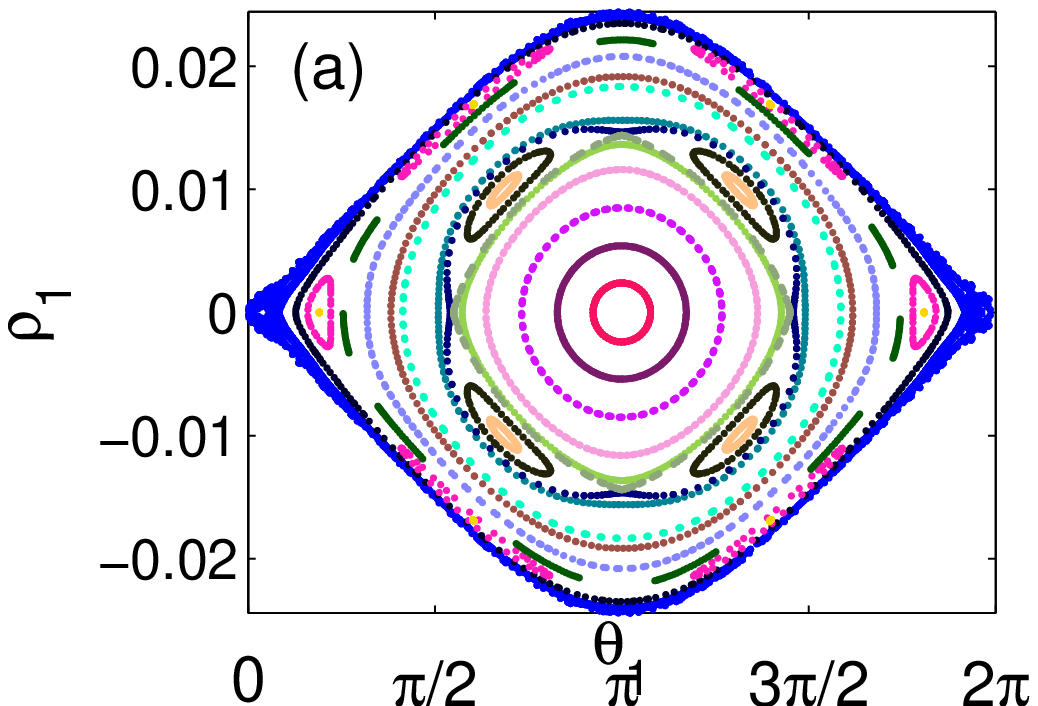}
   \includegraphics[width=6.0cm]{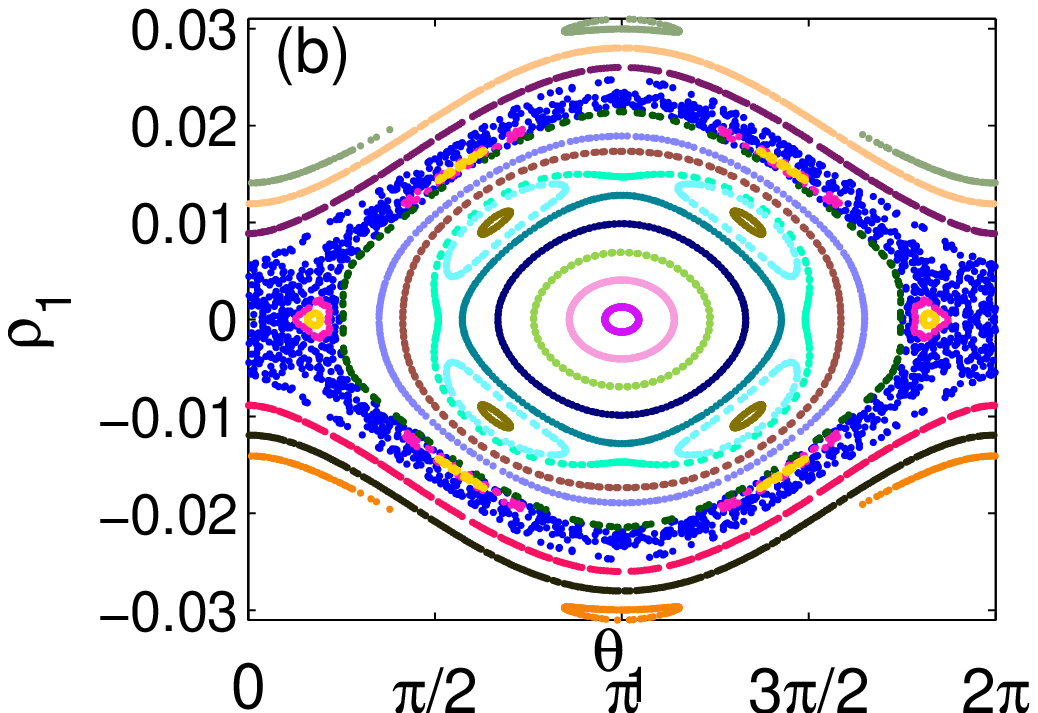}
   \includegraphics[width=6.0cm]{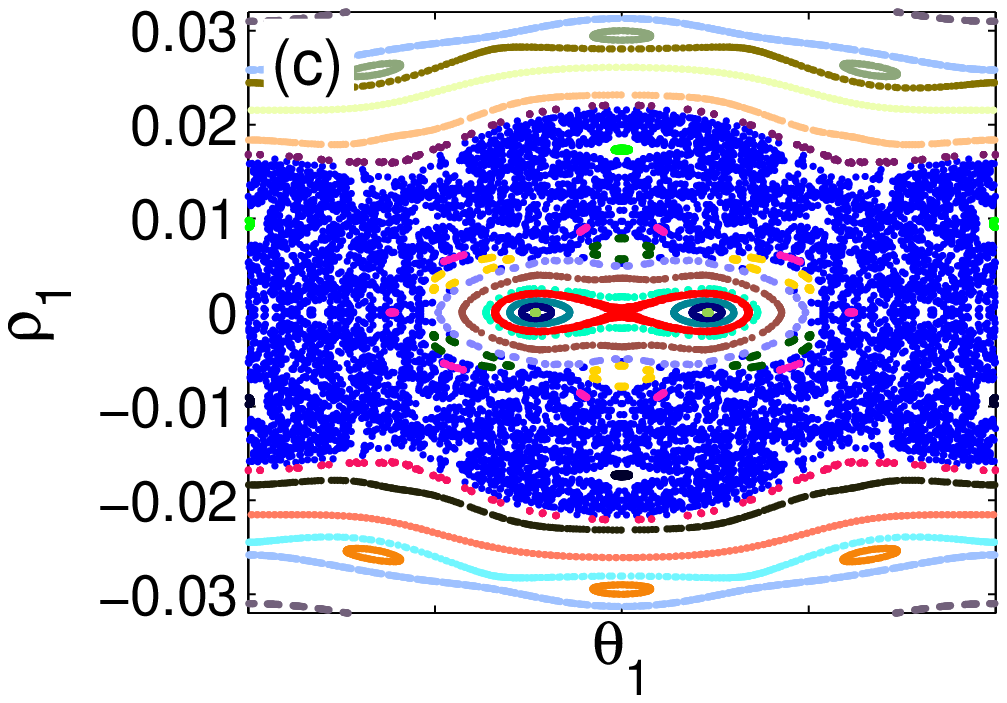}
   \includegraphics[width=6.0cm]{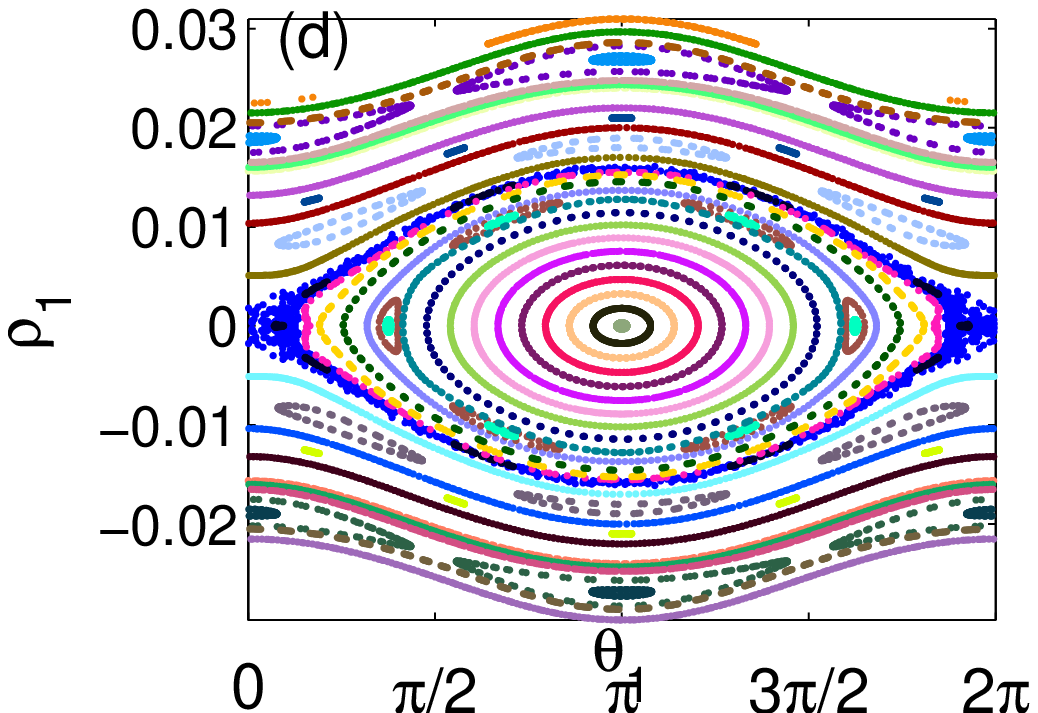}
   \caption{
(color online)
The Poincar\'e sections on the hyperplane $\theta_2\equiv\pi\mod{2\pi}$ with parameter $B=0.22$ in all sections and $\epsilon=0.2$ in all but panel~(d) where $\epsilon=0.1$. 
The sections lie on the energy levels (a)~$H = 0.2246$, 
(b)~$H = 0.23$, (c)~$H = 0.3$, and (d)~$H = 0.23$. 
The value of $H$ restricts the allowable values of $\rho_1$, which is easily 
seen in all panels but~(c).}
\label{fig:poincare}
\end{figure}

Figure~\ref{fig:poincare} shows four  Poincar\'e sections computed  from numerical simulations, with each set of like-colored points coming from the same computed orbit. The first three panels have $\epsilon=0.2$. As the energy is decreased, and $\rho_1$ is held fixed, the accessible region of the section decreases, and for $H \lesssim 0.2245$,
the accessible interval of $\theta_1$ values does not include the entire interval $[-\pi,\pi]$ for any value of $\rho_1$. Panel (a) shows $H=0.2246$ just above $H(\xY)$. The Poincar\'e section looks like a perturbed pendulum phase plane, consistent with the reduction to equation~\eqref{eq:Theta_r}, and the accessible region contains the interior and a small portion of the exterior to the pendulum separatrix. There is a thin region of chaotic dynamics, in blue, containing the separatrix, and several families of period-$n$ points, surrounded by quasi-periodics can be seen, most notably a period-4 orbit and a period-6 orbit. As $H$ is increased in the next two panels, the stochastic layer grows, until in panel~(c) the underlying pendulum dynamics is difficult to discern. More interesting in that panel is the bowtie-shaped region, plotted in red, containing the fixed point $(\pi,0)$. What appears to be a curve has nonzero thickness is actually a narrow band of chaotic trajectories. This point corresponds to a periodic orbit surrounding $\xX$. As $H$ is increased, this periodic orbit undergoes a pitchfork bifurcation, and for energies above this bifurcation value, the periodic orbits are unstable.  The critical value of $H$ at which this instability arises could be found numerically by computing the Floquet multipliers of this family of periodic orbits. Finally panel~(d), which lies on the same energy level as~(b), but with a smaller anisotropy $\epsilon$ has smaller stochastic zones and more regular dynamics, as is to be expected.

\subsection{Bifurcations in the discrete map approximation}
\label{sec:bifurcationdiagram}
\begin{figure}[t] 
   \centering
   \includegraphics[width=\textwidth]{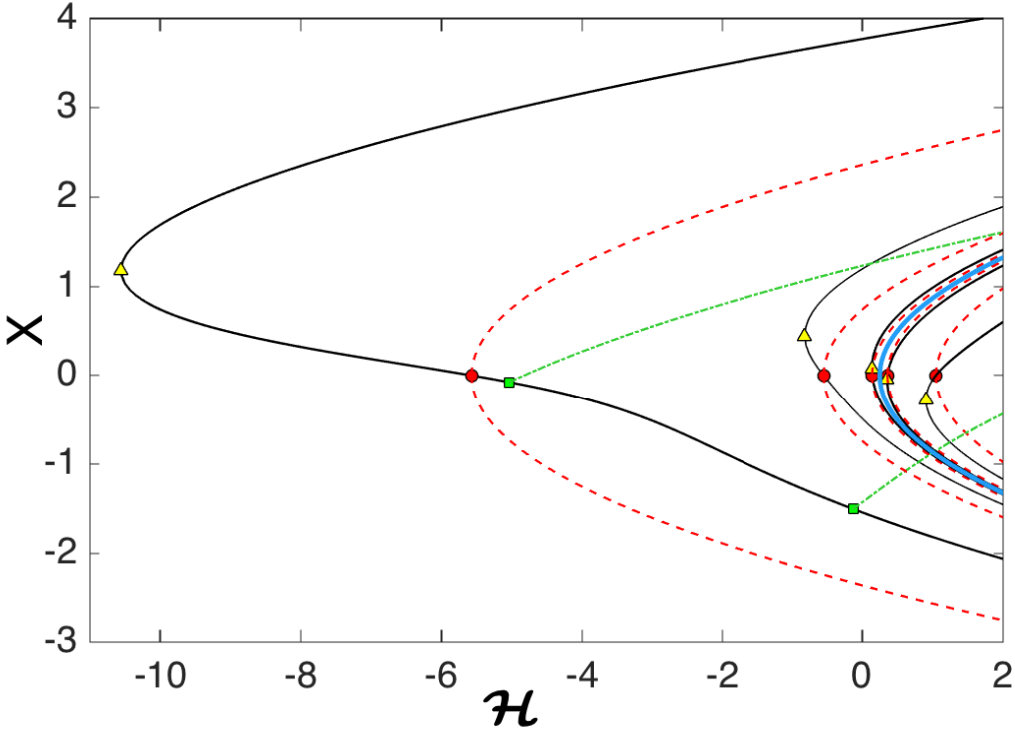}
   \caption{
(color online)
A partial bifurcation diagram for map~\eqref{Zmap}. It shows the separatrix curve $\cS$ (light blue), fixed point branches (black line), type-1 period-two points (green dash-dot, real parts only), type-2 period-two points (red dashed), saddle node bifurcations (yellow triangles), type-1 period-doubling points (green squares), and type-2 period-doubling points (red circles).
}
\label{fig:bifurcationDiagram}
\end{figure}

The separatrix map~\eqref{Zmap} approximates the ODE dynamics in a neighborhood of the separatrix visible in Figure~\ref{fig:poincare}, although it is based on a different method of reduction.
Using the same values of $B$ and $\epsilon$, we include a partial bifurcation diagram for the map~\eqref{Zmap} in Figure~\ref{fig:bifurcationDiagram}, computed using the Matlab continuation package MATCONT~\cite{Dhooge:2003}. It shows five branches of fixed-points, which are real-valued following equation~\eqref{eq:fixpt}. The most salient feature of this diagram is the separatrix curve $\cS$ given by $\cH=X^2 +\frac{1}{4}$, plotted with a light blue line. The fixed points form two countable families of branches that accumulate along $\cS$ from both sides. In addition, there are more solution branches lying outside the curve $\cS$ that cannot be considered physical because they sit too far from the separatrix  to justify the assumption~\eqref{scalingAssumption} under which the map was derived. Points inside the parabola correspond to orbits of system~\eqref{eq:Theta_r_scaled} lying \emph{outside} the pendulum separatrix and points outside the parabola to orbits \emph{inside} the separatrix.

We also plot some of the branches of period-two orbits described in Section~\ref{sec:fixedpts}. The type-1 periodic orbits are complex-valued, but arise in period-doubling bifurcations of real-valued fixed points. The type-2 fixed points are real, and bifurcate wherever a branch of fixed points crosses the axis $X=0$.
The form of the exact period-two points shows that there are no  orbits that move from the inside of the separatrix to the outside. All period-two and period-one orbits either remain on the outside, with the angle $\theta_2$ changing monotonically, or on the inside, with the solutions remaining in an area near the stable equilibrium $\xX$.

\subsection{Systematic enumeration of ODE periodic orbits}

The reasoning from Ref.~\cite{Goodman:2008} can be used to show that the separatrix map is closely connected with the Poincar\'e section of system~\eqref{Hrhotheta} when $\theta_1=\pi \mod{2\pi}$. The real fixed point $\xX$ corresponds to a periodic solution for which $\theta_2=\pi$ along this section. This leads us to the following strategy. We fix the Hamiltonian $H = H^*$ of equation~\eqref{Hrhotheta}. At this value of $H^*$, the action $\rho_1$ can take values on a finite interval $[-\rho_1^*,\rho_1^*]$. This interval is empty for $H^*$ below some critical value. We numerically sweep through values of $\rho_1$, following solutions of system~\eqref{Hrhotheta} with initial conditions $\theta_1=\theta_2=\pi$ and $\rho_2$ chosen to make $H=H^*$. For each $\rho_1 \in (-\rho_1^*,\rho_1^*)$, there exist two such values of $\rho_2$. Call them $\rho_2^+$ and $\rho_2^-$, and assume that $\rho_1<\rho_2^-<\rho_2^+$. We run the ODE simulation until $\theta_1 = \pi \mod2\pi$. If, at this point, $\theta_2 = \pi$, then the numerical solution represents one half of a periodic orbit. In the $(x_j,y_j)$ coordinates, these solutions have $y_1(0)=y_2(0)=0$, and $x_1(0)\cdot x_2(0)<0$.

\begin{figure}[t] 
   \centering
   \includegraphics[width=.75\textwidth]{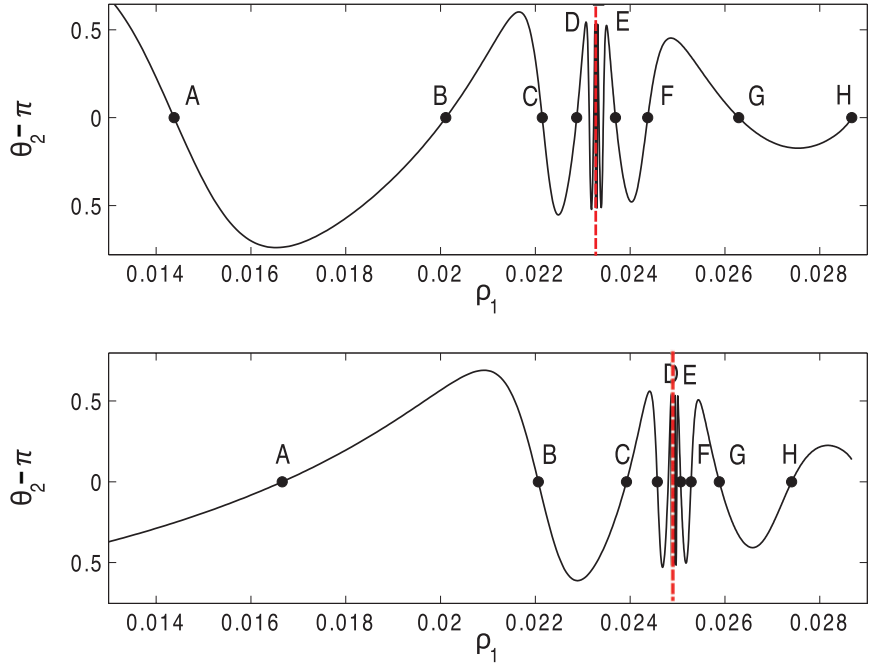}
   \caption{
(color online)
The value of $\theta_2(T)-\pi$ when the solution to system~\eqref{Hrhotheta} is on the Poincar\'e section $\theta_1=\pi$. As described in the text, zeros of this function determine symmetric periodic orbits of the system. Top: initial condition $\rho_2(0) = \rho_2^-$. Bottom: initial condition  $\rho_2(0)=\rho_2^+$. This curve is defined for $-\rho_1*< \rho_1< \rho_1^*$, but is symmetric about $\rho_1=0$, and has no zeros on the interval $(0,0.014)$.
}
\label{fig:in_out_m228}
\end{figure}

We present the results of this experiment beginning in Figure~\ref{fig:in_out_m228}, which shows the output value of $\theta_2$. The top panel shows the results with $\rho_2(0)=\rho_2^+$, and the bottom panel has $\rho_2(0)=\rho_2^-$. These curves oscillate infinitely often as  $\rho_1$ approaches $\rc^{+} \approx 0.0233$ in the upper panel and $\rc^{-} \approx 0.0249$ in the lower.
The  energy level $H^*$ has been chosen large enough such that the branch crosses the critical curve $\cS$ in the bifurcation diagram. Thus, it has infinitely many real fixed points, and two types of periodic orbits: rotations, which lie to the right of $\rc$, and librations, which lie to the left of $\rc$.

\begin{figure}[t] 
   \centering
   \includegraphics[width=3.80cm]{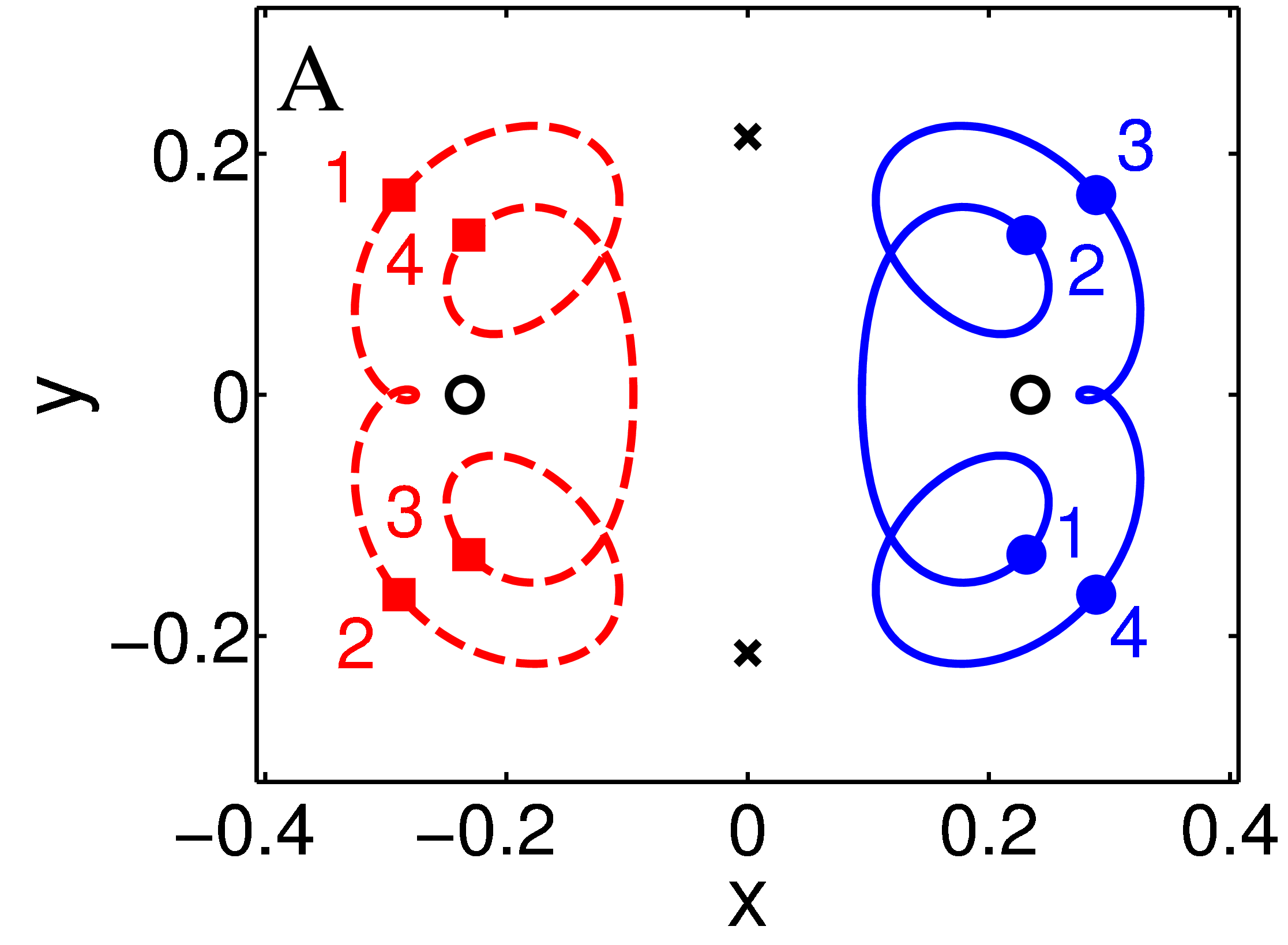}
   \includegraphics[width=3.80cm]{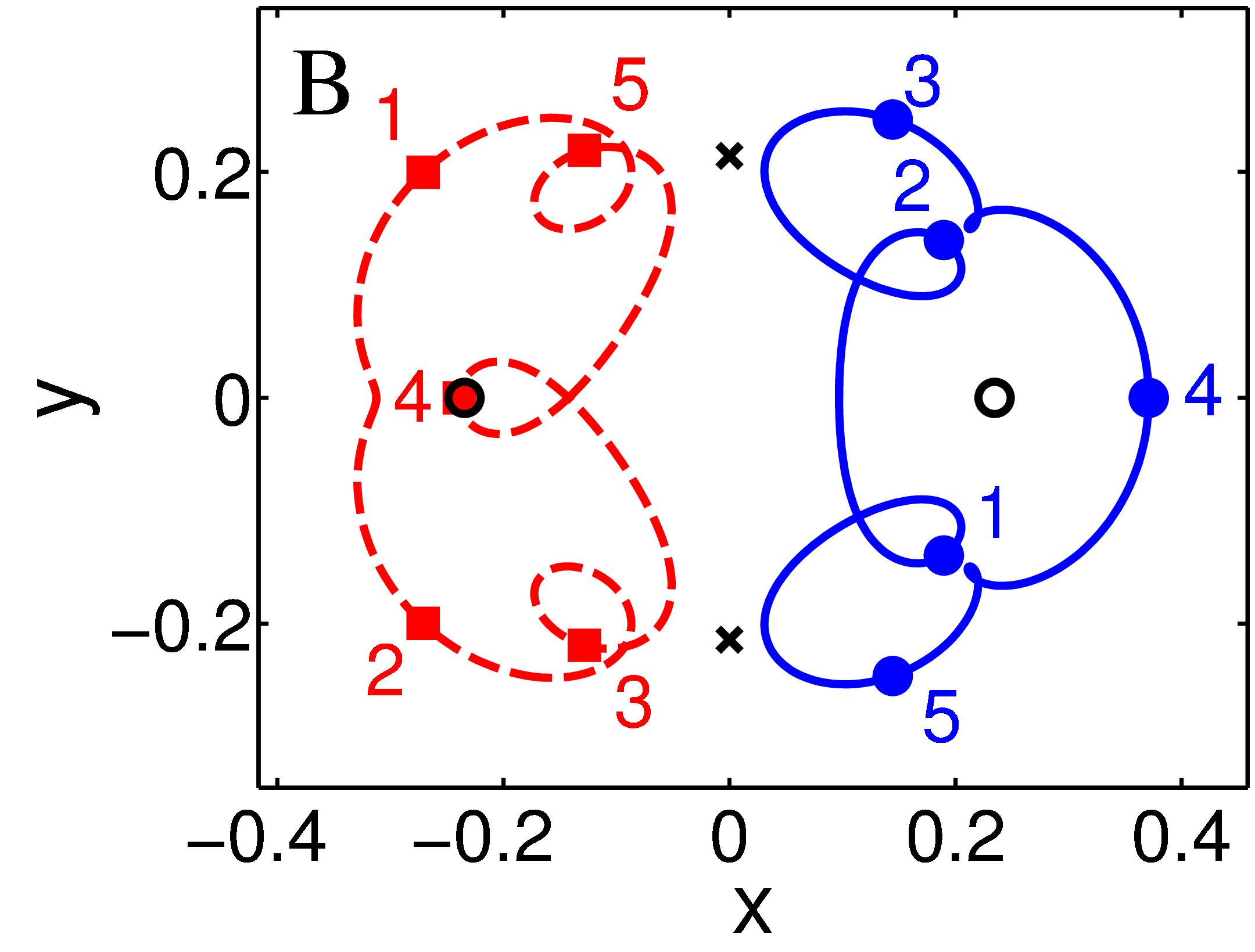}
   \includegraphics[width=3.80cm]{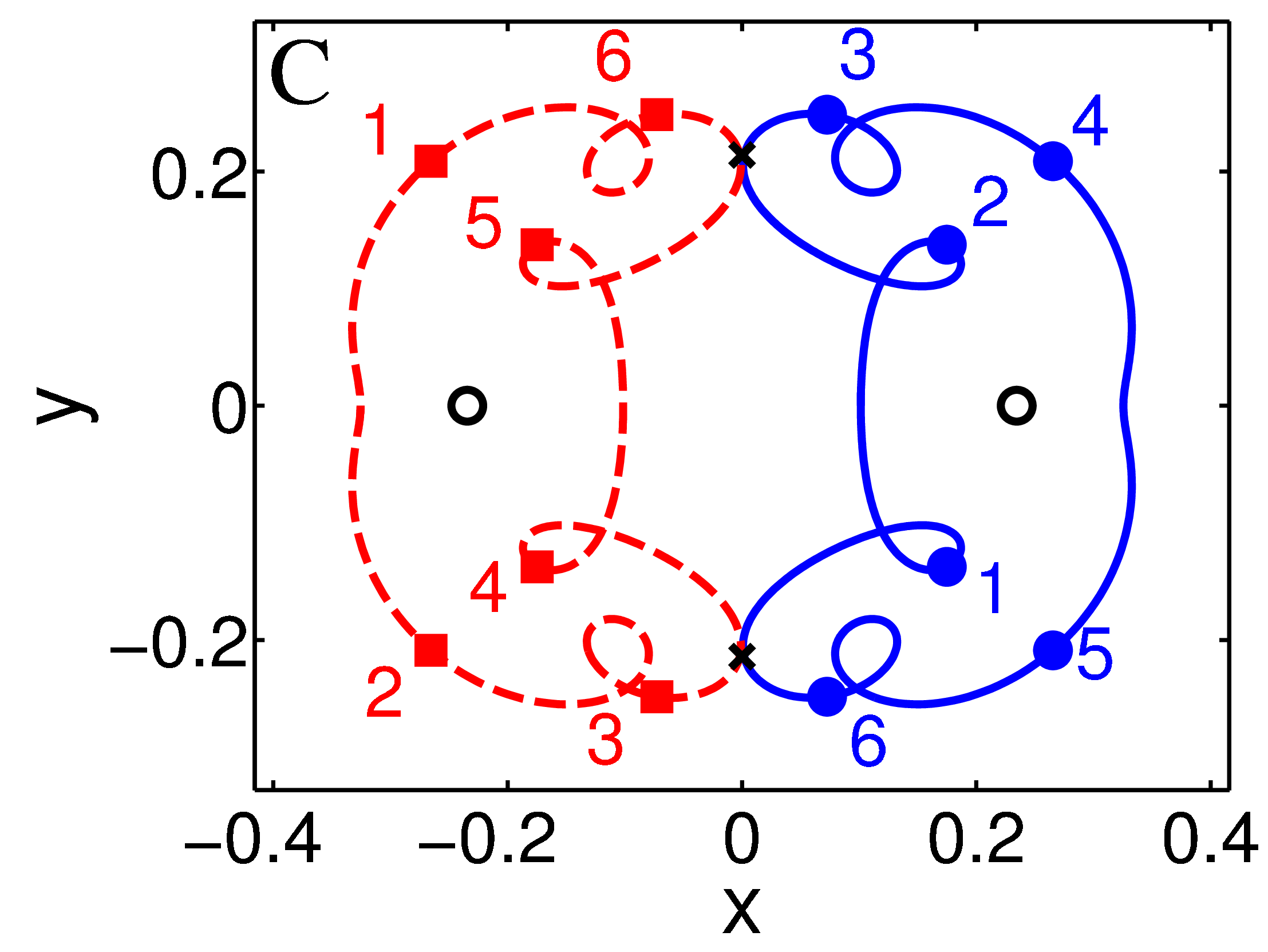}
   \includegraphics[width=3.80cm]{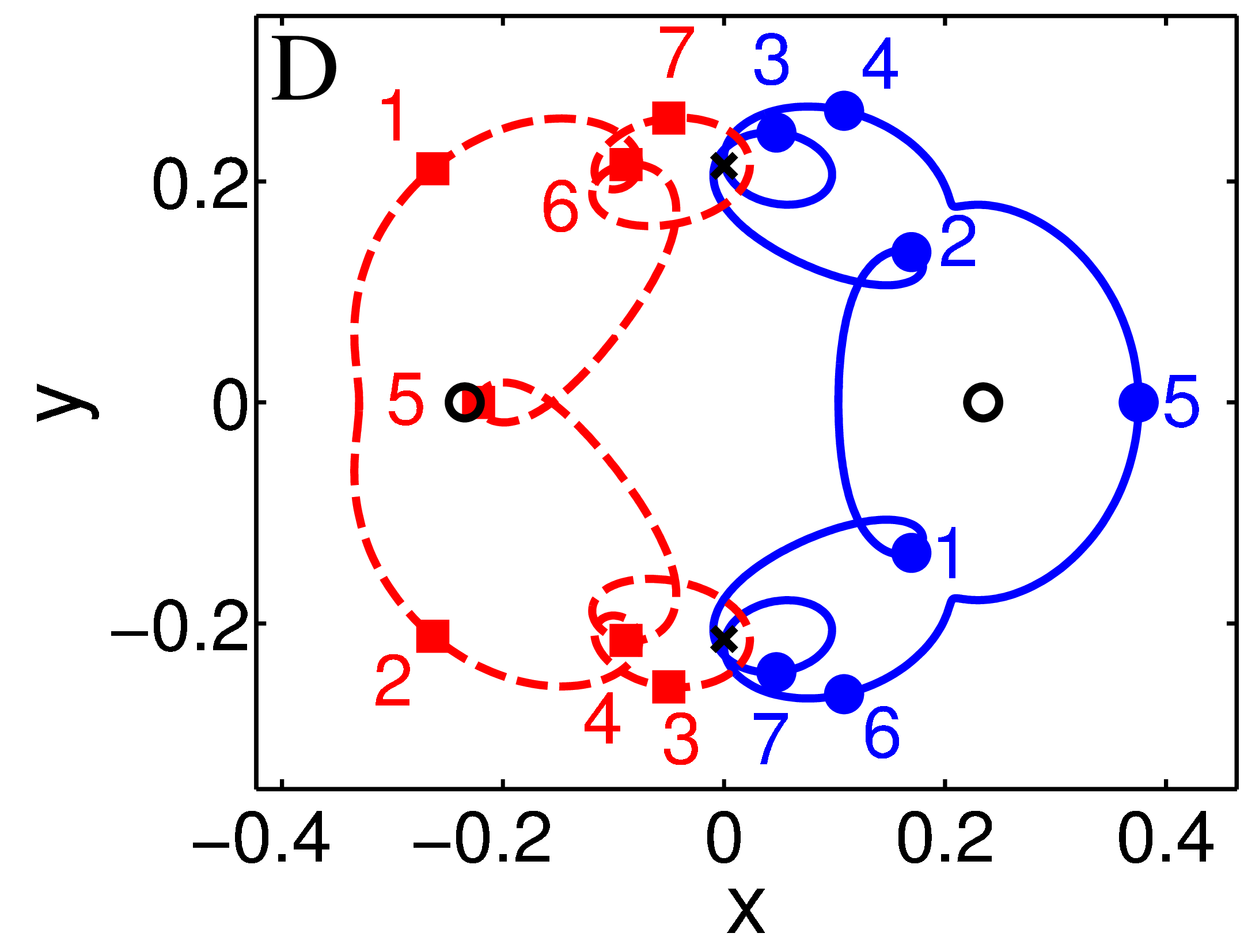} \\
   \includegraphics[width=3.80cm]{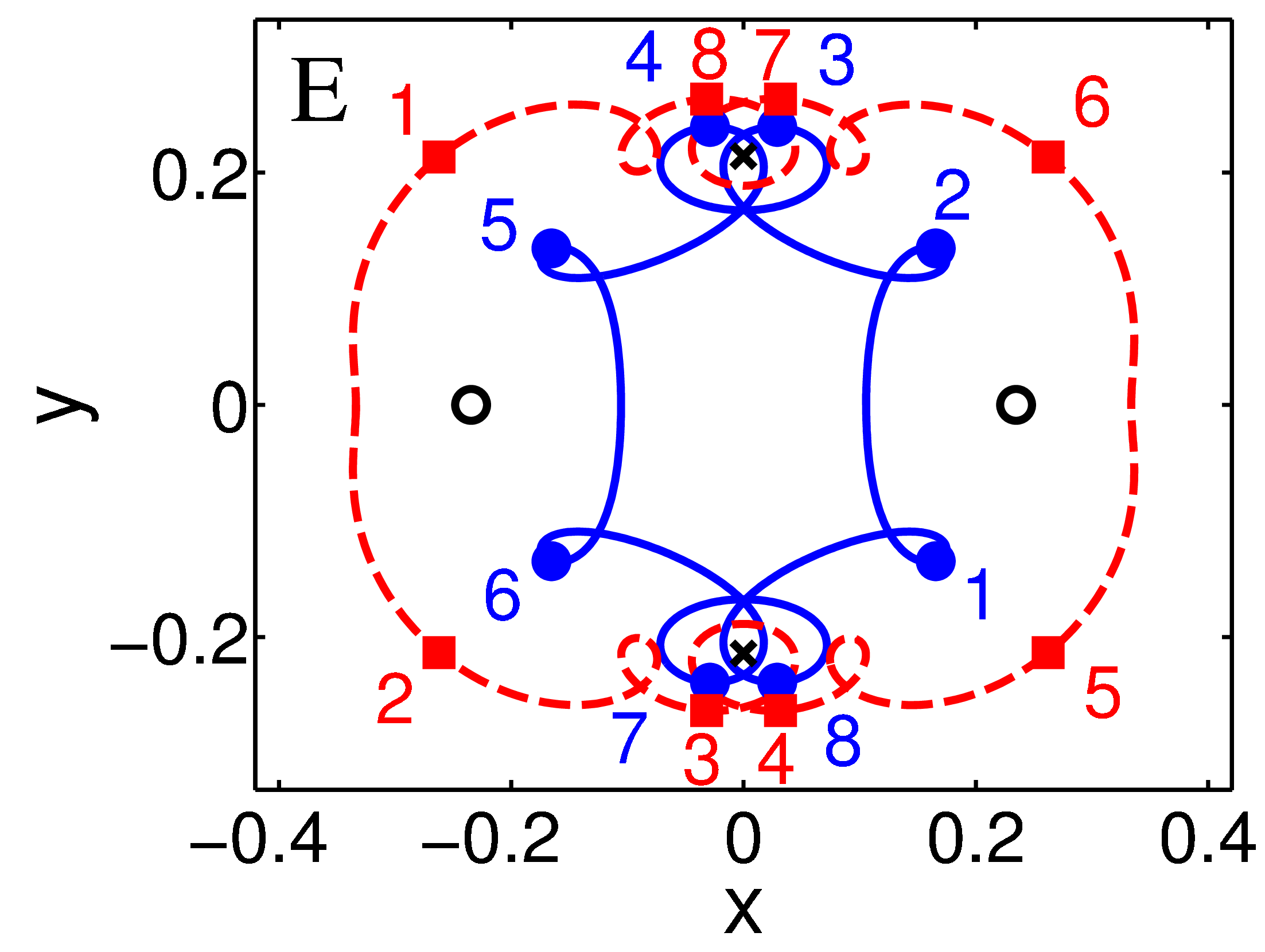}
   \includegraphics[width=3.80cm]{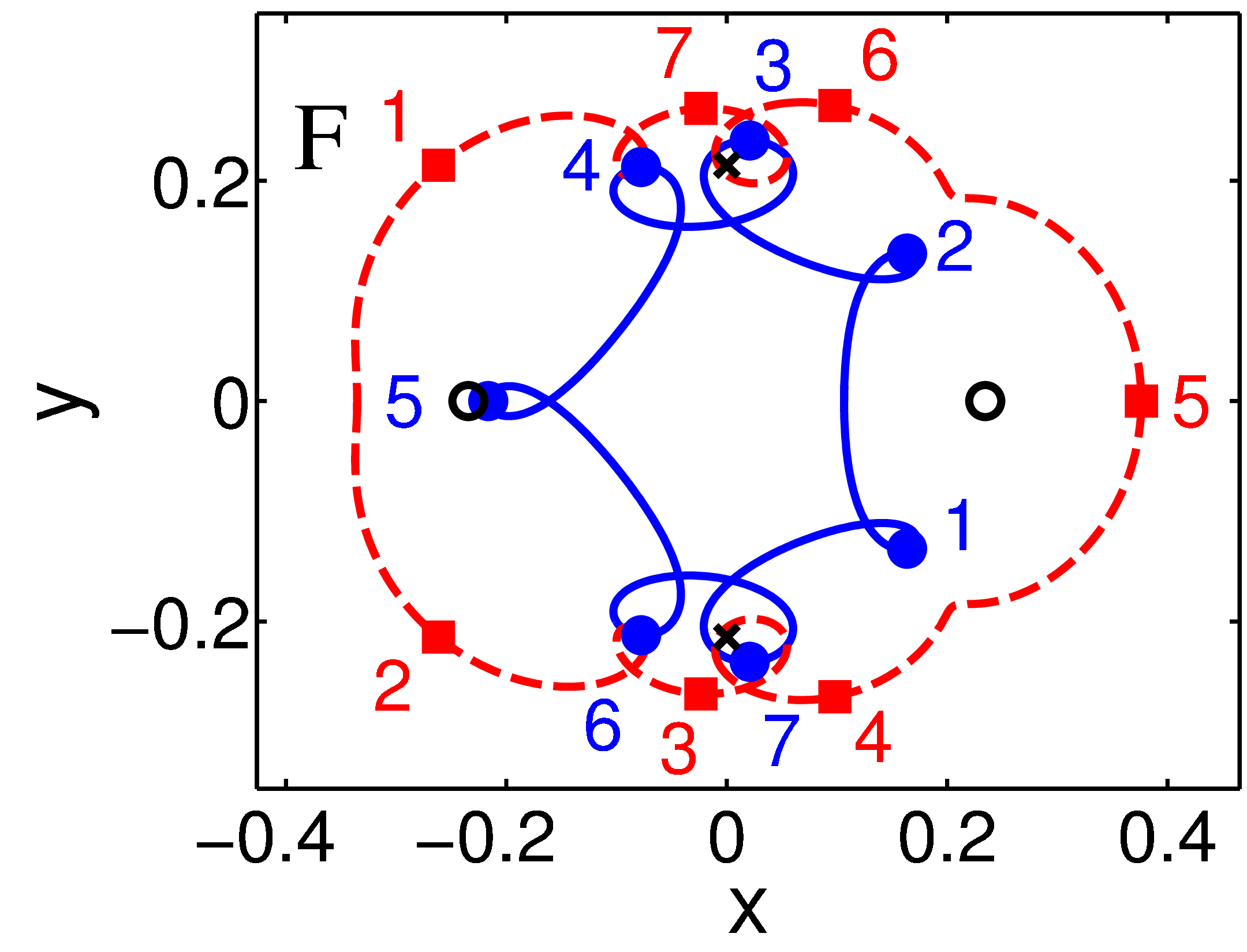}
   \includegraphics[width=3.80cm]{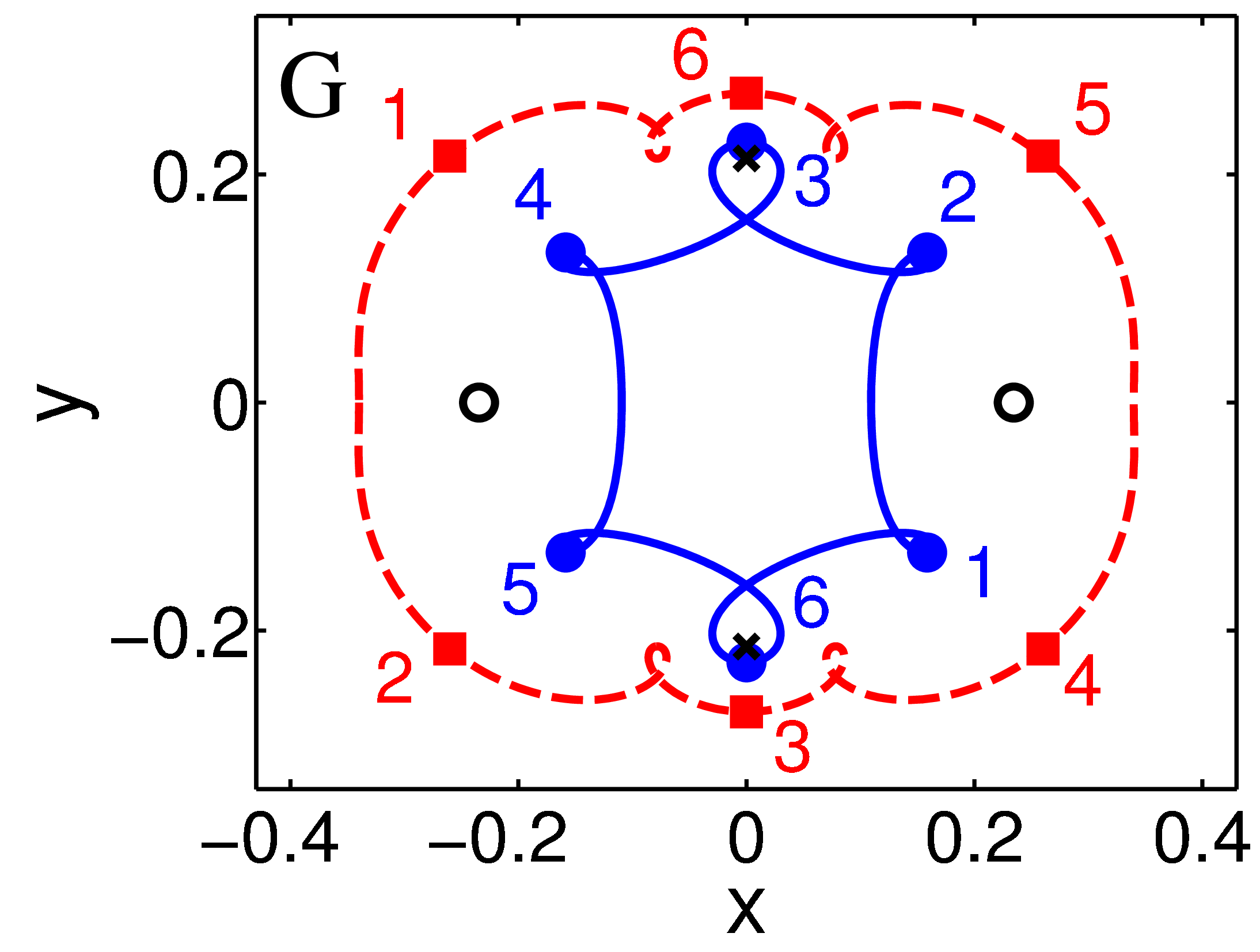}
   \includegraphics[width=3.80cm]{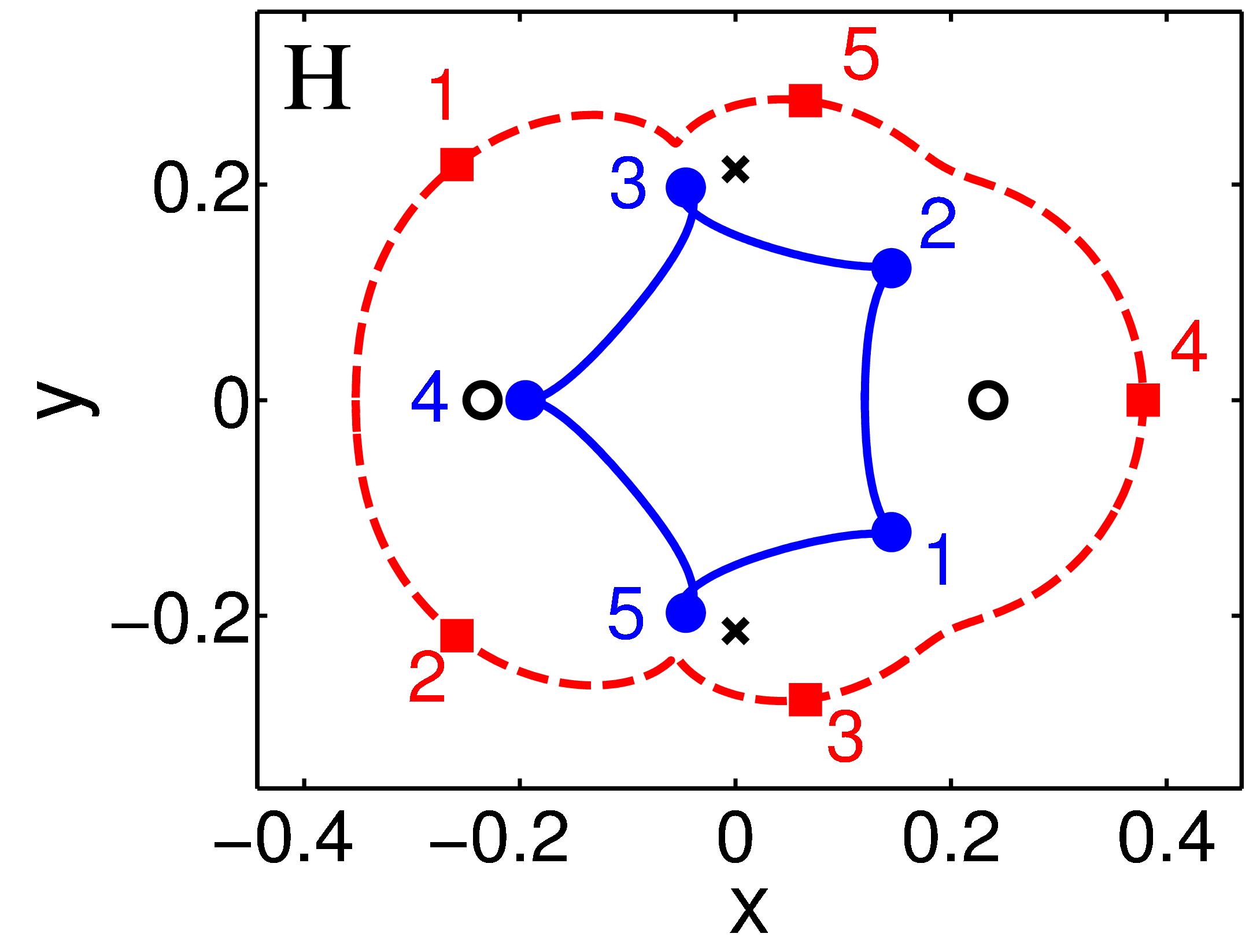}
   \caption{
(color online)
Periodic orbits of system~\eqref{Hrhotheta} with $H^* = 0.228$ and
the larger value of $\rho_2$ chosen as the initial condition. 
Note the $\rho_1$-interval on which this is defined is about $[-0.029,0.029]$,
but the curve is symmetric about zero, and has no roots between
$\rho_1=0$ and the intersections marked `A' (see Figure~\ref{fig:in_out_m228}).
The points depicted by the black empty circles and crosses correspond
to the two pairs of vortex dipole steady state solutions
of equation~(\ref{fixedpoints}).
}
\label{fig:m228}
\end{figure}

Figure~\ref{fig:m228} shows periodic orbits with initial condition $\rho_2^-$. Panels A to D show librations, and panels E through H, rotations.  Plotted on these same axes are the places where the numerical solution intersects the Poincar\'e section. As we move from solution to solution with $\rho_1 \to \rc^-$, the number of intersections with the $\theta_2$ increases by one at each step, and similarly as $\rho_1 \to \rc^+$. The period of the orbit and the number of intersections diverges in this limit. In  simulations
where $\rho_1$ is chosen closer to $\rc$ (not shown), the trajectories spend more and more time close to the unstable fixed points.

In panels A, C, E, and G, the number of points of section is even, while in the other four, the number of intersections is odd. In the librations with an even number of intersections (A and G), the two vortices execute orbits that are identical up to a $180^{\circ}$ rotation and a $180^{\circ}$ phase shift. These orbits correspond to fixed points~\eqref{eq:fixpt} of the map~\eqref{Zmap}. Periodic orbits with odd numbers correspond to real period-two points of the same map. For these orbits, the trajectories of the two vortices differ substantially.

\begin{figure}[htbp] 
   \centering
   \includegraphics[width=3.80cm]{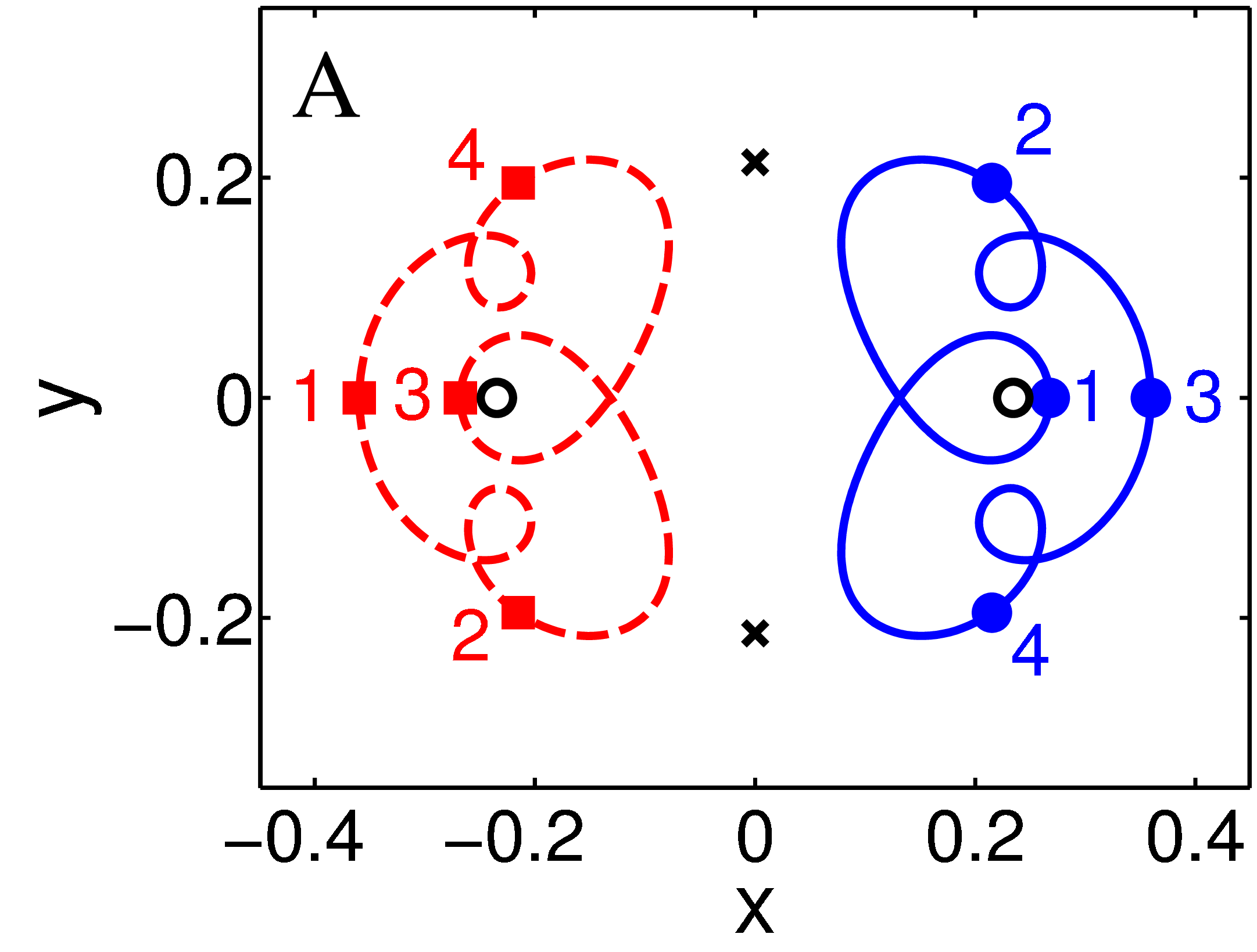}
   \includegraphics[width=3.80cm]{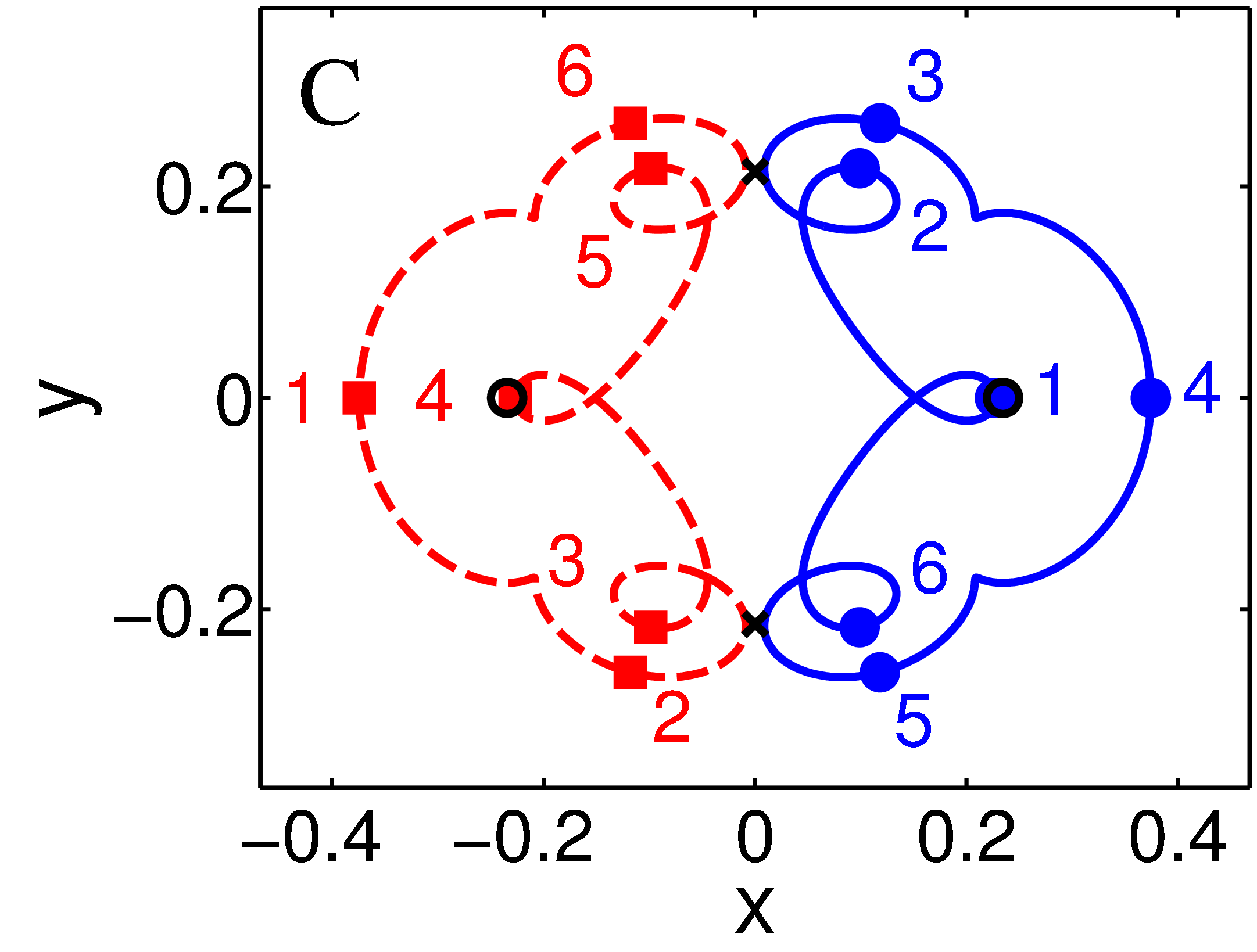}
   \includegraphics[width=3.80cm]{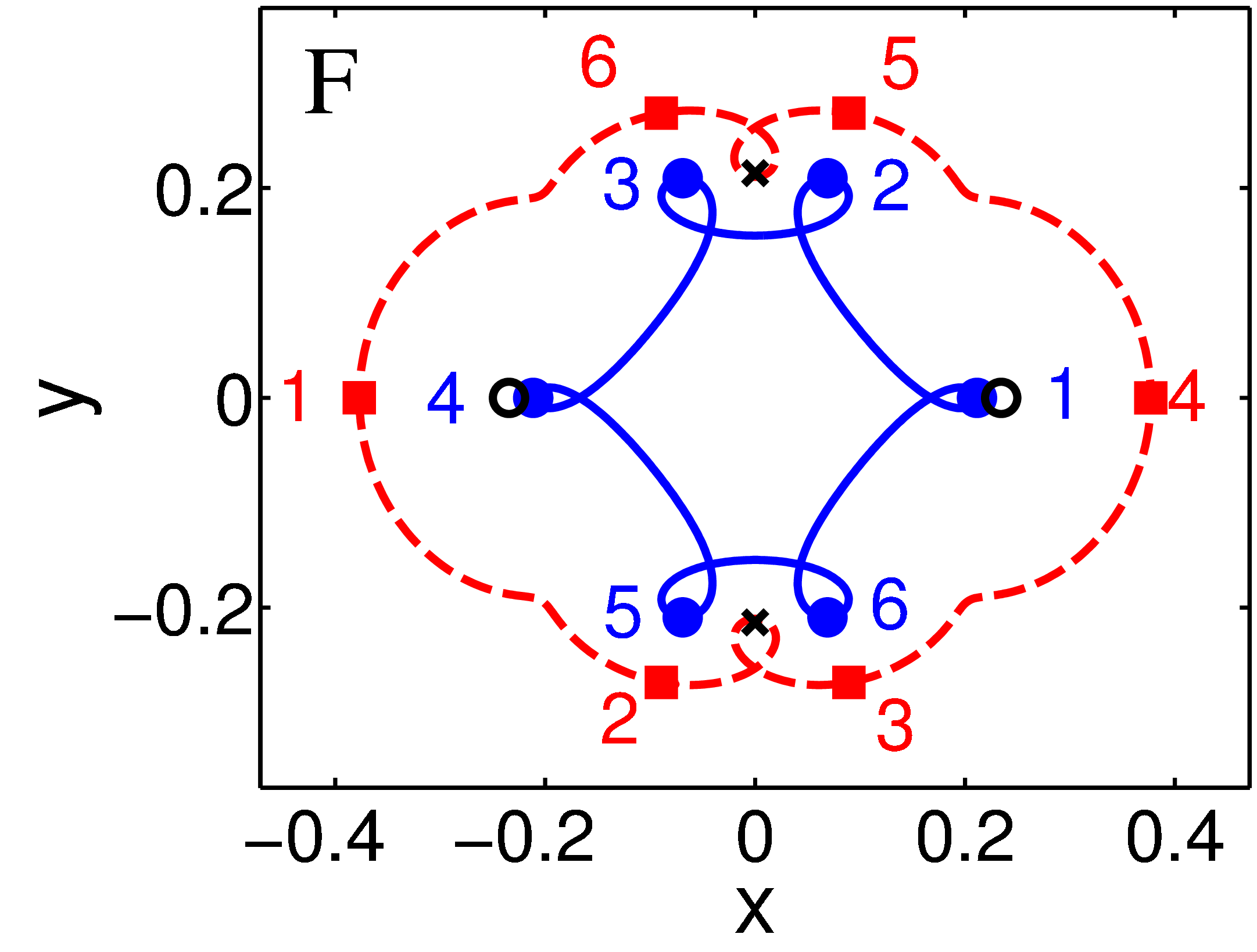}
   \includegraphics[width=3.80cm]{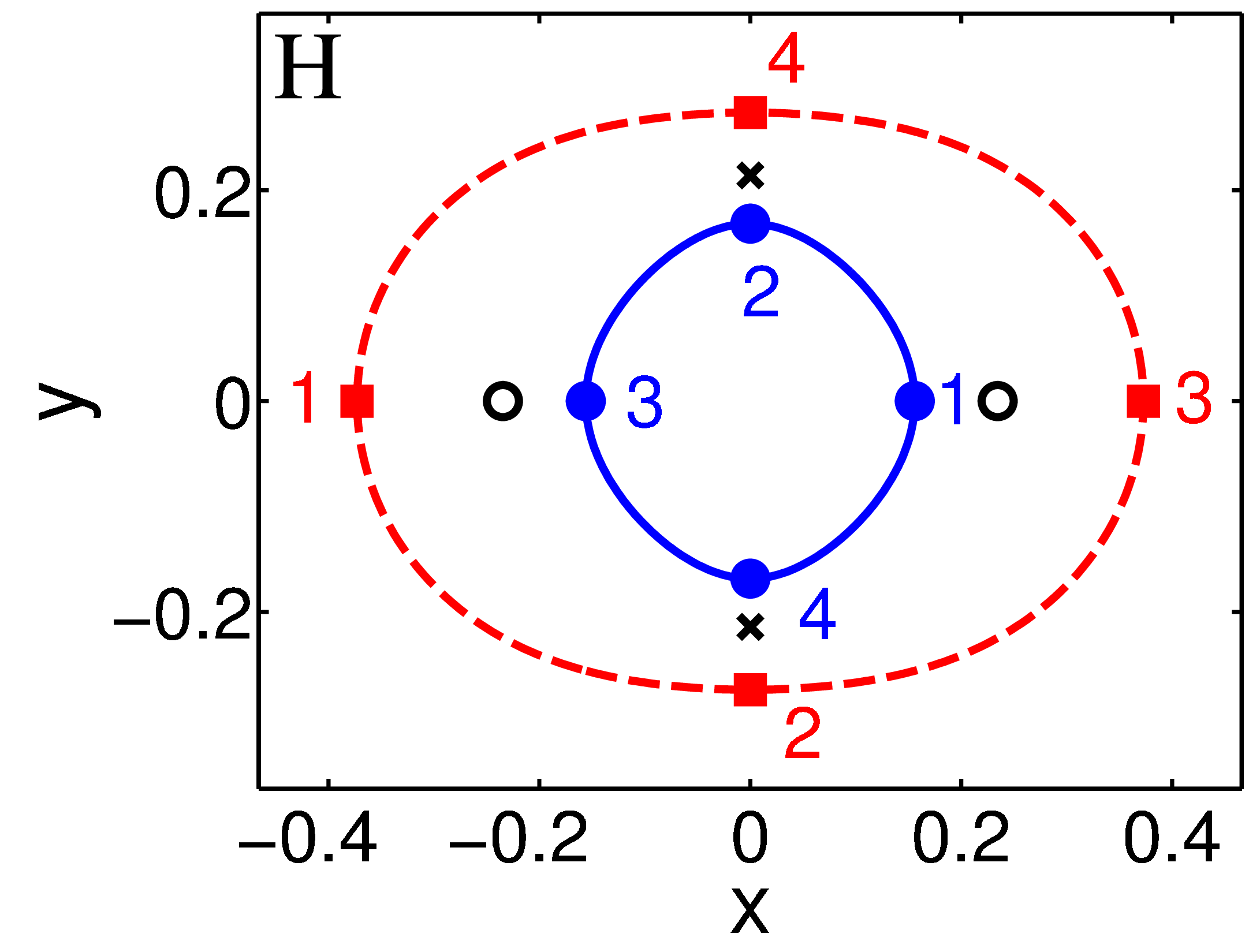}
   \caption{
(color online)
The solutions with initial condition $\rho_2^+$. Only those that intersect the Poincar\'e section an even number of times are shown.}
\label{fig:p228}
\end{figure}

Figure~\ref{fig:p228} shows the same picture for initial conditions with initial condition~$\rho_2^+$. All the comments about the previous figure apply here. We have not plotted the solutions with an odd number of intersections, which 
 include similar panels as shown in the previous figure, up to interchanging the vortices or reflecting the figures across the $y$-axis.  In particular~\ref{fig:m228}B$\leftrightarrow$\ref{fig:p228}B,~\ref{fig:m228}D$\leftrightarrow$\ref{fig:p228}D,~\ref{fig:m228}F$\leftrightarrow$\ref{fig:p228}E, and~\ref{fig:m228}H$\leftrightarrow$\ref{fig:p228}G.

Of interest is how the picture changes as the energy surface $H= H^*$ is varied. As $H^*$ is decreased, the value $\rc$ moves toward the right endpoint of the interval at $\rho_*$. To the right of $\rc$ are the rotational periodic orbits, and these get pushed out of the allowable interval starting with those that have a small number of Poincar\'e intersections. Eventually, the accumulation point $\rc$ exits too, eliminating all the rotational motions. Then, the librations with large numbers of Poincar\'e intersections disappear, followed by those with small numbers of intersection.  So, for example, when $H=0.223$, only the solutions corresponding to A, B, and C in Figure~\ref{fig:in_out_m228} remain.

\begin{figure}[htbp] 
   \centering
   \includegraphics[width=3.80cm]{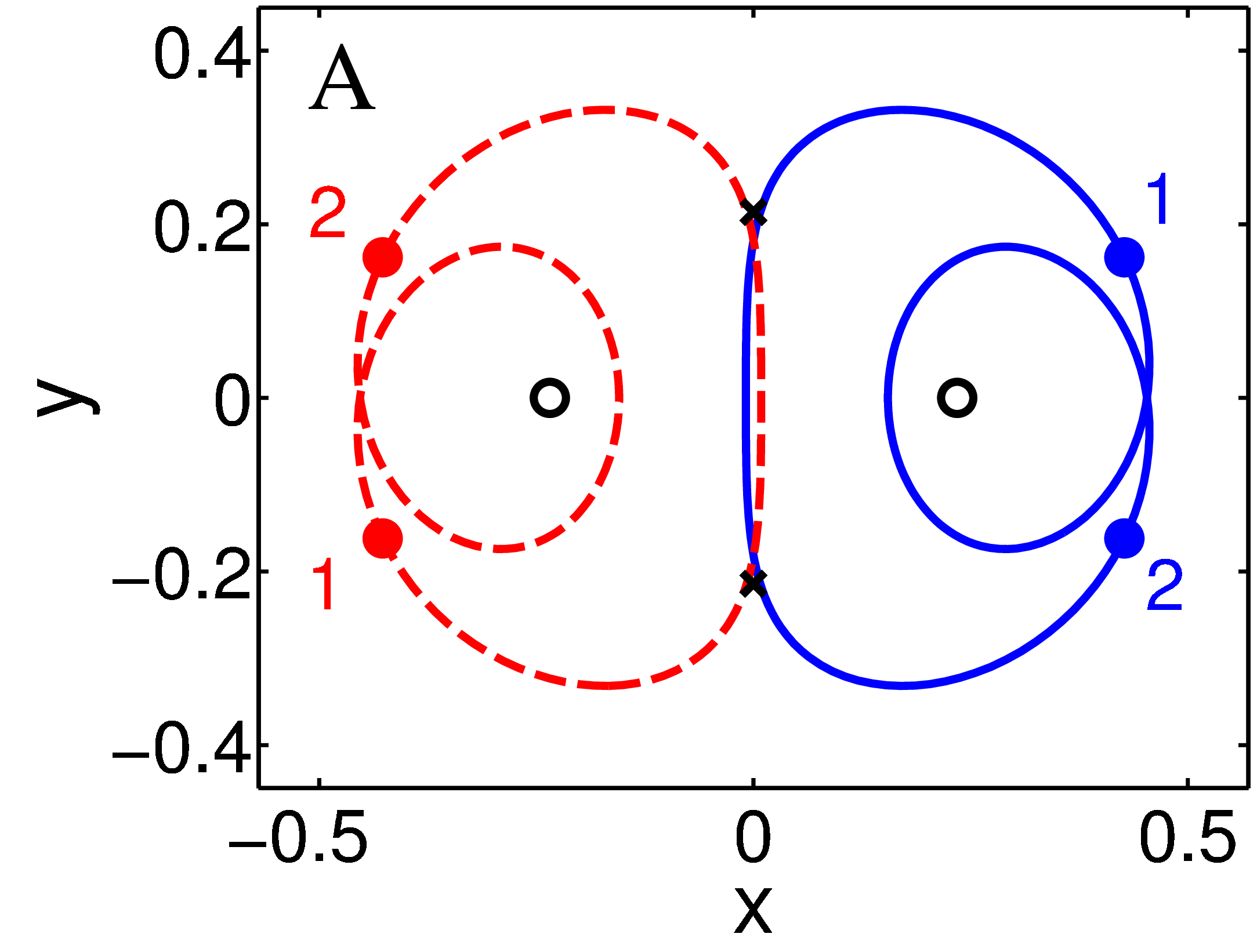}
   \includegraphics[width=3.80cm]{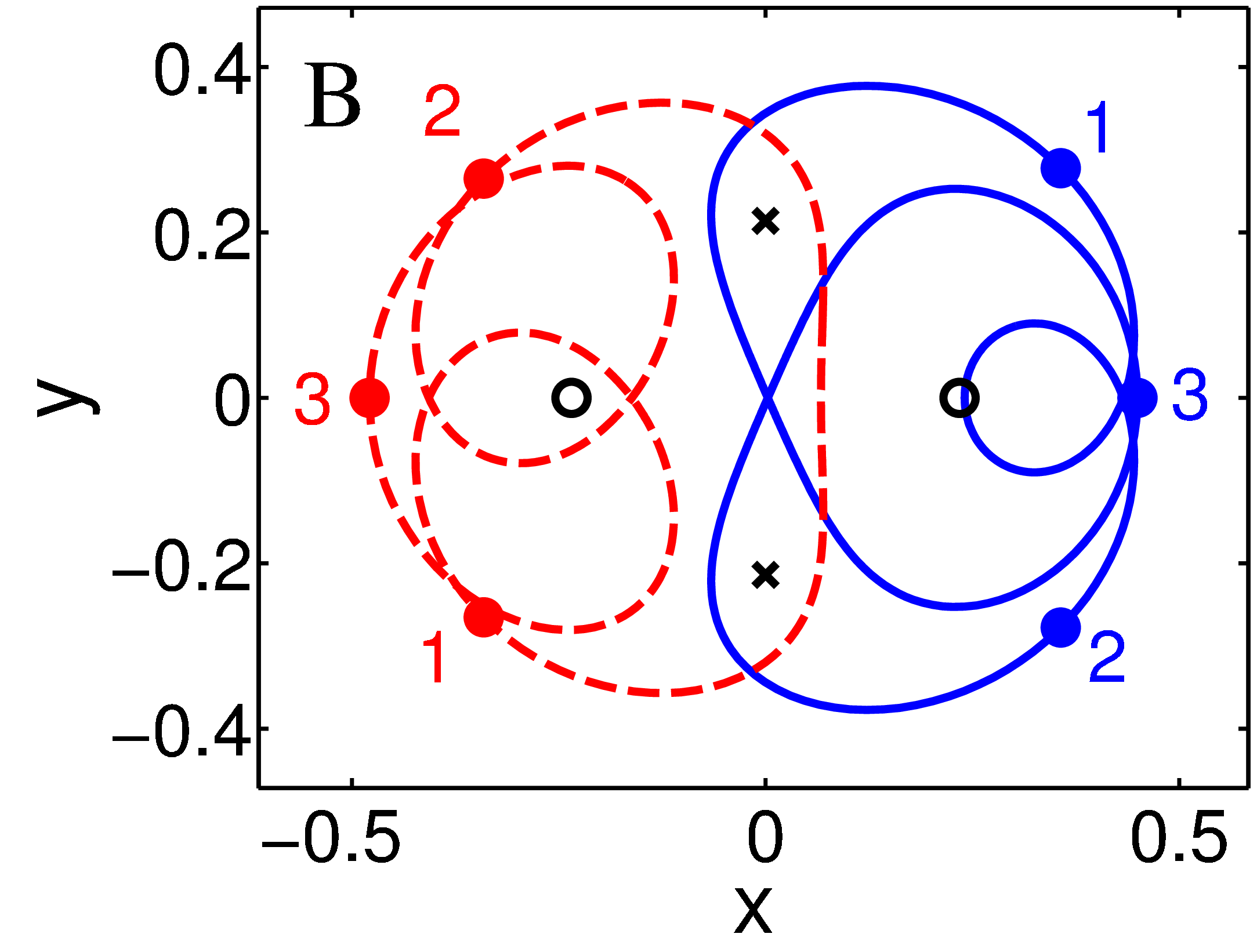}
   \includegraphics[width=3.80cm]{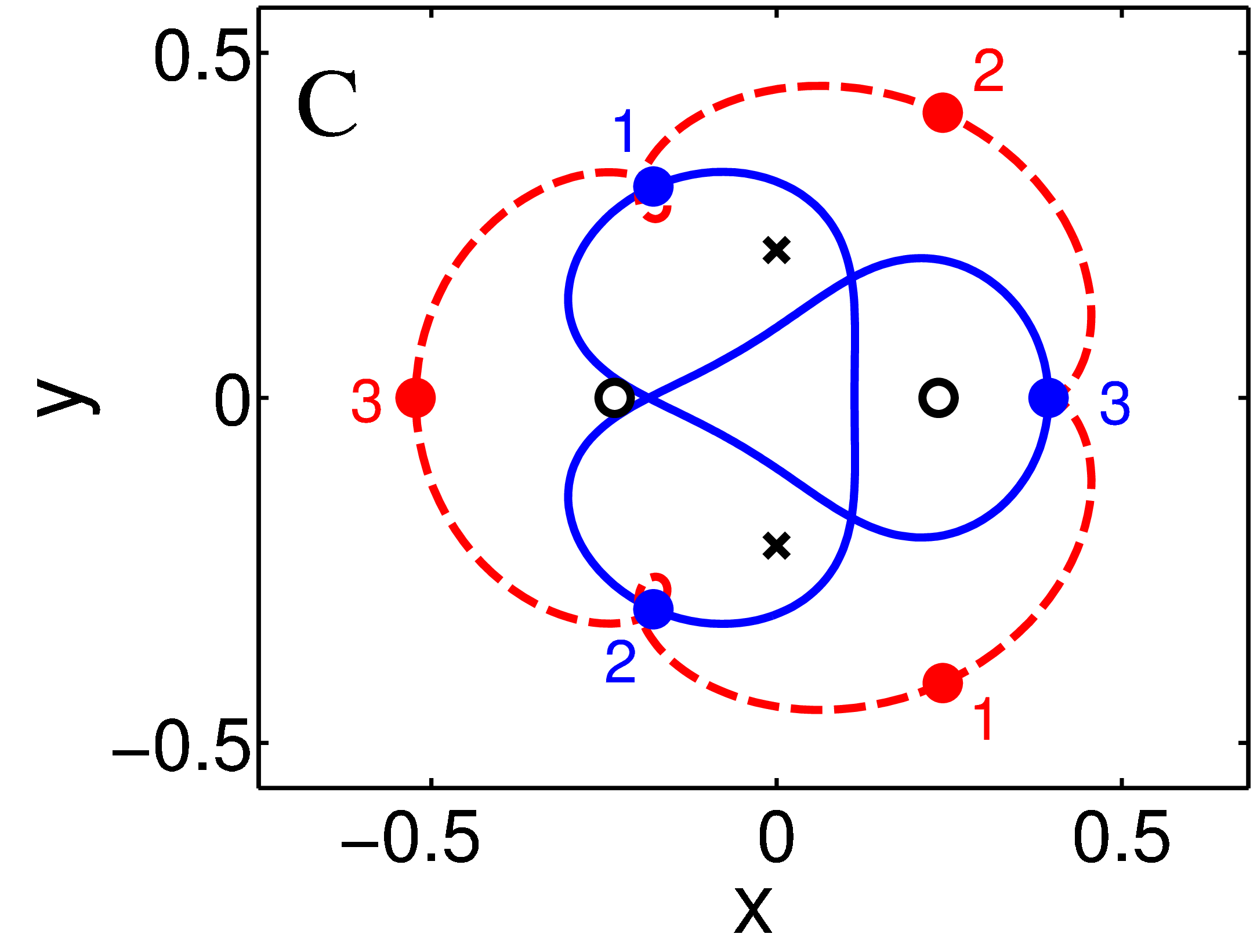}
   \includegraphics[width=3.80cm]{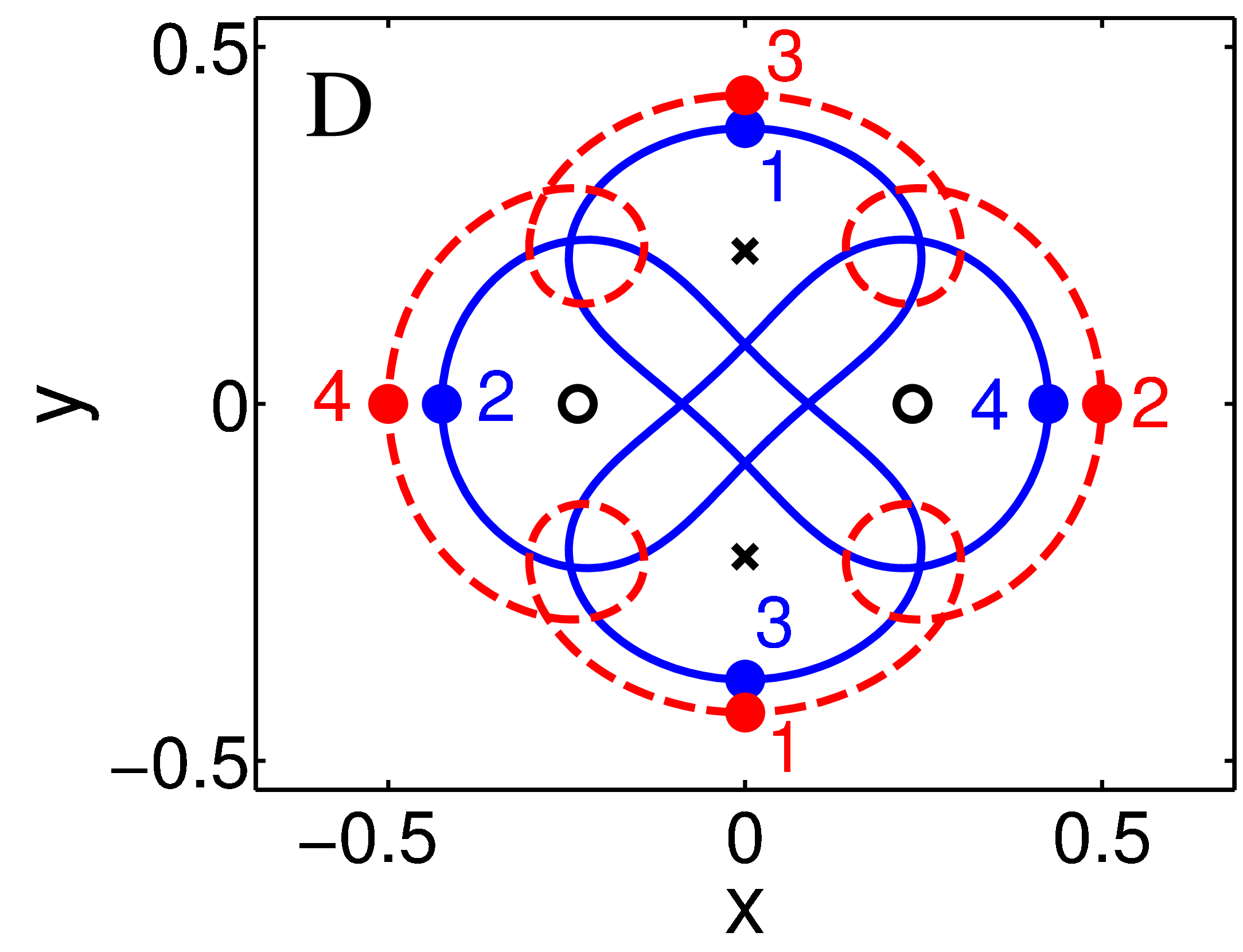}
   \caption{
(color online)
Periodic vortex paths corresponding with Hamiltonian $H=0.3$, with fewer Poincar\'e sections than seen in previous figure.}
\label{fig:vortexPaths}
\end{figure}

Conversely, by increasing $H$, additional rotational orbits bifurcate in from the right endpoint. In fact, some libratory orbits bifurcate from $\rho_1 =0$. Figure~\ref{fig:vortexPaths} shows four additional periodic orbits that exist for $H=0.3$ but not for $H=0.228$ that is plotted above. One of these has two Poincar\'e intersections, two more have three and a fourth has four. The last one is included for comparison with PDE simulations of the next section.

\section{Numerical solutions to the Gross-Pitaevskii equation}
\label{sec:numericalGP}

In this section we corroborate the results obtained from the
reduced system of ODEs (\ref{eq:xy}) for the original
GP equation~\eqref{GP}, showcasing in this way that the
relevant periodic orbits identified by the map approach,
albeit of considerable complexity, are especially relevant
for observation in experimentally accessible BEC settings.
The parameter $B=0.22$ used in the previous sections was earlier used
to theoretically mirror the experimental dynamics of an isotropic
BEC containing approximately 500,000 $^{87}$Rb atoms trapped
in the radial and axial directions with trap frequencies
$(\omega_r,\omega_z)/(2\pi) = (35.8,101.2)$Hz~\cite{Navarro:2013uv}.
These values for the physical parameters translate into a
GP equation (\ref{GP}) with the effective quasi-two-dimensional
(isotropic) potential
\begin{equation}
\label{eq:pot}
V(x,y) =\frac{1}{2}{\left(\frac{\omega_r}{\omega_z}\right)}^2 r^2
       =\frac{1}{2} \Omega^2 (x^2+y^2)
\end{equation}
with an effective trap strength of $\Omega={\omega_r}/{\omega_z}=0.3538$
and an adimensional chemical potential $\mu=16.69$ so that
the stationary state solution, $v(x,y)$, for equation~(\ref{GP}) can be
written as $u(x,y,t) = v(x,y)e^{-i\mu t}$.
We introduce the anisotropy in the potential by replacing the
potential in equation~(\ref{eq:pot}) by
\begin{equation}
\label{eq:pot:ani}
V(x,y) = \frac{1}{2} \Omega^2 \left[ x^2+  (1+\epsilon) y^2 \right],
\end{equation}
where the anisotropy parameter $\epsilon$ corresponds to
(a) $\epsilon=0$ for the isotropic case and
(b) $\epsilon=0.2$ for the anisotropic case studied in the
previous sections.

\begin{figure}[htb]
   \centering
   \includegraphics[height=4.6cm]{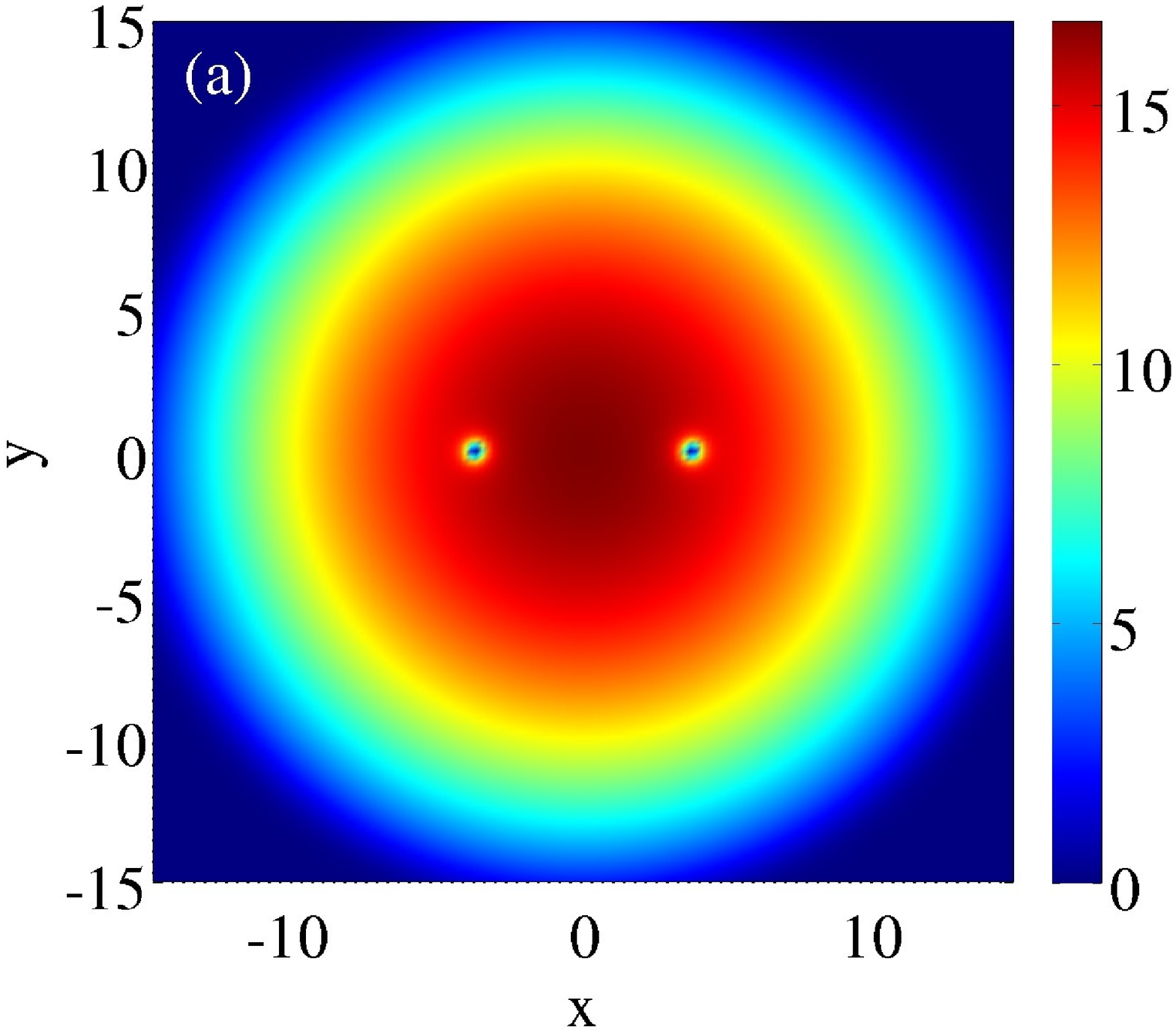}
   \includegraphics[height=4.6cm]{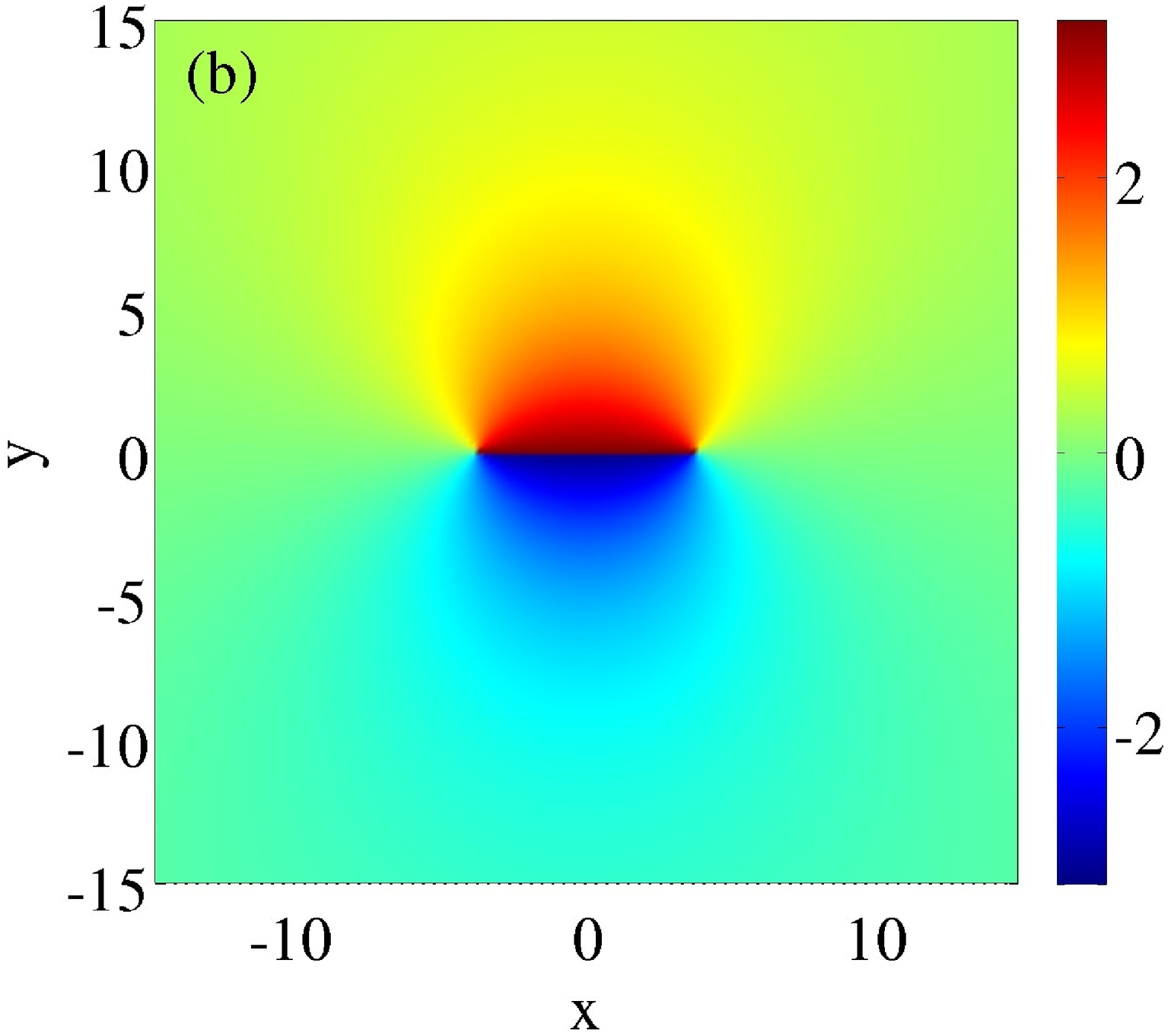}
   \includegraphics[height=4.6cm]{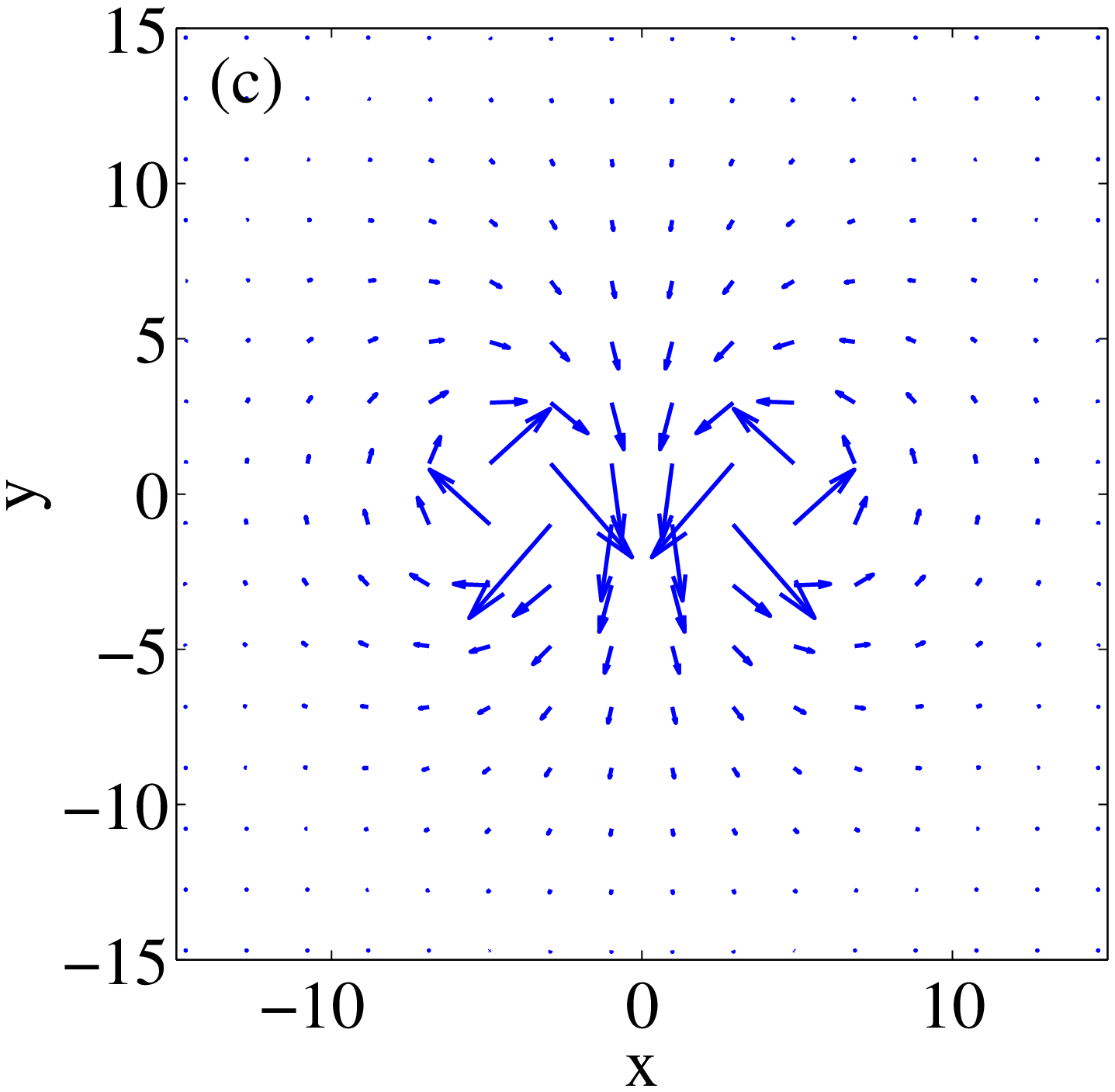}\\
   \includegraphics[height=4.6cm]{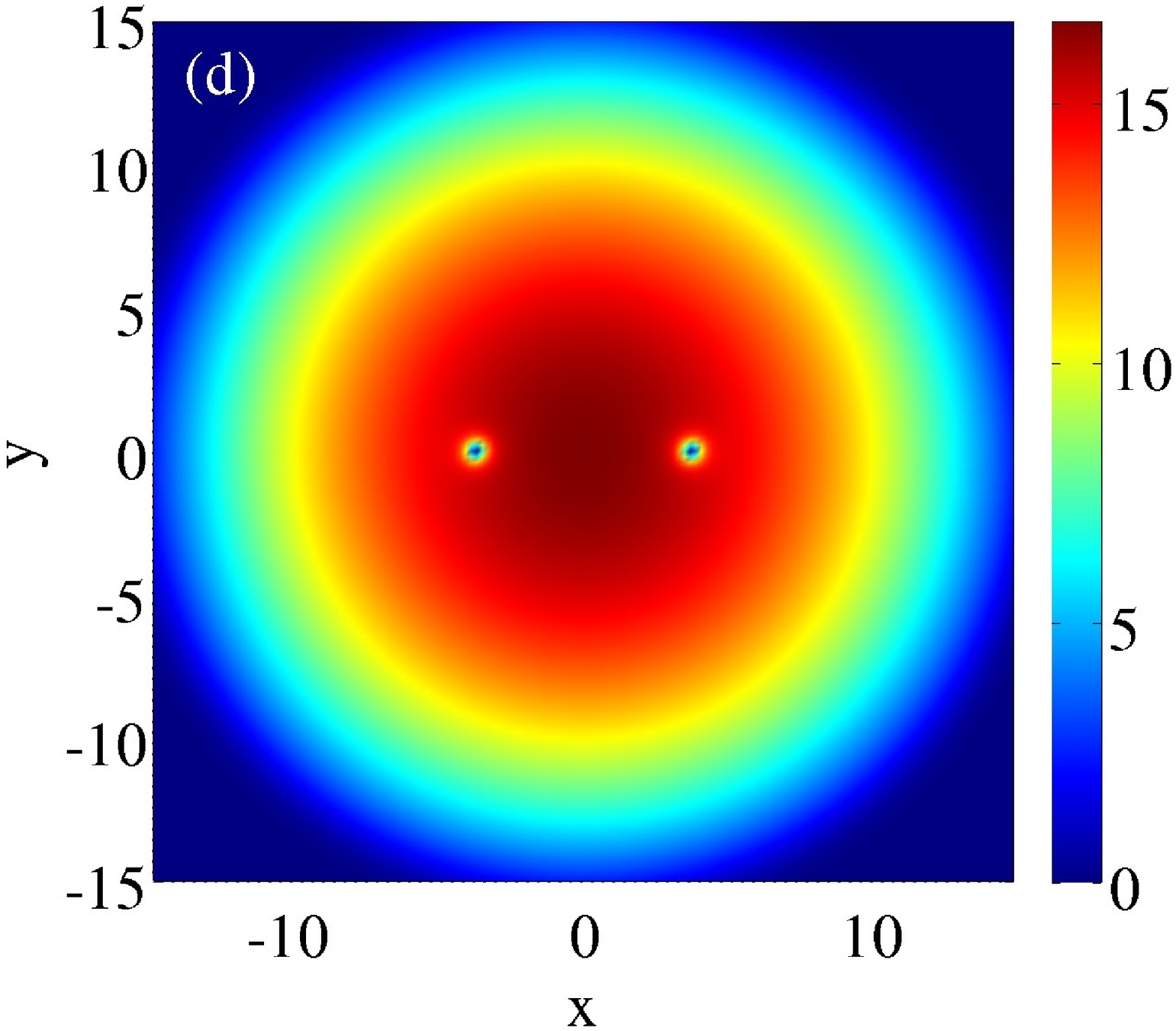}
   \includegraphics[height=4.6cm]{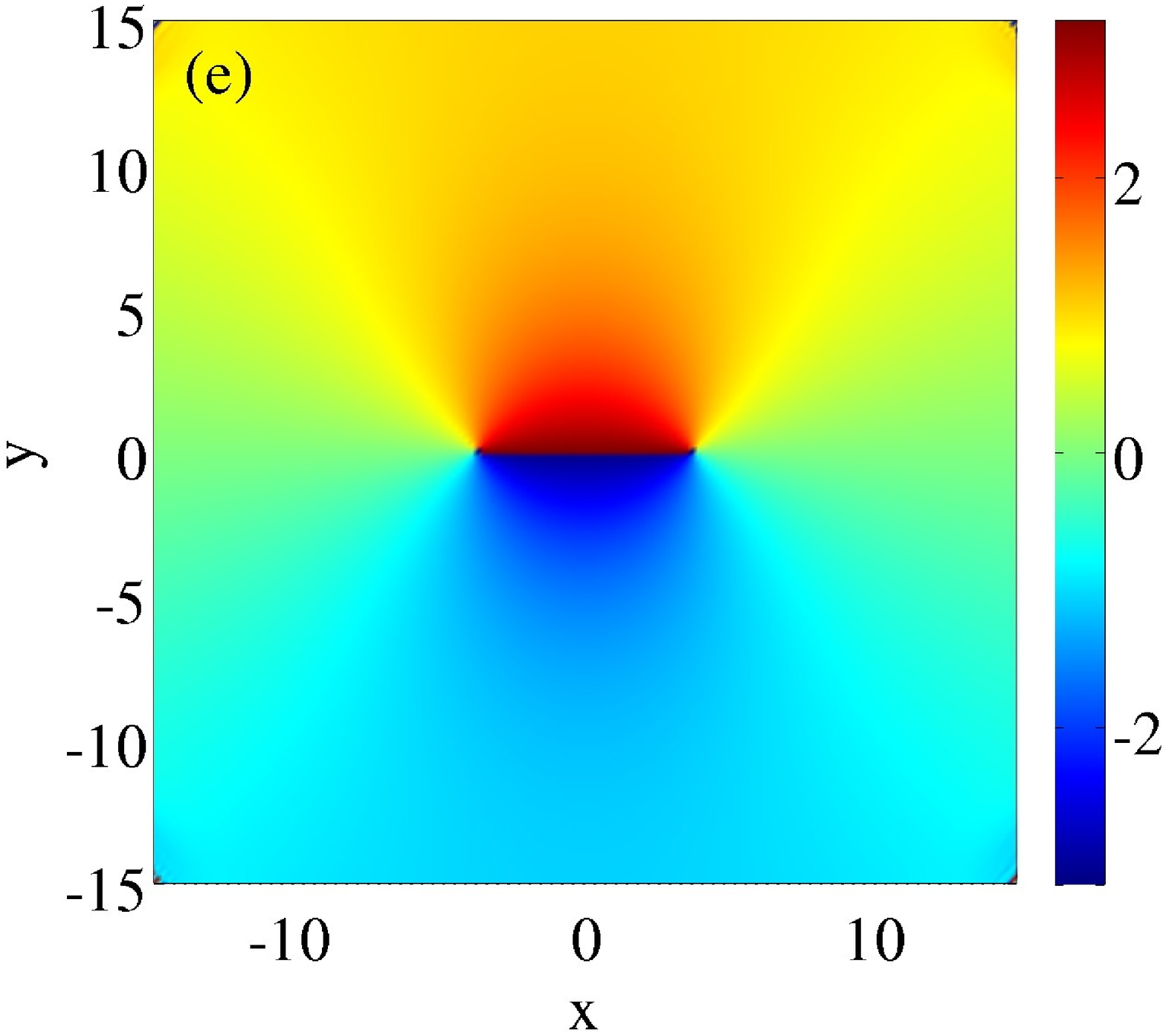}
   \includegraphics[height=4.6cm]{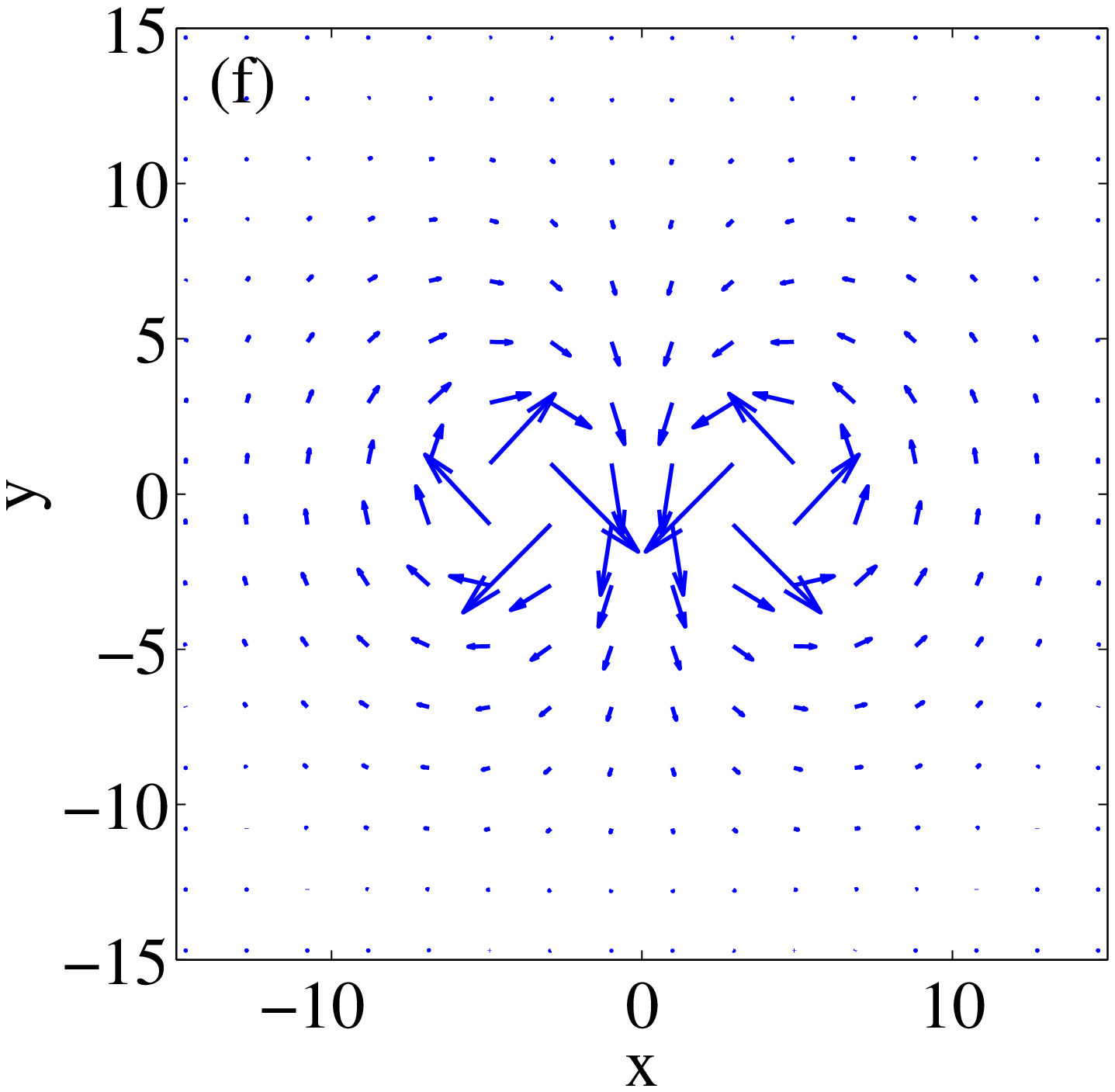}\\
   \caption{
(color online)
Steady state dipole configurations.
The top row depicts (a) density, (b) phase, and (c)
fluid velocity for a vortex dipole seeded, using
the approximate solution (\ref{eq:vortex_xy}), at the steady
state position (\ref{fixedpoints}) on an isotropic BEC.
Similarly, the bottom row depicts the actual steady state solution
for the stationary vortex dipole using a Newton-type
fixed point iteration method.
The fluid velocity corresponds to the gradient of the phase.
}
\label{fig:slosh}
\end{figure}

To numerically investigate the dynamics of vortex dipoles in
the GP equation (\ref{GP}) we first find the steady state
solution, namely the ground (vortex-free) state of the system, $v(x,y)$,
by using $u(x,y,t) = v(x,y)e^{-i\mu t}$ into equation~(\ref{GP})
and solving the ensuing time-independent problem with a Newton-type
fixed point iterative method.
Once the steady state is found, we need to seed the vortex dipole
into this ground state. To do so, we first extract the vortex
profile by solving the homogeneous ($V=0$) problem (\ref{GP})
in radial coordinates $(r,\theta)$ for a vortex solution
of charge $S=1$ centered at the origin:
$u(r,\theta,t)=f(r) e^{-i\mu t} e^{iS \theta}$.
The ensuing boundary value problem with boundary values
$f(r=0)=0$ and $f(r=\infty)=\sqrt{\mu}$ is numerically
solved using the standard boundary value problem solver {\tt bvp4c}
in Matlab.
Now, equipped with the numerically exact radial profile $f(r)$
we proceed to imprint a vortex of charge $S=\pm 1$ at any
desired location ${\bf r_0}=(x_0,y_0)$ within the BEC
by multiplying the steady
solution $v(x,y)$ found above by a normalized vortex
profile:
\begin{equation}
\label{eq:vortex_xy}
u_0(x,y) = v(x,y) \times \frac{f({\bf r}-{\bf r_0})}{\sqrt{\mu} }
\times e^{-i S \theta_0},
\end{equation}
where $\theta_0$ is the polar angle of any point $(x,y)$ within
the BEC measured from the center of the vortex $(x_0,y_0)$.
This initial condition provides a vortex configuration
that is very close to the exact profile. Our approximation
relies on the assumption that the background profile for
the steady state varies slowly over the vortex core size.
In this manner, our approximation using the homogeneous
background yields a close approximation to the actual vortex
profile that sheds very little radiation when numerically
integrated.
It is worth mentioning at this stage that the typical
vortex width for the GP numerics (and the associated
laboratory experiment described in Ref.~\cite{Navarro:2013uv})
is small ($\approx 1/30$th) compared to the width of the BEC cloud
(see for example the two vortices in the left panels
of Figure~\ref{fig:slosh}).

\begin{figure}[t]
   \centering
   \includegraphics[height=6.5cm]{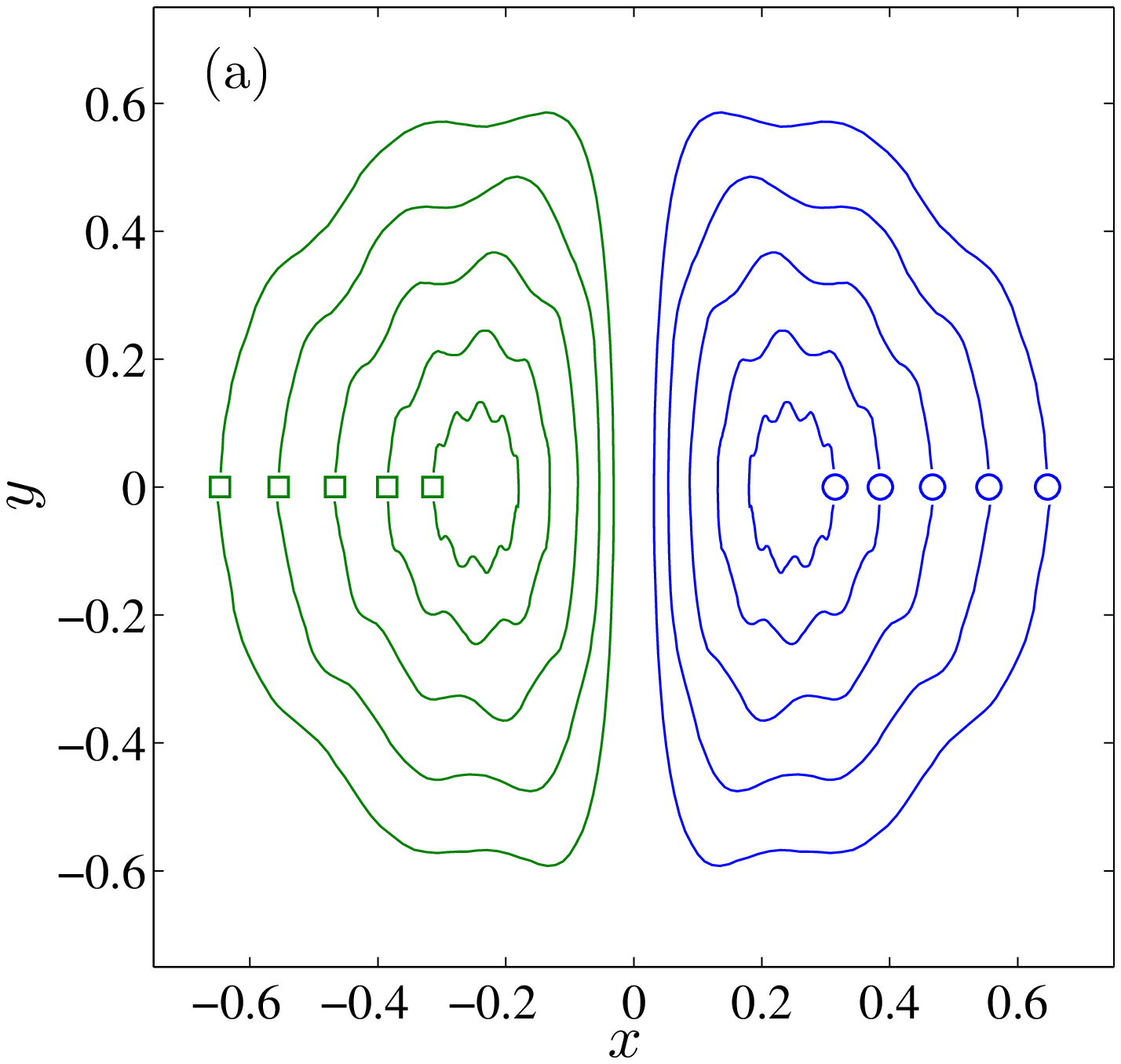}
   \includegraphics[height=6.5cm]{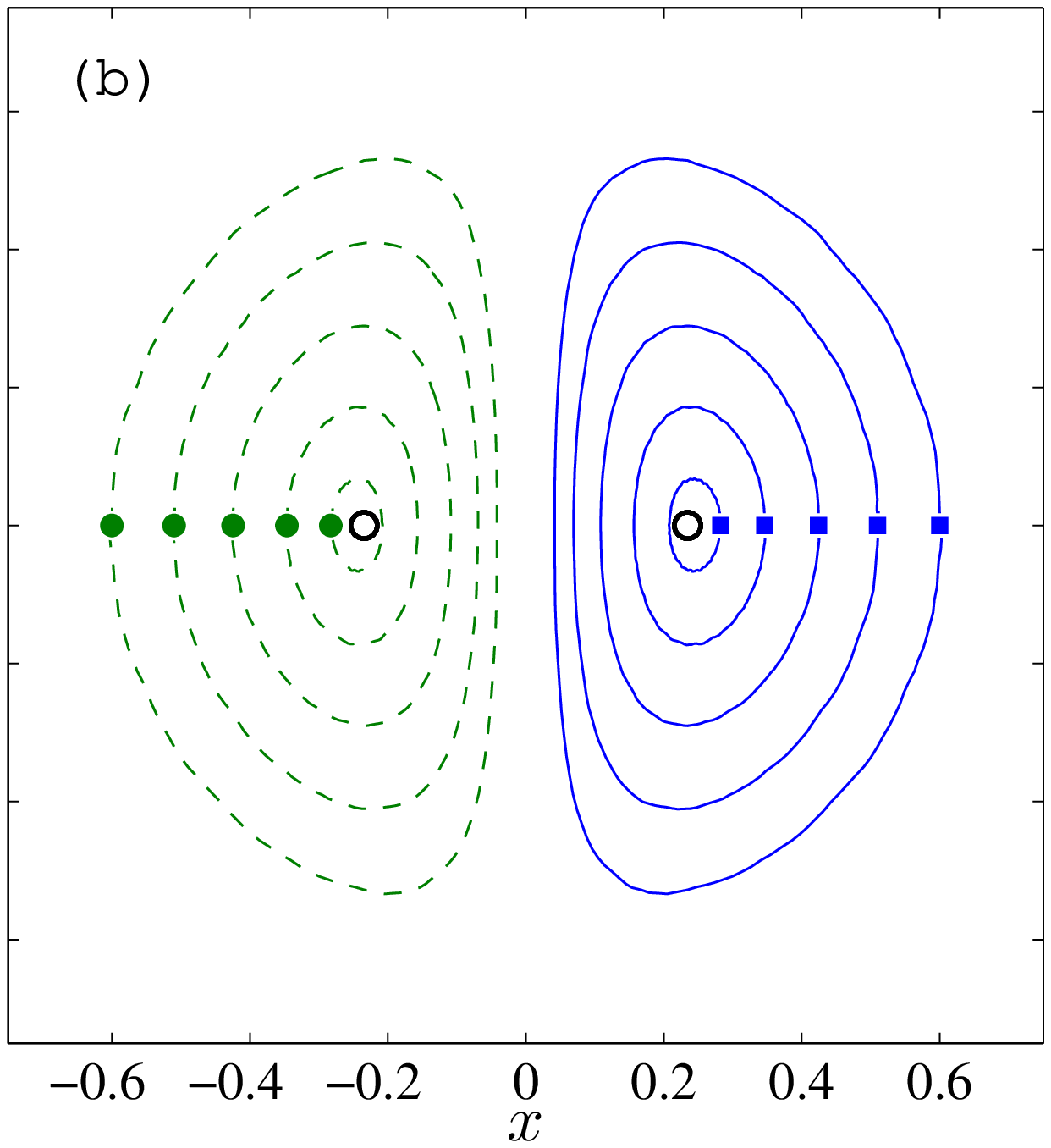}
   \caption{
(color online)
(a) Orbits generated by the full GP dynamics of a vortex dipole seeded in
the BEC using equation~(\ref{eq:vortex_xy}). The orbits are similar
to the ones obtained using the ODE reduction as in
Figure~\ref{fig:particle_paths_reduced}a. However, notice the
spurious ``wiggles'' present in all GP dynamics.
(b) Corresponding orbits after subtracting the linear momenta
of the initial configuration (\ref{eq:vortex_xy}). Note that the
wiggles are no longer present.
}
\label{fig:wiggles}
\end{figure}

\begin{figure}[t]
   \centering
   \includegraphics[height=5.5cm]{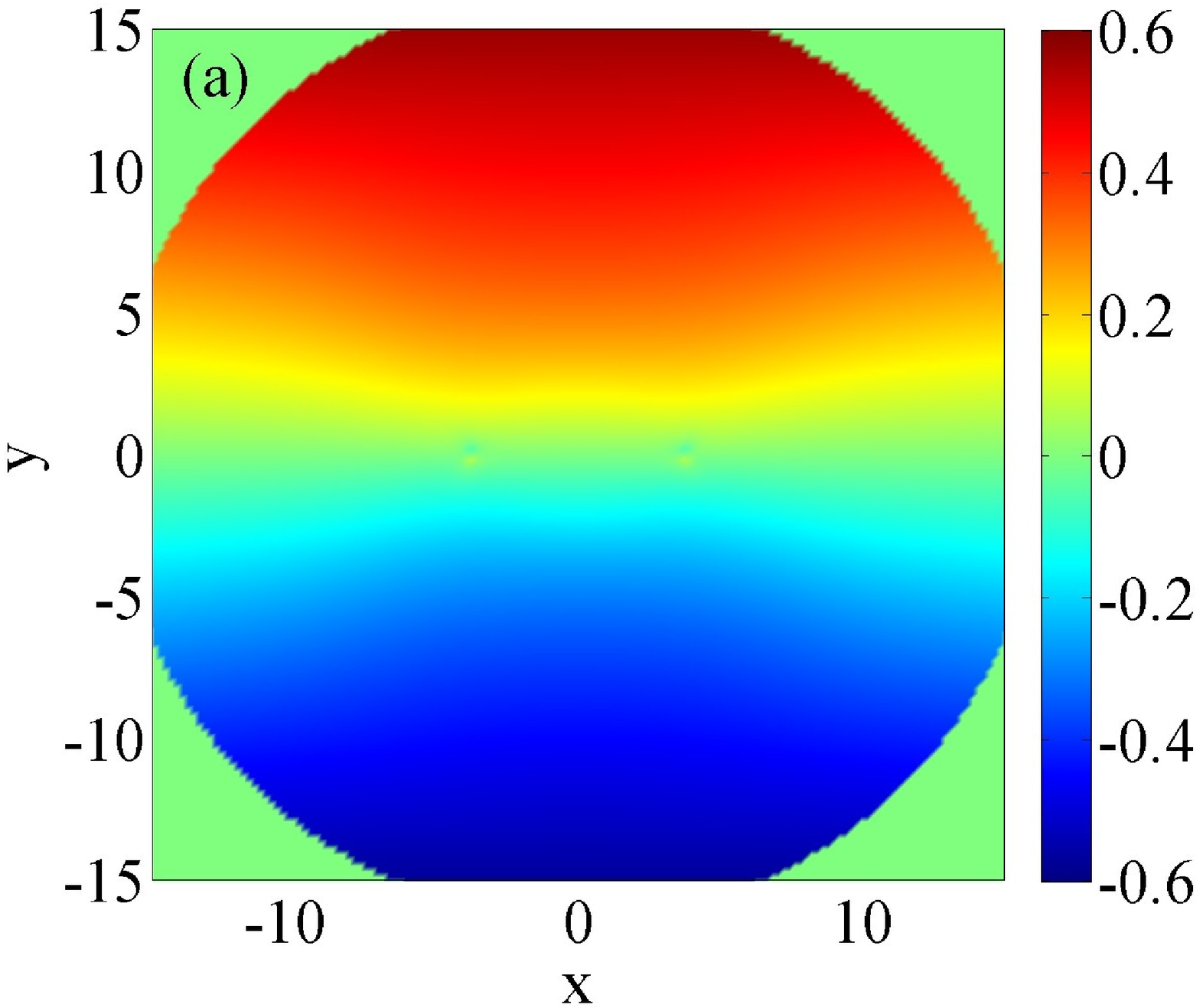}
   \includegraphics[height=5.5cm]{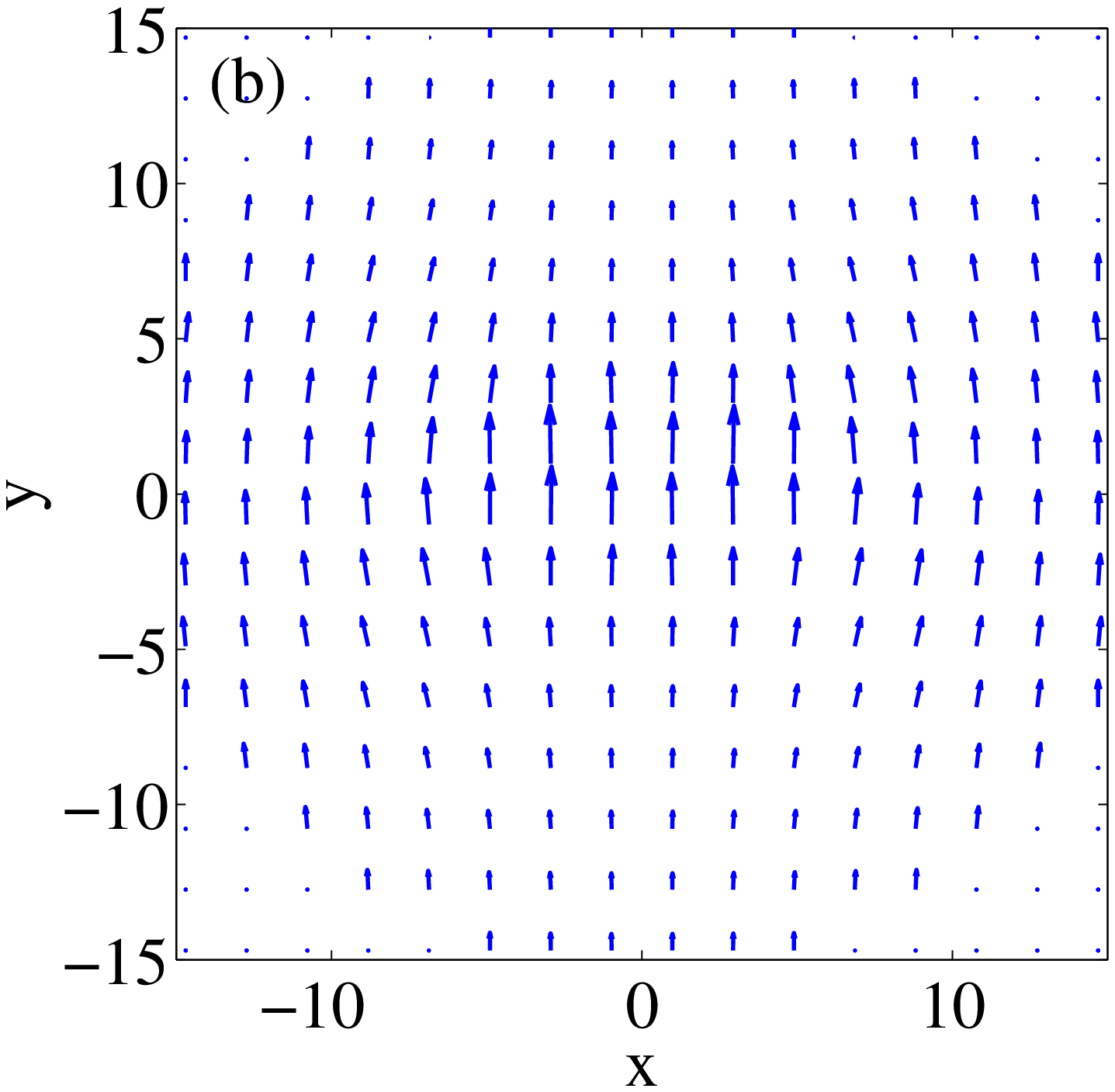}
   \caption{
(color online)
Difference between the steady state dipole vortex
configuration seeded using equation~(\ref{eq:vortex_xy})
(see top row of panels of Figure~\ref{fig:slosh}) and the numerically
exact solution (see bottom row of panels of Figure~\ref{fig:slosh}).
Panels (a) and (b) depict, respectively, the difference between
the phases and the fluid velocities.
}
\label{fig:slosh_diff}
\end{figure}

The top row of panels in Figure~\ref{fig:slosh} shows, respectively, from left to
right, the density, phase, and fluid velocity associated with
a vortex dipole initiated, using the above method, at the
steady state positions (\ref{fixedpoints}) on an isotropic
($\epsilon=0$) BEC.
Seeding the GP numerics using the approximate solution (\ref{eq:vortex_xy})
has an unexpected spurious effect as it can be observed from the
extracted vortex orbits depicted in Figure~\ref{fig:wiggles}a.
The figure shows, contrary to the smooth orbits from the effective
ODE dynamics depicted in Figure~\ref{fig:particle_paths_reduced},
spurious up-down ``wiggles''.
Close inspection of these wiggles reveals that they originate
from an undesired perturbation of the sloshing (back-and-forth)
mode of the steady state background.
This can be confirmed by comparing the seeded solution
using equation~(\ref{eq:vortex_xy}) to the true steady state
when the dipole is placed at the fixed point (\ref{fixedpoints}).
The bottom panels of Figure~\ref{fig:slosh} depict, respectively,
the density, phase, and fluid velocity of the true steady state
dipole found through a Newton-type fixed point iteration method.
It is interesting to compare our approximate seeded solution
(top panels) to the numerical exact solution (bottom panels).
In fact, the density and fluid velocity seem quite close for
both cases. However, it is evident that the phase distribution
is indeed different.
To elucidate the effects of the discrepancy between these two
configurations we depict in Figure~\ref{fig:slosh_diff} their
differences. The panels correspond to
the difference between the (a) phases and
(b) the fluid velocities. 
As it becomes clear from this figure, the phase difference between the two
cases shows a clear vertical gradient (see Figure~\ref{fig:slosh_diff}a)
that, in turn, is reflected in a vertical fluid velocity
(see Figure~\ref{fig:slosh_diff}b).
This upward fluid  velocity is then responsible for perturbing the
sloshing (up-down) mode of the background cloud resulting
in the spurious wiggles present in the orbits depicted in
Figure~\ref{fig:wiggles}a.

\begin{figure}[t]
   \centering
   \includegraphics[height=3.5cm]{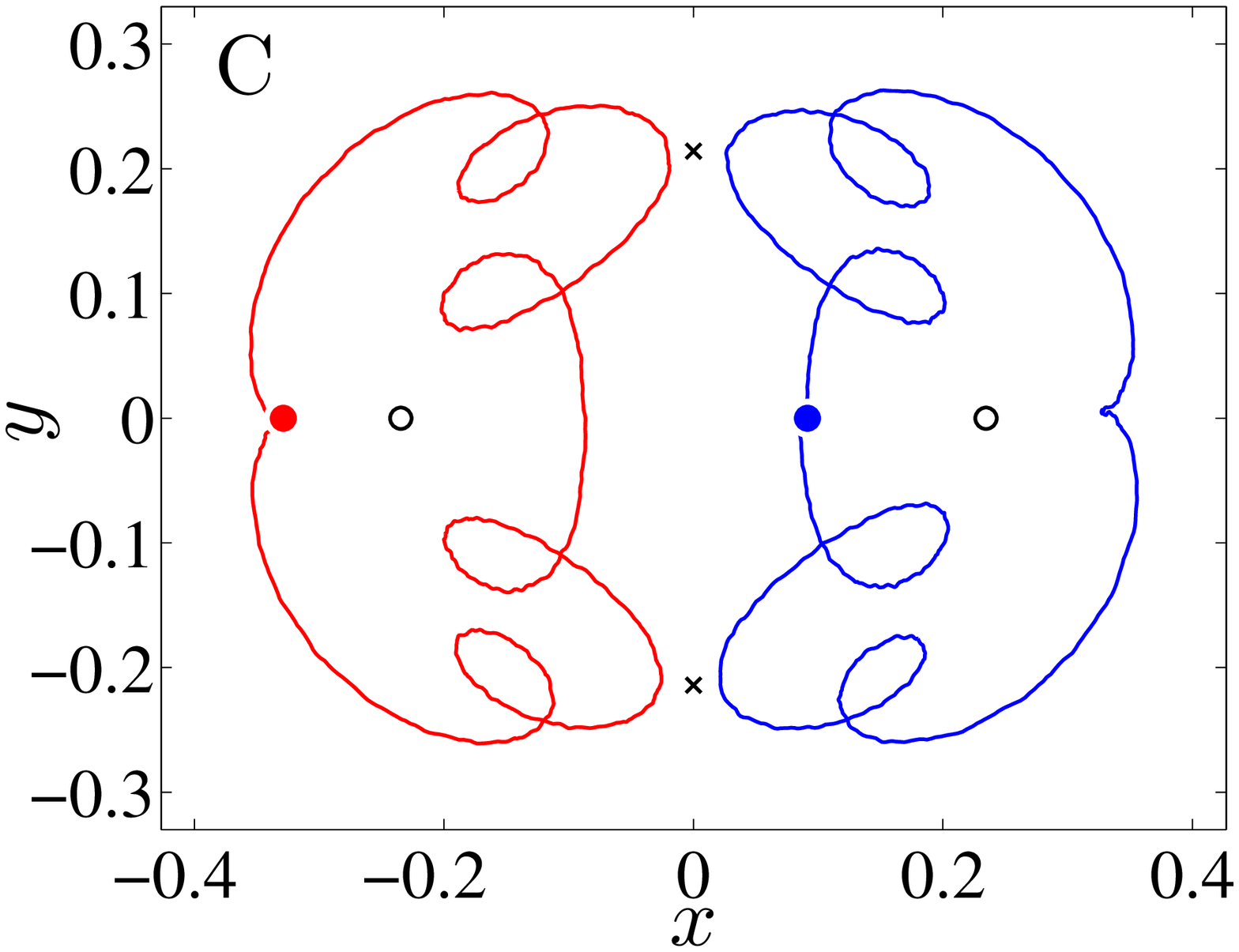}
   \includegraphics[height=3.5cm]{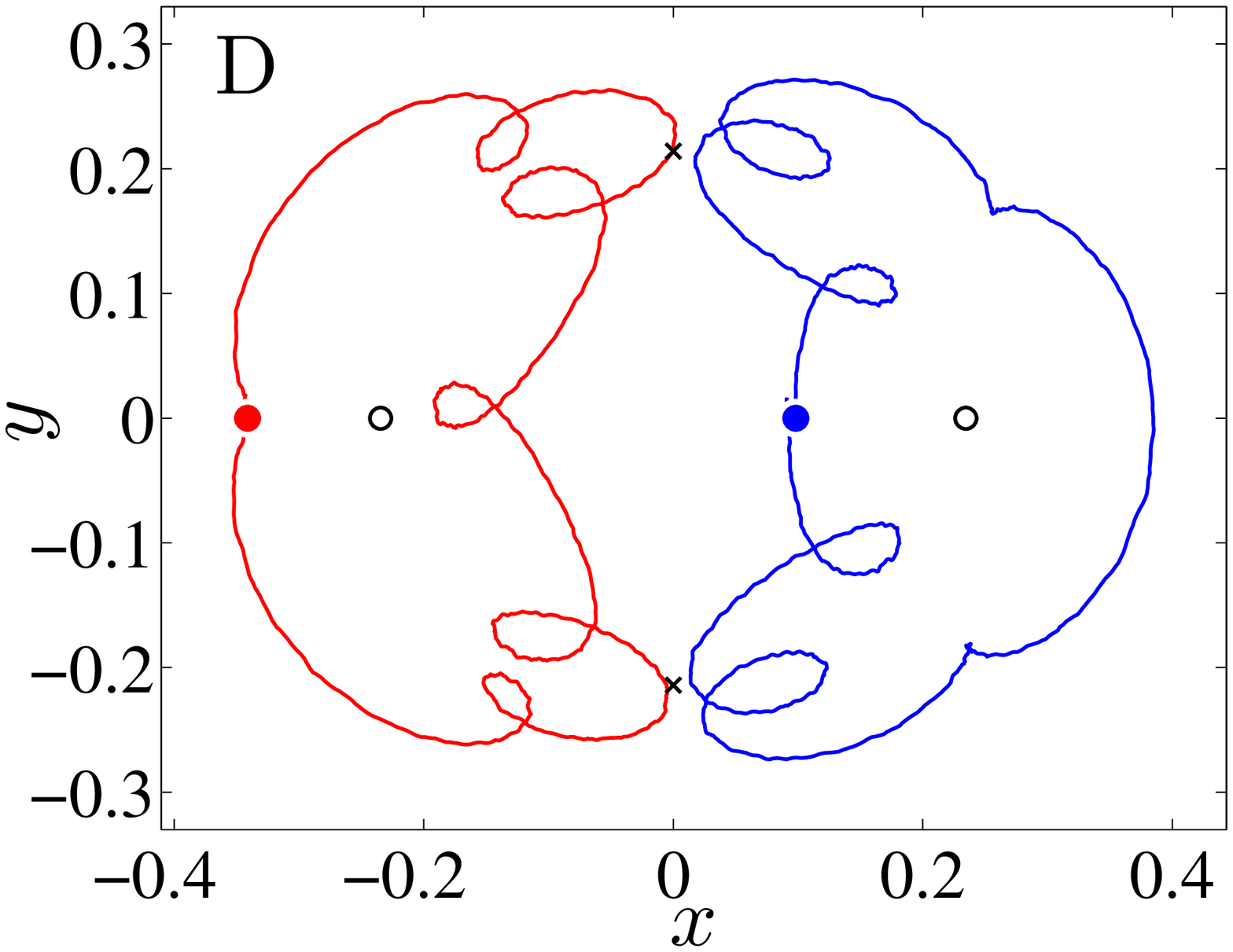}
   \includegraphics[height=3.5cm]{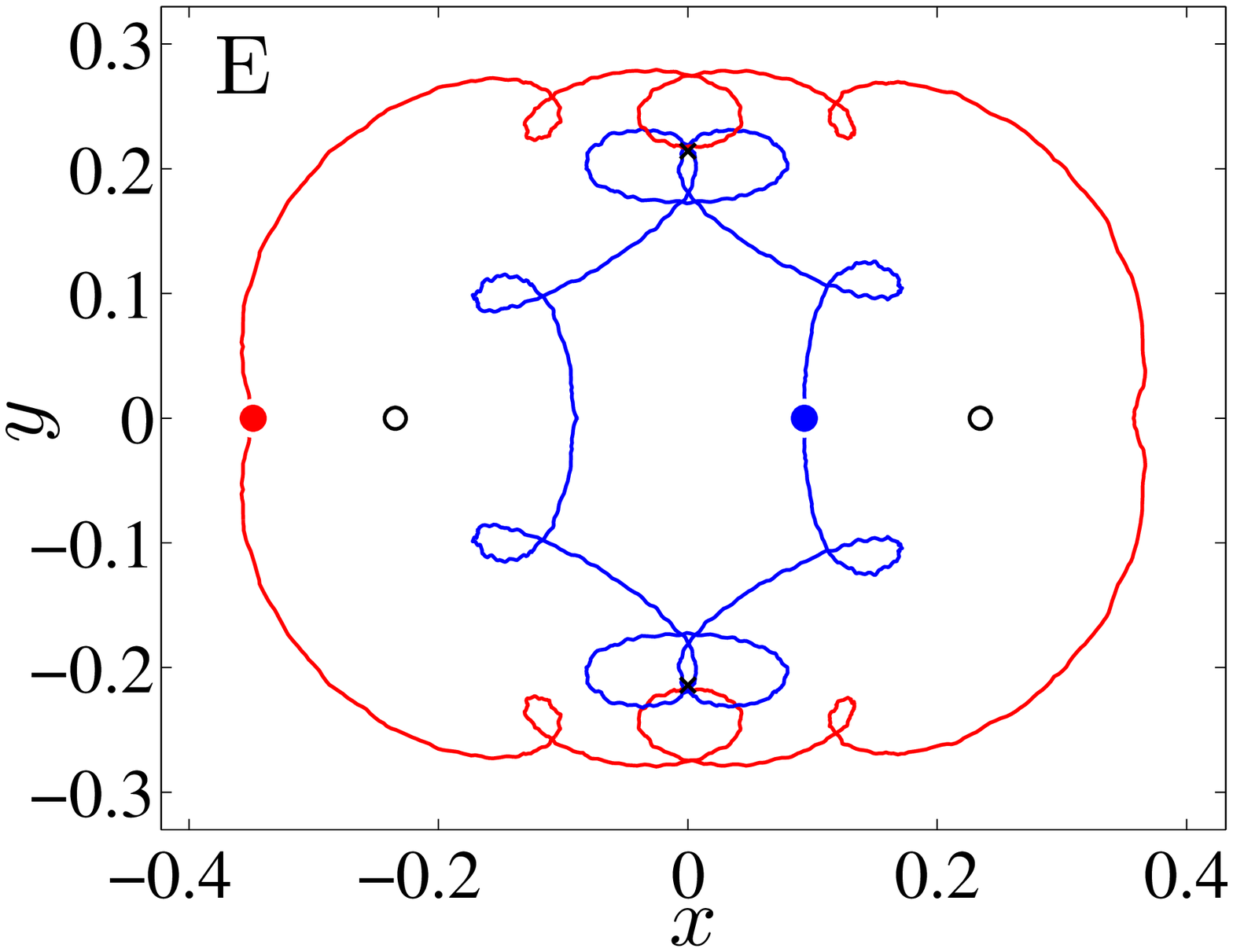}
   \includegraphics[height=3.5cm]{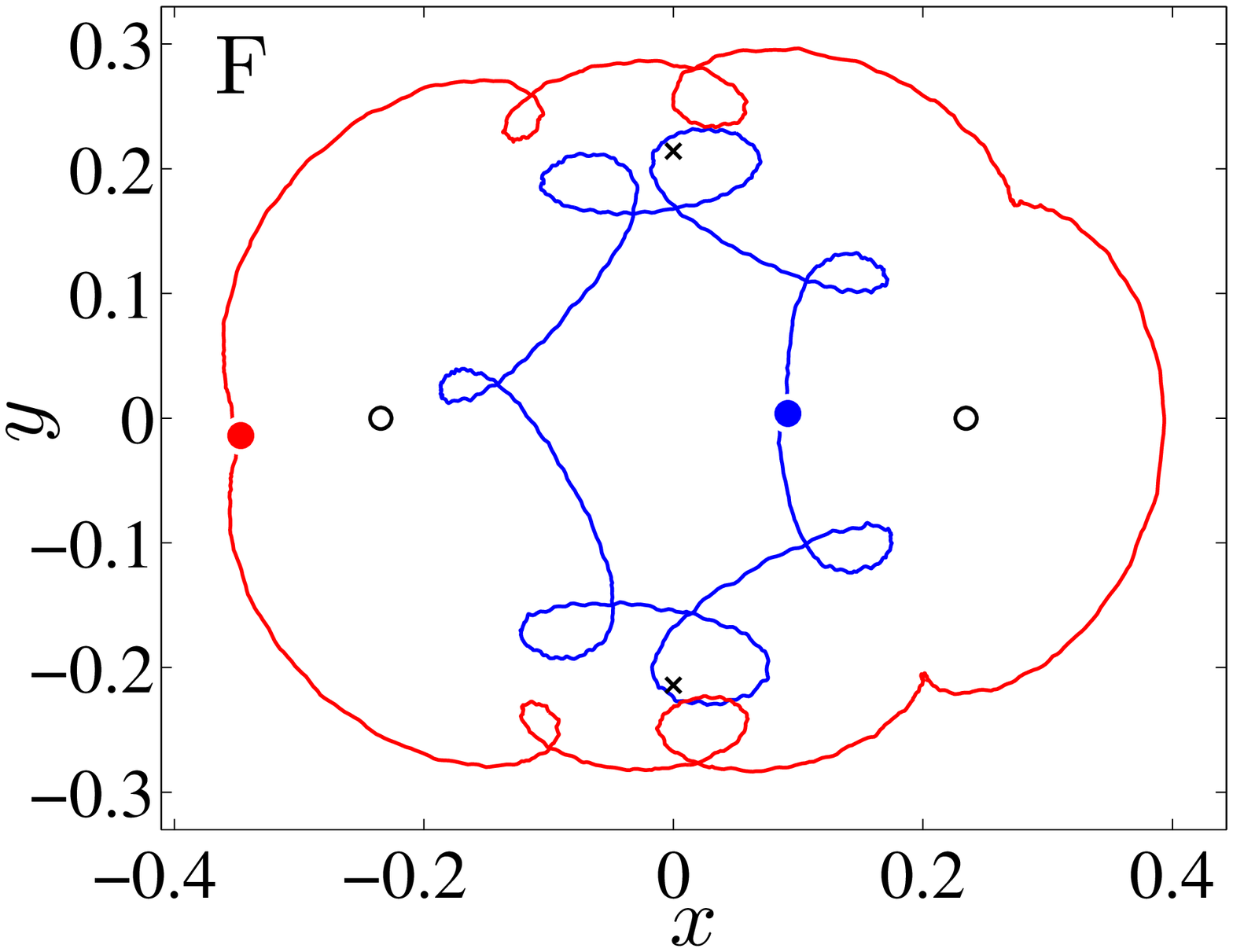}
   \includegraphics[height=3.5cm]{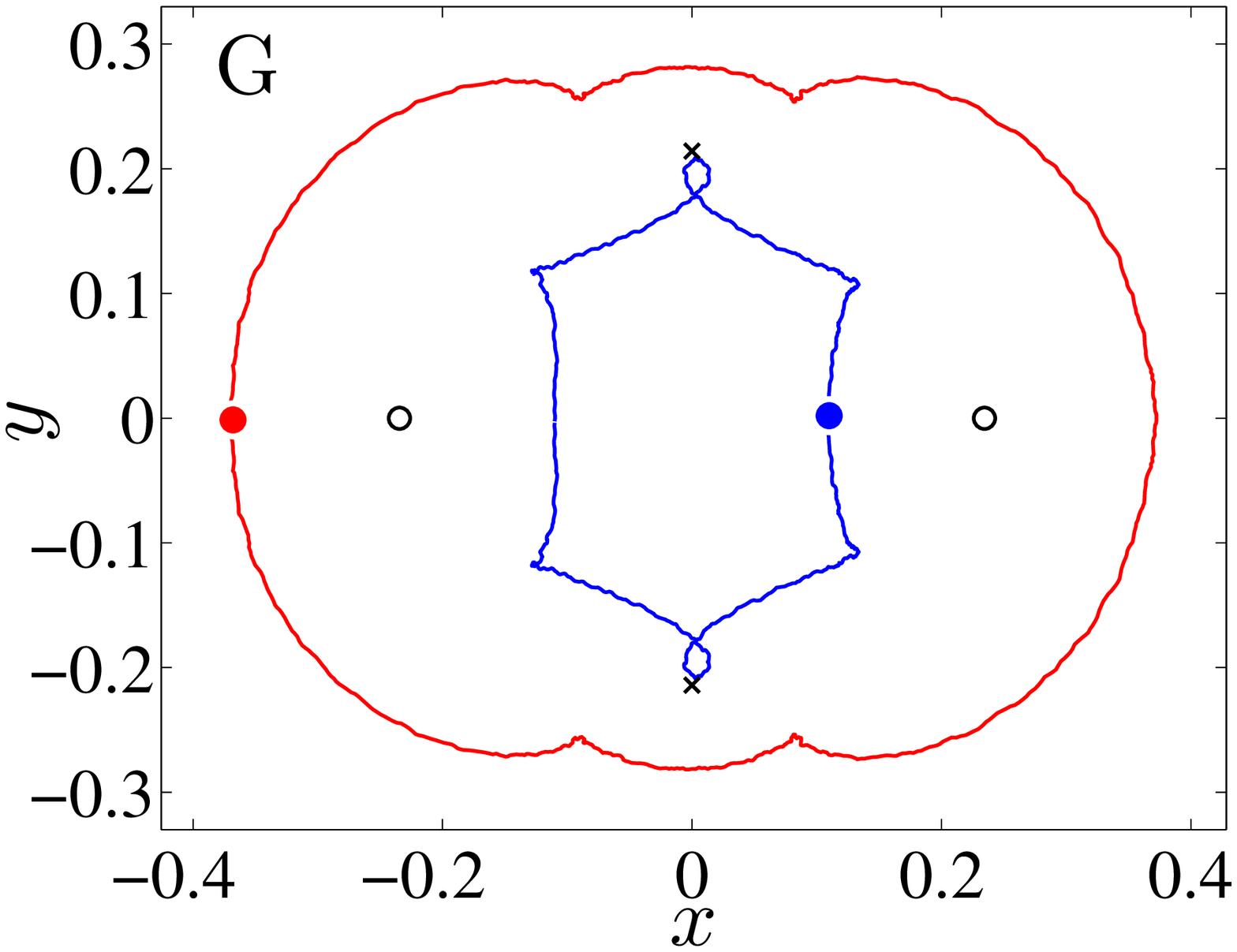}
   \includegraphics[height=3.5cm]{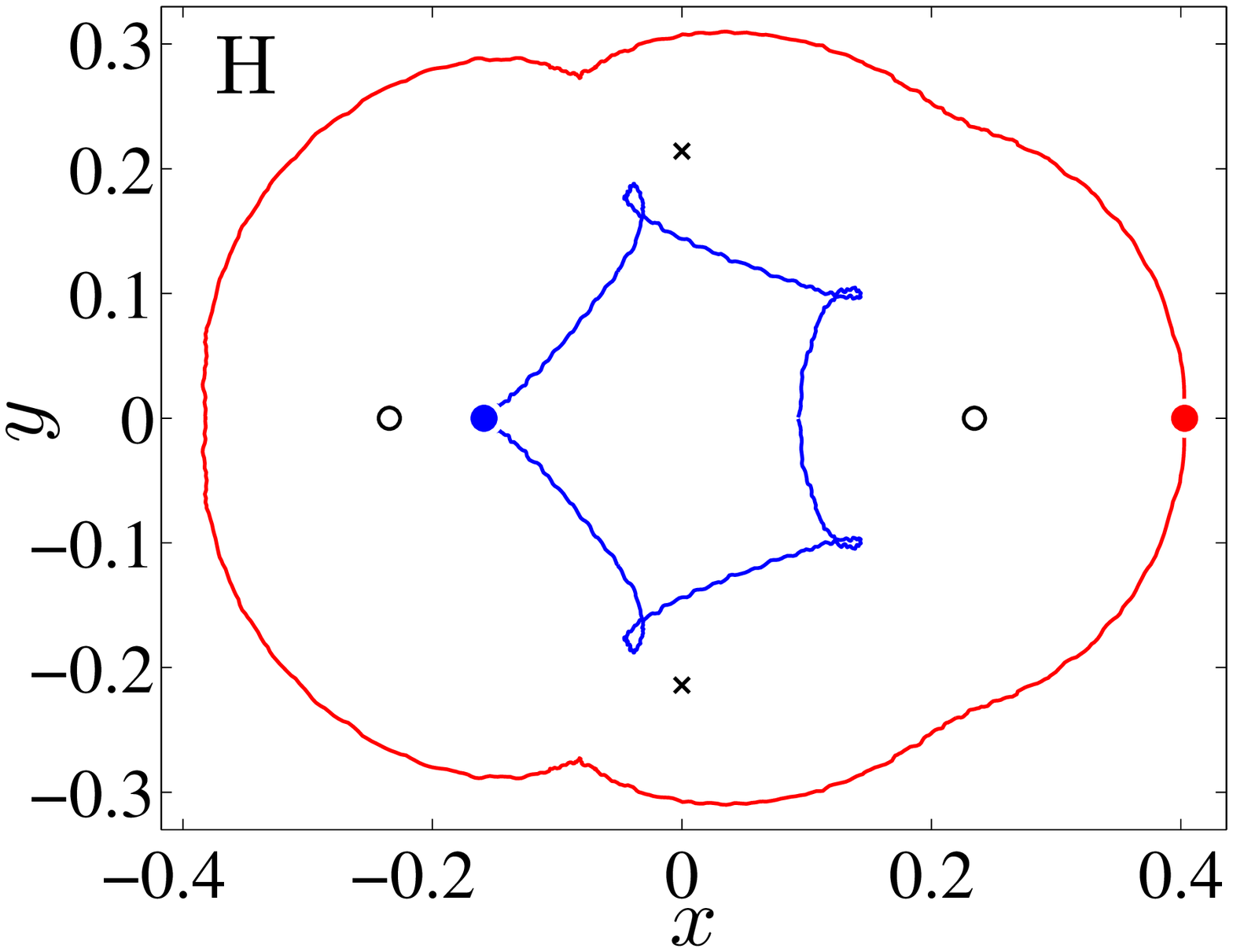}
   \caption{
(color online)
Periodic orbits of the full GP model (\ref{GP}) with an anisotropic
potential (\ref{eq:pot:ani}) with $\epsilon=0.2$. The initial
conditions where found by using the initial conditions corresponding
to the periodic orbits of the reduced ODE system (\ref{eq:xy})
as seeds for a local search in initial condition space
(see text for details).
The different panels are labelled using the same notation
as the respective panels in Figure~\ref{fig:m228} and the
top panel of Figure~\ref{fig:in_out_m228}.
Namely, for an energy of $H=0.228$.
}
\label{fig:periodic_PDE10}
\end{figure}

Having detected the origin of the spurious wiggles when naively
using the seed given by equation~(\ref{eq:vortex_xy}), we proceed
to remove the effects of the undesired perturbation of the sloshing
mode. A possible way to eliminate the sloshing from the dynamics
is to use initial conditions in which the projection of  the ground state onto the
sloshing mode is 
subtracted off.  This would require us to obtain the eigenfunctions of the
linearized steady state (ground state) configuration.
Instead, we opt here to use a simpler and more direct method.
We simply compute the linear momenta, in both $x$ and $y$,
of the seeded initial vortex configuration and imprint the
opposite momenta into the initial configuration to cancel any
possible sloshing.
Thus, after obtaining the desired combination of seeded
vortices using equation~(\ref{eq:vortex_xy}), we compute the linear
momenta:
\begin{equation*}
P_{z} = i \iint{\left(u_0 \frac{d\bar{u}_0}{dz}-\bar{u}_0 \frac{d{u_0}}{dz}
\right) dx\,dy},
\end{equation*}
where $z$ needs to be replaced by $x$ or $y$ for the respective
horizontal and vertical linear momenta.
Then, the ``distilled'' initial configuration is obtained by adding
the corresponding opposite momenta to cancel any residual
linear momenta that may be responsible for the appearance
of the sloshing mode, i.e.,
\begin{equation*}
u_{\rm ini} = u_0 \times e^{i k_x x} \times e^{i k_y y},
\end{equation*}
where the ``kicks'' $k_x$ and $k_y$ are chosen so that the
new linear momenta of $u_{\rm ini}$ are zero.
The advantage of using this method is that, in principle, it works
for any vortex configuration with any number of vortices.
In Figure~\ref{fig:wiggles}b we depict the corresponding orbits
after removal of the spurious sloshing mode. As it is clear
from the figure, the spurious wiggles are no longer present
since we are not perturbing the sloshing mode any longer.
From now on we use the above subtraction of the
sloshing mode for all the seeding of our initial vortex dipole
configurations in the GP model.

\begin{figure}[t]
   \centering
   \includegraphics[height=3.5cm]{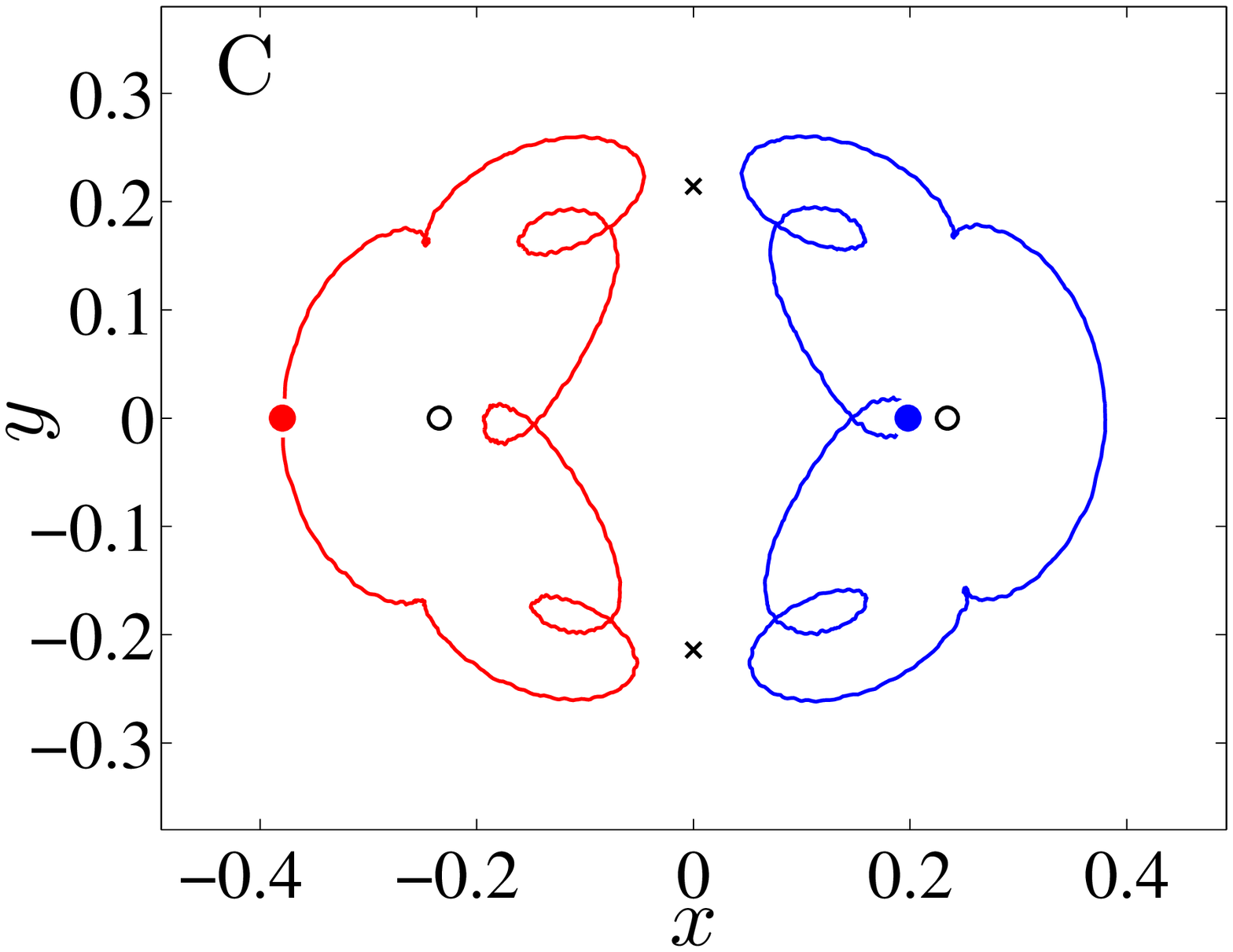}
   \includegraphics[height=3.5cm]{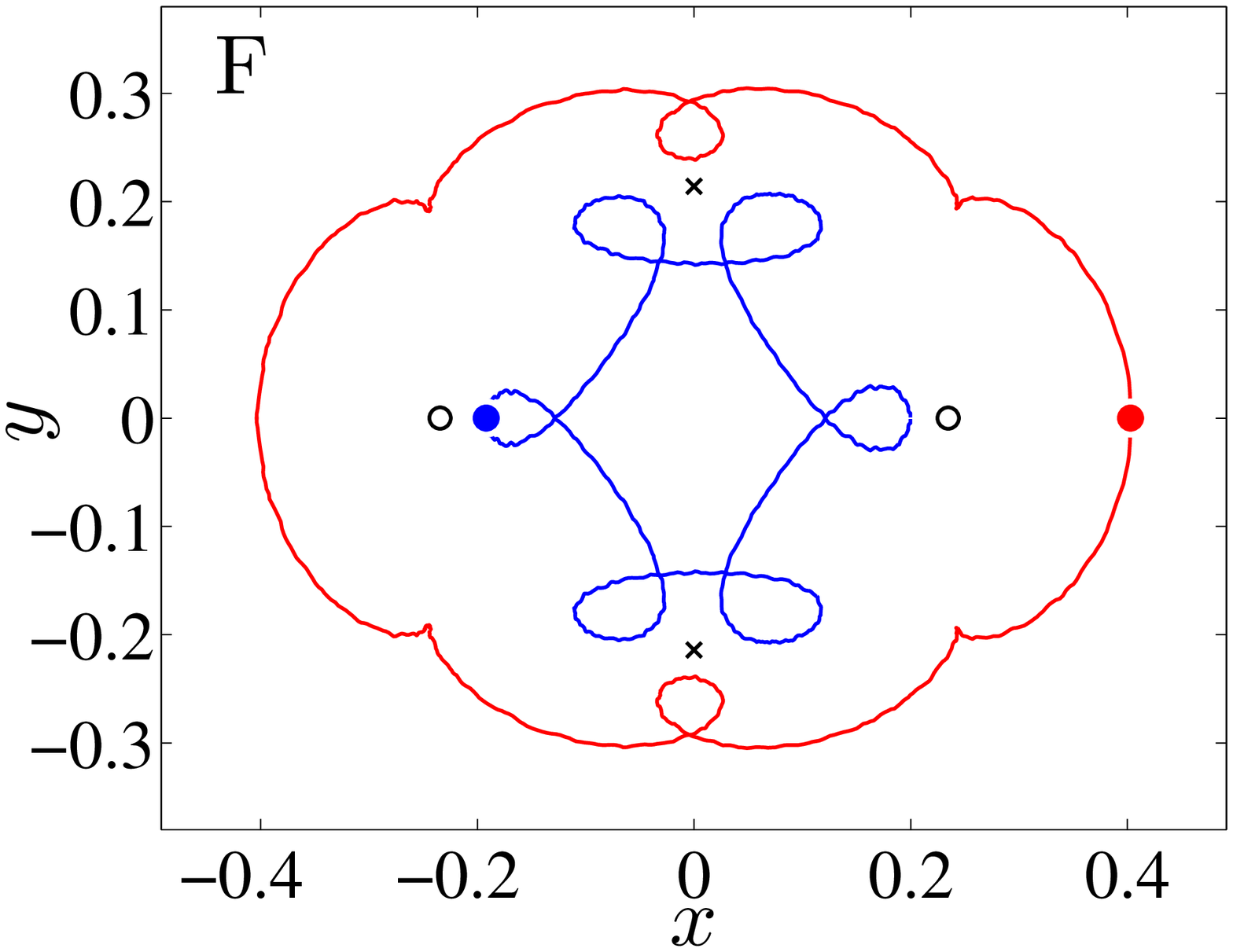}
   \includegraphics[height=3.5cm]{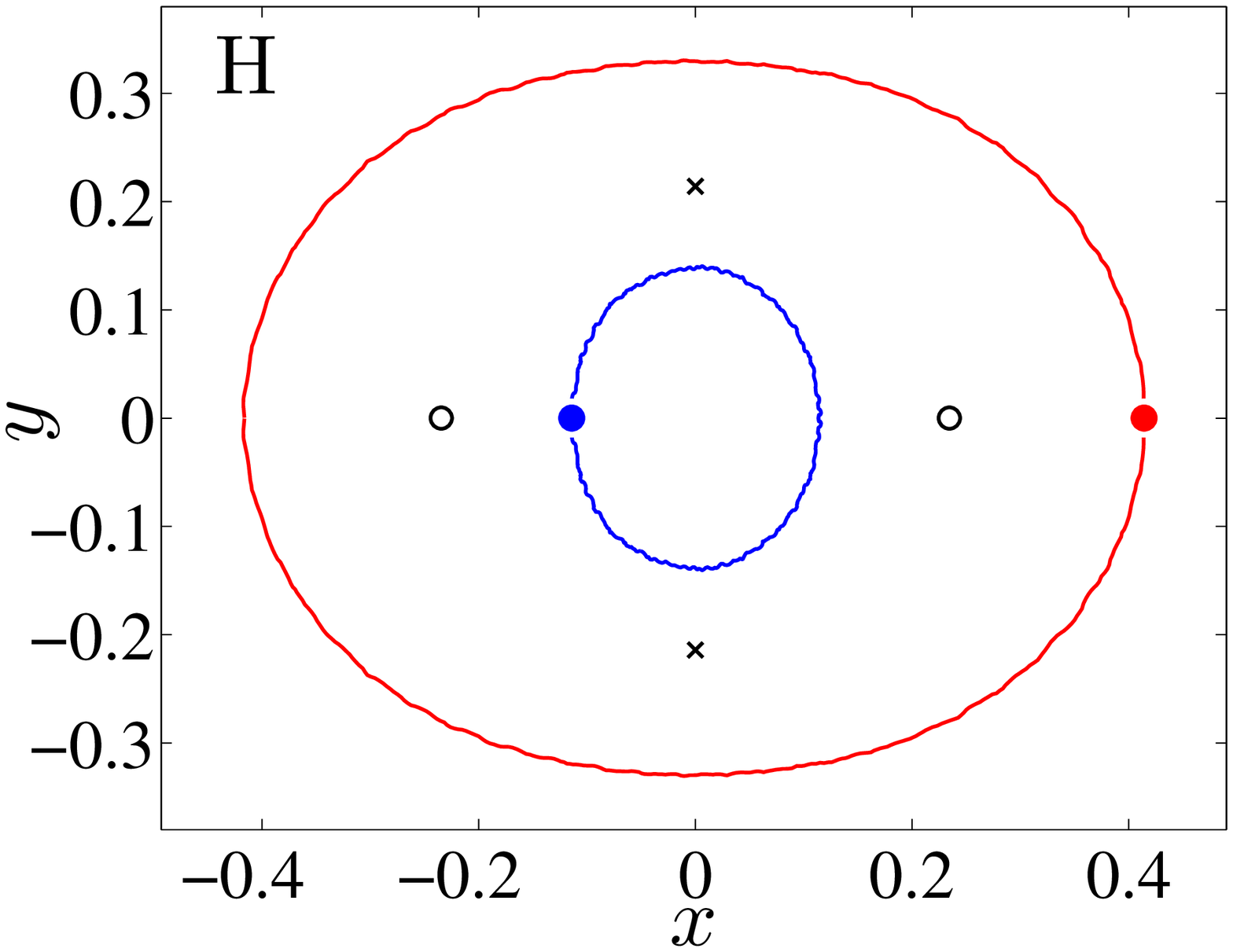}
   \caption{
(color online)
Same as in Figure~\ref{fig:periodic_PDE11} for the
respective panels of Figure~\ref{fig:p228} and the
bottom panel of Figure~\ref{fig:in_out_m228}.
Namely, for an energy of $H=0.228$.
}
\label{fig:periodic_PDE11}
\end{figure}

\begin{figure}[t]
   \centering
   \includegraphics[height=3.5cm]{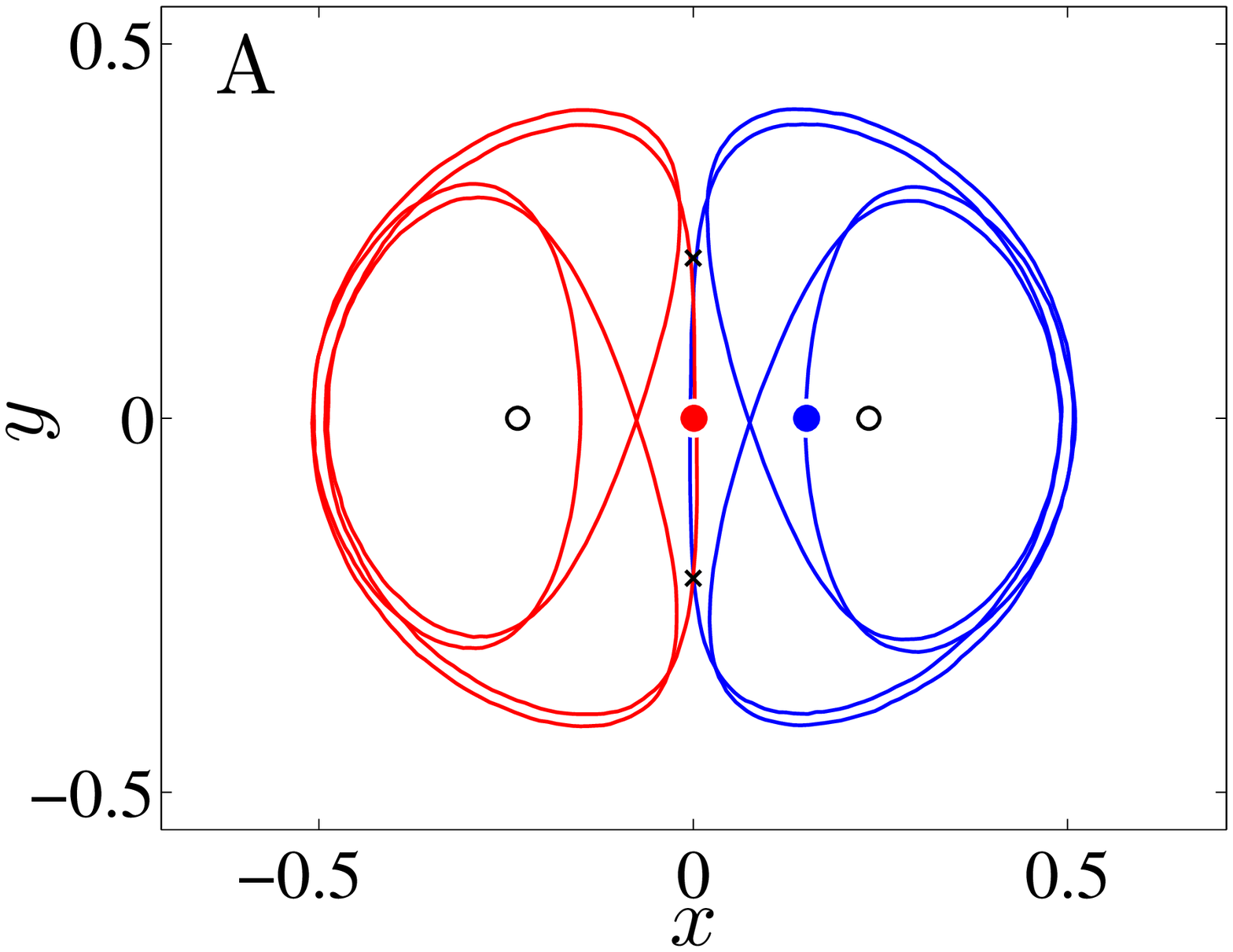}
   \includegraphics[height=3.5cm]{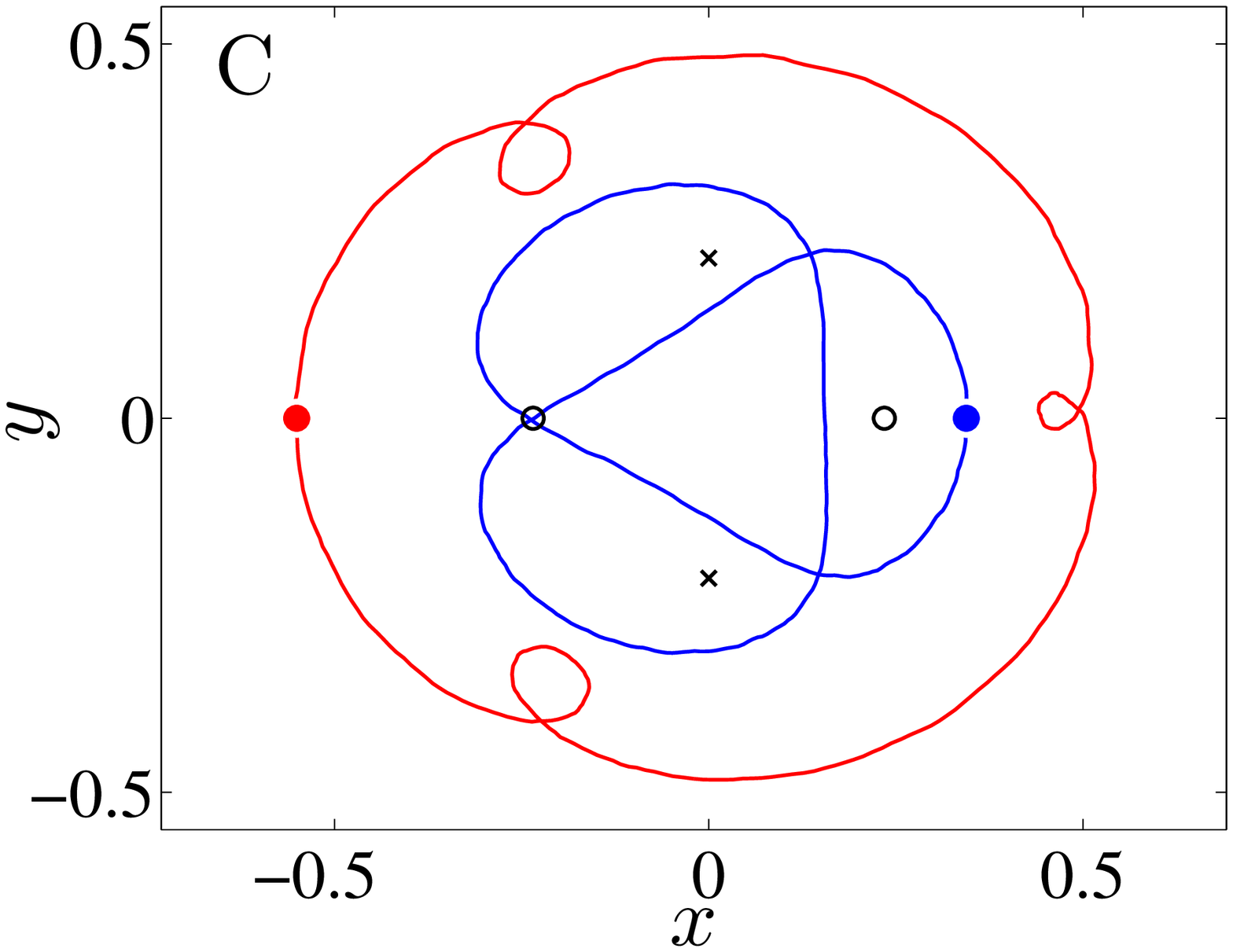}
   \includegraphics[height=3.5cm]{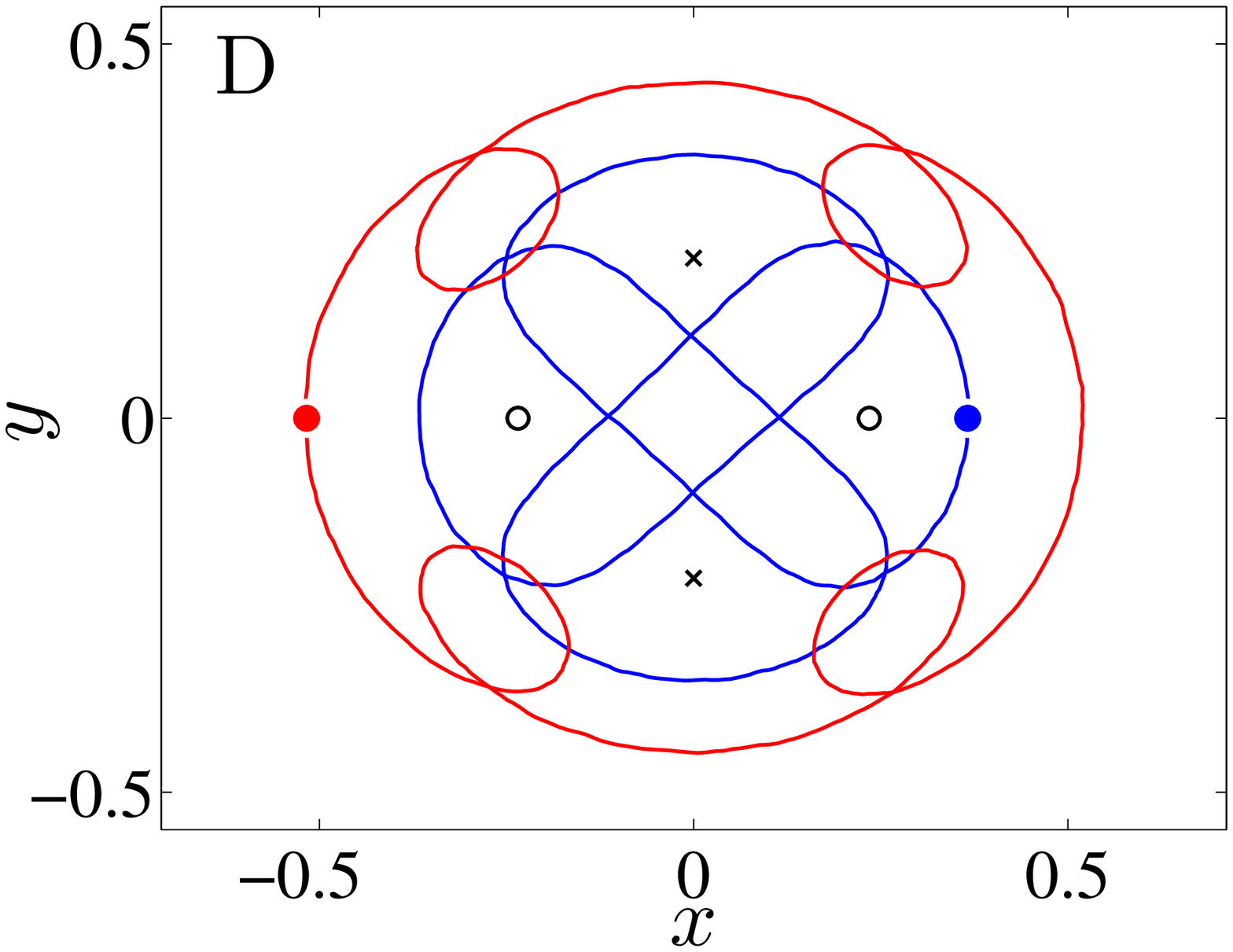}
   \caption{
(color online)
Same as in Figure~\ref{fig:periodic_PDE12} for the
respective panels of Figure~\ref{fig:vortexPaths}.
Namely, for an energy of $H=0.3$.
}
\label{fig:periodic_PDE12}
\end{figure}

We now proceed to find the periodic orbits on the full GP
model (\ref{GP}) corresponding to the ODE orbits described
in the previous section. It is important to mention that since
the ODE reduction for a vortex dipole is an approximation of
the corresponding full GP dynamics, initial conditions yielding
periodic orbits for the ODE model do not exactly correspond
to initial conditions yielding periodic orbits of the full GP model.
However, since the ODE reduction provides a reasonable
approximation, we  use the initial conditions corresponding
to periodic orbits for the ODE as starting seeds for a local
(in initial condition space) search of the full GP periodic orbits.
The local search for initial conditions $(x_1,y_1,x_2,y_2)$
corresponding to periodic orbits is carried by setting
$y_1=y_2=0$ and performing a local 2D parameter sweep in $(x_1,x_2)$
space around the corresponding ODE initial condition.
In this manner, using a $10\times 10$ grid in $(x_1,x_2)$
space with a width of 0.01, we are able to find approximate
periodic orbits for the full GP model. In some instances,
a second parameter sweep on a new $10\times 10$ grid with spacing
0.001, centered about the most promising initial
condition of the previous grid, was used to refine the
initial condition for better convergence to the periodic orbit.
The resulting periodic orbits of the full GP model are depicted
in Figs.~\ref{fig:periodic_PDE10}--\ref{fig:periodic_PDE12}.
The depicted orbits correspond, respectively, to most of the ODE
periodic orbits depicted in Figs.~\ref{fig:m228}--\ref{fig:vortexPaths},
including both librations (such as C, D) and rotations (such as
E--H).
As it is clear from the figures, the periodic orbits of the
reduced ODE model, found with the Poincar\'e sections described
in the previous section, have equivalent orbits in the original
GP model.
It is remarkable that such a reduction is able to capture
this type of complex periodic orbits, revealing their numerous twists
which are, in turn, associated as indicated in the previous
section, to their number of intersections with the Poincar{\'e}
section.

\begin{figure}[t]
   \centering
   \includegraphics[width=14cm]{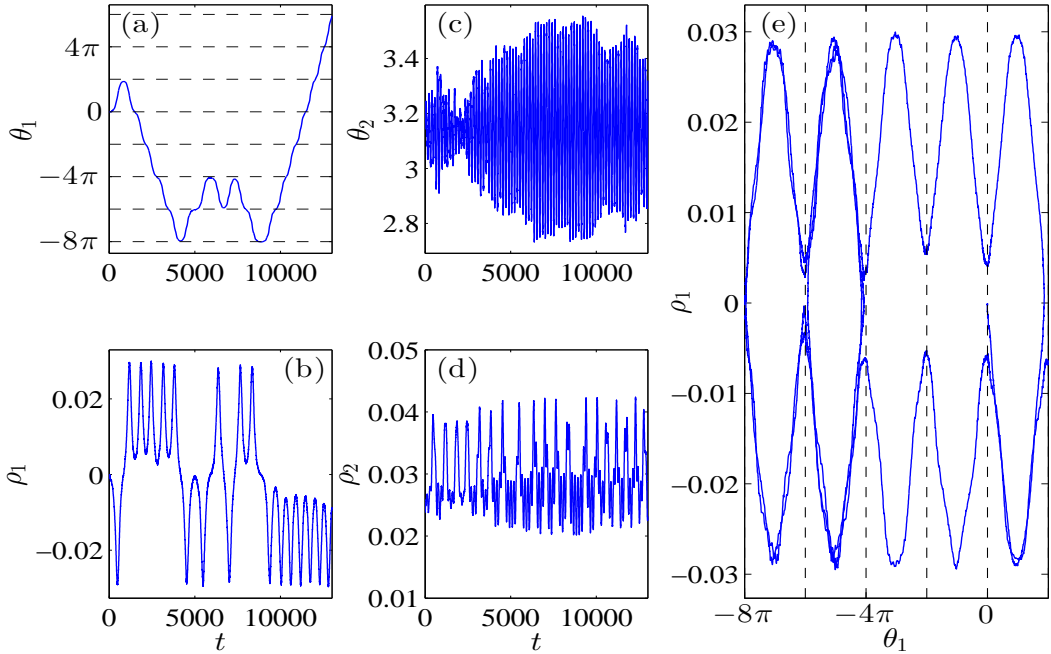}
   \caption{
(color online)
Same as in Figure~\ref{fig:nearPendulumSeparatrix} but for the
full GP model (\ref{GP}). Same layout and notation as in
Figure~\ref{fig:nearPendulumSeparatrix}.
}
\label{fig:PDE_near_guide_flips}
\end{figure}

Finally, to highlight the correspondence between the full GP
dynamics and the reduced ODE model, we searched for initial
conditions, using the seeding and the local search method explained
above, that reproduce the chaotic behavior observed near the
unstable guiding-center fixed point displaying a pendulum
separatrix-type behavior (see Figure~\ref{fig:nearPendulumSeparatrix}).
An example of this behavior for the full GP model is depicted
in Figure~\ref{fig:PDE_near_guide_flips} using the same layout
as in Figure~\ref{fig:nearPendulumSeparatrix}. As the figure shows,
there is a striking similarity between the reduced ODE
dynamics and the original PDE model including the chaotic switching
between left and right rotations, the relation between the
sign of $\rho_1$ and the rotational direction of $\theta_1$ and
the overall proximity of the $(\rho_1,\theta_1)$ plane to
the corresponding pendulum separatrix.

\section{Discussion}
\label{sec:discussion}

While previous studies have chiefly focused on the fixed-point type
orbits of a vortex-dipole (and their linear stability)
in an anisotropic trap, we have used techniques
of Hamiltonian mechanics to explore, in a detailed way, the dynamics of
such an anisotropic system. Arguably, this is one of the simplest
vortex systems where the lack of integrability can be seen to
be responsible for complex and chaotic dynamics, in addition to
periodic and quasi-periodic orbits.
In the process, we have discovered and enumerated a large number
of families of  periodic orbits and regions of chaotic dynamics.

This approach should be useful in uncovering the dynamics of many related systems. A straightforward example would be to change the sign of one of the vortices to make them co-rotating. This has been considered for the case of an isotropic trap in Refs.~\cite{Kolokolnikov,Navarro:2012vn,Navarro:2013uv,zampetaki}. The isotropic case shows more interesting dynamics than what was shown in Section~\ref{sec:unperturbed} of the present case, including pitchfork bifurcations leading to asymmetric time-dependent motions (although these are critically seeded
by the radial dependence of the precession frequency not considered herein).
How such dynamics interacts with an imposed anisotropy would be an
interesting followup question. While this theme has already been touched
upon in the physics community~\cite{McEndoo:2009tp,mcendoo2}, it still
lacks a systematic characterization based on the techniques presented
above.

The isotropic case examples of the co-rotating~\cite{Navarro:2013uv}
and of the counter-rotating~\cite{dshall1} dipoles have been
experimentally explored.
Yet, it should be especially
interesting to look for the behavior found here in similar laboratory
experiments. This may not be easy. The imaging procedure used in the
experiments causes the atom number to decrease over time (as a small fraction
is transferred to a different hyperfine state in order to be imaged).
Consequently, a more realistic model of the motion may include
dissipation which, in turn, has a non-trivial impact on the dynamics,
leading individual vortices to a spiraling out (rather than rotational)
motion; see, e.g., the recent discussion of Ref.~\cite{spirn2} and references
therein.
Additionally, many of the phenomena described require a long time series,
and the time series obtainable in the 
laboratory experiments may be too short to observe them.

The related system consisting of three vortices and an isotropic trap, either with all three co-rotating or else with one vortex of opposite sign, is known from numerical experiments to exhibit chaotic dynamics~\cite{Chang:2002}, and an
attempt has been made to understand these systems using finite-dimensional
reductions and index-based tools that enable the quantification of
the potential chaoticity of the
orbits~\cite{Koukouloyannis:2013,Kyriakopoulos:2013}. For systems with more vortices, the reductions carried out here may be less helpful as the number of dimensions in the reduced system may still be too large to say very much about.  In addition, such systems may lack the small parameter necessary to apply perturbation methods. Here, the energy-momentum bifurcation diagram  as applied in Ref.~\cite{Shlizerman:2005tt},  may be a useful tool, as might symmetry-based methods~\cite{Bountis:2011wl,Montaldi:1988wq,Montaldi:1990kg,Montaldi:1990db}.

In general, there exists a wide body of literature in nonlinear waves looking at special solutions such as periodic orbits and solitary waves, and examining their stability. Much of this work makes the implicit assumption that only stable solutions are experimentally important, since unstable solutions would be unobservable in practice. This is contrary to the lessons from finite-dimensional dynamical systems that unstable solutions are important because their stable
and unstable manifolds separate regions of phase space and provide skeletons that organize the dynamics, as we have also seen in this paper. There is a more recent trend recognizing this fact and exploring related phenomena in nonlinear wave systems, e.g.~Refs.~\cite{Goodman:2011,Goodman:2014,Marzuola:2010,Pelinovsky:2012}, and there exist many problems which might be further understood using such an approach. However, again an emerging crucial step will be the ability to
overcome the limitation of very low-dimensional systems. E.g., in the realm
of vortices, it is especially relevant to build 
a progressive understanding of small
or intermediate clusters of (3 or 4 and up
to 11 or so~\cite{Navarro:2013uv}) vortices and even of larger clusters
in the form of vortex lattices~\cite{Raman}
which are of particular experimental interest.

\section*{Acknowledgments}
RHG gratefully acknowledges the support of NSF-DMS-0807284.
PGK and RCG gratefully acknowledge the support of NSF-DMS-0806762
and NSF-DMS-1312856.
PGK also acknowledges support from
the Alexander von Humboldt Foundation, the Binational Science
Foundation under grant 2010239, NSF-CMMI-1000337,
FP7, Marie Curie Actions, People, International Research
Staff Exchange Scheme (IRSES-606096)
and from  the US-AFOSR under grant FA9550-12-10332.
Finally, PGK acknowledges the hospitality of the Center for NonLinear
Studies at the Los Alamos National Laboratory during the final stages
of this work; his work at the Los Alamos National Laboratory is supported
in part by the U.S. Department of Energy.

\bibliographystyle{siam}
\bibliography{anisotropic_bec}

\begin{thebibliography}{10}

\bibitem{Bountis:2011wl}
{\sc T.~Bountis, G.~Chechin, and V.~Sakhnenko}, {\em {Discrete Symmetry and
  Stability in Hamiltonian Dynamics}}, Int. J. Bifurcat. Chaos, 21 (2011),
  pp.~1539--1582.

\bibitem{Camassa:1998tc}
{\sc R.~Camassa, G.~Kovacic, and S.~K. Tin}, {\em {A Melnikov method for
  homoclinic orbits with many pulses}}, Arch. Ration. Mech. An., 143 (1998),
  pp.~105--193.

\bibitem{castin}
{\sc Y.~Castin and R.~Dum}, {\em {{Bose}-{Einstein} condensates with vortices
  in rotating traps}}, Eur. Phys. J. D,  (1999), pp.~399--412.

\bibitem{Chang:2002}
{\sc S.M. Chang, W.W. Lin, and T.C. Lin}, {\em {Dynamics of vortices in
  two-dimensional {Bose}-{Einstein} condensates}}, Int. J. Bifurcat. Chaos, 12
  (2002), pp.~739--764.

\bibitem{Dhooge:2003}
{\sc A.~Dhooge, W.~Govaerts, and Y.~A. Kuznetsov}, {\em {MATCONT: a MATLAB
  package for numerical bifurcation analysis of ODEs}}, ACM T. Math. Software,
  29 (2003), pp.~141--164.

\bibitem{fetter}
{\sc A.L. Fetter and A.A. Svidzinsky}, {\em {Vortices in a trapped dilute
  {Bose}-{Einstein} condensate}}, J. Phys.-Condens. Mat.,  (2001),
  pp.~R135--R194.

\bibitem{fetter1}
{\sc A.~L. Fetter}, {\em Rotating trapped {Bose}-{Einstein} condensates}, Rev.
  Mod. Phys., 81 (2009), pp.~647--691.

\bibitem{fetter2}
\leavevmode\vrule height 2pt depth -1.6pt width 23pt, {\em Vortex dynamics in
  spin-orbit-coupled {Bose}-{Einstein} condensates}, Phys. Rev. A, 89 (2014),
  p.~023629.

\bibitem{dshall}
{\sc D.~V. Freilich, D.~M. Bianchi, A.~M. Kaufman, T.~K. Langin, and D.~S.
  Hall}, {\em Real-time dynamics of single vortex lines and vortex dipoles in a
  {Bose}-{Einstein} condensate}, Science, 329 (2010), pp.~1182--1185.

\bibitem{Goldstein:2001}
{\sc H.~Goldstein, C.~Poole, and J.~Safko}, {\em Classical Mechanics}, Addison
  Wesley, 3rd~ed., 2001.

\bibitem{Goodman:2008}
{\sc R.~H. Goodman}, {\em {Chaotic scattering in solitary wave interactions: A
  singular iterated-map description}}, Chaos, 18 (2008), p.~023113.

\bibitem{Goodman:2011}
\leavevmode\vrule height 2pt depth -1.6pt width 23pt, {\em {Hamiltonian Hopf
  bifurcations and dynamics of NLS/GP standing-wave modes}}, J. Phys. A: Math.
  Theor., 44 (2011), p.~425101.

\bibitem{Goodman:2005vv}
{\sc R.~H. Goodman and R.~Haberman}, {\em {Kink-antikink collisions in the
  $\phi^4$ equation: The $n$-bounce resonance and the separatrix map}}, SIAM J.
  Appl. Dyn. Syst., 4 (2005), pp.~1195--1228.

\bibitem{Goodman:2007wj}
\leavevmode\vrule height 2pt depth -1.6pt width 23pt, {\em {Chaotic scattering
  and the $n$-bounce resonance in solitary-wave interactions}}, Phys. Rev.
  Lett., 98 (2007), p.~104103.

\bibitem{Goodman:2014}
{\sc R.~H. Goodman, J.~L. Marzuola, and M.~I. Weinstein}, {\em {Self-trapping
  and Josephson tunneling solutions to the nonlinear Schr{\"o}dinger /
  Gross-Pitaevskii Equation}}, Discrete. Contin. Dyn. S., 35 (2015),
  pp.~225--246.

\bibitem{Holmes:1990kz}
{\sc P.~Holmes}, {\em {Poincar\'e, celestial mechanics, dynamical-systems
  theory and ``chaos''}}, Phys. Rep., 193 (1990), pp.~137--163.

\bibitem{jerrard}
{\sc R.L. Jerrard and D.~Smets}, {\em {Vortex dynamics for the two dimensional
  non homogeneous Gross-Pitaevskii equation}}, 1301.5213,  (2013), pp.~1--32.

\bibitem{emergent}
{\sc P.G. Kevrekidis, D.J. Frantzeskakis, and R.~Carretero-Gonz{\'a}lez}, {\em
  Emergent Nonlinear Phenomena in {Bose}-{Einstein} Condensates},
  Springer-Verlag, Heidelberg, 2008.

\bibitem{Kolokolnikov}
{\sc T.~Kolokolnikov, P.~G. Kevrekidis, and R.~Carretero-Gonz\'{a}lez}, {\em A
  tale of two distributions: from few to many vortices in quasi-two-dimensional
  {Bose}-{Einstein} condensates}, P. Roy. Soc. A.-Math. Phy., 470 (2014).

\bibitem{Koukouloyannis:2013}
{\sc V.~Koukouloyannis, G.~Voyatzis, and P.~G. Kevrekidis}, {\em Dynamics of
  three noncorotating vortices in {Bose}-{Einstein} condensates}, Phys. Rev. E,
  89 (2014), p.~042905.

\bibitem{mikko2}
{\sc P.~Kuopanportti, J.~A.~M. Huhtam{\"a}ki, and M.~M{\"o}tt{\"o}nen}, {\em
  Size and dynamics of vortex dipoles in dilute {Bose}-{Einstein} condensates},
  Phys. Rev. A, 83 (2011), p.~011603.

\bibitem{spirn}
{\sc M.~Kurzke, C.~Melcher, R.~Moser, and D.~Spirn}, {\em {Dynamics for
  Ginzburg-Landau Vortices under a Mixed Flow}}, Indiana Univ. Math. J.,
  (2009), pp.~2597--2621.

\bibitem{Kyriakopoulos:2013}
{\sc N.~Kyriakopoulos, V.~Koukouloyannis, C.~Skokos, and P.~G. Kevrekidis},
  {\em Chaotic behavior of three interacting vortices in a confined
  {Bose}-{Einstein} condensate}, Chaos, 24 (2014), p.~024410.

\bibitem{S2Ket}
{\sc A.~E. Leanhardt, A.~G{\"o}rlitz, A.~P. Chikkatur, D.~Kielpinski, Y.~Shin,
  D.~E. Pritchard, and W.~Ketterle}, {\em Imprinting vortices in a
  {Bose}-{Einstein} condensate using topological phases}, Phys. Rev. Lett., 89
  (2002), p.~190403.

\bibitem{komineas}
{\sc W.~Li, M.~Haque, and S.~Komineas}, {\em Vortex dipole in a trapped
  two-dimensional {Bose}-{Einstein} condensate}, Phys. Rev. A, 77 (2008),
  p.~053610.

\bibitem{Lichtenberg:1992}
{\sc A.~J. Lichtenberg and M.~A. Lieberman}, {\em Regular and Chaotic
  Dynamics}, Springer-Verlag, New York, 1992.

\bibitem{Lorenz:1986}
{\sc E.~N. Lorenz}, {\em On the existence of a slow manifold}, J. Atmos. Sci,
  43 (1986), pp.~1547--1557.

\bibitem{Madison01}
{\sc K.~W. Madison, F.~Chevy, V.~Bretin, and J.~Dalibard}, {\em Stationary
  states of a rotating {Bose}-{Einstein} condensate: Routes to vortex
  nucleation}, Phys. Rev. Lett., 86 (2001), pp.~4443--4446.

\bibitem{Madison00}
{\sc K.~W. Madison, F.~Chevy, W.~Wohlleben, and J.~Dalibard}, {\em Vortex
  formation in a stirred {Bose}-{Einstein} condensate}, Phys. Rev. Lett., 84
  (2000), pp.~806--809.

\bibitem{Marzuola:2010}
{\sc J.~L. Marzuola and M.~I. Weinstein}, {\em Long time dynamics near the
  symmetry breaking bifurcation for nonlinear
  {S}chr{\"o}dinger/{G}ross-{P}itaevskii equations}, Discrete. Contin. Dyn. S.,
  28 (2010), pp.~1505--1554.

\bibitem{Matthews99}
{\sc M.~R. Matthews, B.~P. Anderson, P.~C. Haljan, D.~S. Hall, C.~E. Wieman,
  and E.~A. Cornell}, {\em Vortices in a {Bose}-{Einstein} condensate}, Phys.
  Rev. Lett., 83 (1999), pp.~2498--2501.

\bibitem{McEndoo:2009tp}
{\sc S.~McEndoo and T.~Busch}, {\em {Small numbers of vortices in anisotropic
  traps}}, Phys. Rev. A, 79 (2009), p.~053616.

\bibitem{mcendoo2}
{\sc S.~McEndoo and Th. Busch}, {\em Vortex dynamics in anisotropic traps},
  Phys. Rev. A, 82 (2010), p.~013628.

\bibitem{dshall1}
{\sc S.~Middelkamp, P.~J. Torres, P.~G. Kevrekidis, D.~J. Frantzeskakis,
  R.~Carretero-Gonz\'alez, P.~Schmelcher, D.~V. Freilich, and D.~S. Hall}, {\em
  Guiding-center dynamics of vortex dipoles in {Bose}-{Einstein} condensates},
  Phys. Rev. A, 84 (2011), p.~011605.

\bibitem{Montaldi:1988wq}
{\sc J.~A. Montaldi, R.~M. Roberts, and I.~N. Stewart}, {\em {Periodic
  solutions near equilibria of symmetric Hamiltonian systems}}, Philos. T. R.
  Soc. A, 325 (1988), pp.~237--293.

\bibitem{Montaldi:1990kg}
\leavevmode\vrule height 2pt depth -1.6pt width 23pt, {\em {Existence of
  nonlinear normal modes of symmetric Hamiltonian systems}}, Nonlinearity, 3
  (1990), pp.~695--730.

\bibitem{Montaldi:1990db}
\leavevmode\vrule height 2pt depth -1.6pt width 23pt, {\em {Stability of
  nonlinear normal modes of symmetric Hamiltonian systems}}, Nonlinearity, 3
  (1990), pp.~731--772.

\bibitem{mikko1}
{\sc M.~M{\"o}tt{\"o}nen, S.~M.~M. Virtanen, T.~Isoshima, and M.~M. Salomaa},
  {\em Stationary vortex clusters in nonrotating {Bose}-{Einstein}
  condensates}, Phys. Rev. A, 71 (2005), p.~033626.

\bibitem{Navarro:2012vn}
{\sc R.~Navarro, R.~Carretero-Gonz{\'a}lez, P.~J. Torres, P.~G. Kevrekidis,
  D.~J. Frantzeskakis, and D.~S. Hall}, {\em {Normal Modes of Interacting
  Vortices in a Trapped {Bose}-{Einstein} Condensate}}, preprint,  (2012),
  pp.~1--22.

\bibitem{Navarro:2013uv}
{\sc R.~Navarro, R.~Carretero-Gonz{\'a}lez, P.~J. Torres, P.~G. Kevrekidis,
  D.~J. Frantzeskakis, M.~W. Ray, E.~Altunta\c{s}, and D.~S. Hall}, {\em
  {Dynamics of Few Co-rotating Vortices in {Bose}-{Einstein} Condensates}},
  Phys. Rev. Lett., 110 (2013), p.~225301.

\bibitem{BPA3}
{\sc T.~W. Neely, E.~C. Samson, A.~S. Bradley, M.~J. Davis, and B.~P.
  Anderson}, {\em Observation of vortex dipoles in an oblate {Bose}-{Einstein}
  condensate}, Phys. Rev. Lett., 104 (2010), p.~160401.

\bibitem{kett99}
{\sc R.~Onofrio, C.~Raman, J.~M. Vogels, J.~R. Abo-Shaeer, A.~P. Chikkatur, and
  W.~Ketterle}, {\em Observation of superfluid flow in a {Bose}-{Einstein}
  condensed gas}, Phys. Rev. Lett., 85 (2000), pp.~2228--2231.

\bibitem{peli}
{\sc D.E. Pelinovsky and P.~G. Kevrekidis}, {\em {Variational approximations of
  trapped vortices in the large-density limit}}, Nonlinearity,  (2011),
  pp.~1271--1289.

\bibitem{Pelinovsky:2012}
{\sc D.~Pelinovsky and T.~Phan}, {\em Normal form for the symmetry-breaking
  bifurcation in the nonlinear {S}chr{{\"o}}dinger equation}, J. Diff. Eq., 253
  (2012), pp.~2796--2824.

\bibitem{becbook1}
{\sc C.J. Pethick and H.~Smith}, {\em {Bose}-{Einstein} condensation in dilute
  gases}, Cambridge University Press, Cambridge, 2002.

\bibitem{becbook2}
{\sc L.P. Pitaevskii and S.~Stringari}, {\em {Bose}-{Einstein} Condensation},
  Oxford University Press, Oxford, 2003.

\bibitem{Poincare:1890}
{\sc H.~Poincar{\'e}}, {\em {Sur les \'equations de la dynamique et le
  prob\`eme des trois corps}}, Acta. Math., 13 (1890), pp.~1--270.

\bibitem{Raman}
{\sc C.~Raman, J.~R. Abo-Shaeer, J.~M. Vogels, K.~Xu, and W.~Ketterle}, {\em
  Vortex nucleation in a stirred {Bose}-{Einstein} condensate}, Phys. Rev.
  Lett., 87 (2001), p.~210402.

\bibitem{Recati01}
{\sc A.~Recati, F.~Zambelli, and S.~Stringari}, {\em Overcritical rotation of a
  trapped {Bose}-{Einstein} condensate}, Phys. Rev. Lett., 86 (2001),
  pp.~377--380.

\bibitem{BPAPRL}
{\sc D.~R. Scherer, C.~N. Weiler, T.~W. Neely, and B.~P. Anderson}, {\em Vortex
  formation by merging of multiple trapped {Bose}-{Einstein} condensates},
  Phys. Rev. Lett., 98 (2007), p.~110402.

\bibitem{tripole}
{\sc J.~A. Seman, E.~A.~L. Henn, M.~Haque, R.~F. Shiozaki, E.~R.~F. Ramos,
  M.~Caracanhas, P.~Castilho, C.~Castelo~Branco, P.~E.~S. Tavares, F.~J.
  Poveda-Cuevas, G.~Roati, K.~M.~F. Magalh\~aes, and V.~S. Bagnato}, {\em
  Three-vortex configurations in trapped {Bose}-{Einstein} condensates}, Phys.
  Rev. A, 82 (2010), p.~033616.

\bibitem{Shlizerman:2005tt}
{\sc E.~Shlizerman and V.~Rom-Kedar}, {\em {Hierarchy of bifurcations in the
  truncated and forced nonlinear Schr{{\"o}}dinger model}}, Chaos, 15 (2005),
  p.~013107.

\bibitem{Sinha01}
{\sc Y.~Sinha, SA.nd~Castin}, {\em Dynamic instability of a rotating
  {Bose}-{Einstein} condensate}, Phys. Rev. Lett., 87 (2001), p.~190402.

\bibitem{Stockhofe:2011}
{\sc J.~Stockhofe}, {\em {Vortex cluster and dark-bright ring solitons in
  {Bose}-{Einstein} condensates}}, master's thesis, Univ. Hamburg., 2011.

\bibitem{Stockhofe:2013}
{\sc J.~Stockhofe, P.~G. Kevrekidis, and P.~Schmelcher}, {\em {Existence,
  stability and nonlinear dynamics of vortices and vortex clusters in
  anisotropic {Bose}-{Einstein} condensates}}, in Spontaneous Symmetry
  Breaking, Self-Trapping, and Josephson Oscillations, B.~Malomed, ed.,
  Progress in Optical Science and Photonics, Springer, 2013, pp.~543--581.

\bibitem{Stockhofe:2011df}
{\sc J.~Stockhofe, S.~Middelkamp, P.~G. Kevrekidis, and P.~Schmelcher}, {\em
  {Impact of anisotropy on vortex clusters and their dynamics}}, Euro. Phys.
  Lett., 93 (2011), p.~20008.

\bibitem{Torres:2011fp}
{\sc P.~J. Torres, P.~G. Kevrekidis, D.~J. Frantzeskakis,
  R.~Carretero-Gonz{\'a}lez, P.~Schmelcher, and D.~S. Hall}, {\em {Dynamics of
  vortex dipoles in confined {Bose}--{Einstein} condensates}}, Phys. Lett. A,
  375 (2011), pp.~3044--3050.

\bibitem{BPA}
{\sc C.N. Weiler, T.W. Neely, D.R. Scherer, A.S. Bradley, M.J. Davis, and B.P.
  Anderson}, {\em {Spontaneous vortices in the formation of {Bose}-{Einstein}
  condensates}}, Nature, 455 (2008), pp.~948--951.

\bibitem{Williams99}
{\sc J.E. Williams and M.J. Holland}, {\em {Preparing topological states of a
  {Bose}-{Einstein} condensate}}, Nature, 401 (1999), pp.~568--572.

\bibitem{spirn2}
{\sc D.~Yan, R.~Carretero-Gonz{\'a}lez, D.~J. Frantzeskakis, P.~G. Kevrekidis,
  N.~P. Proukakis, and D.~Spirn}, {\em Exploring vortex dynamics in the
  presence of dissipation: Analytical and numerical results}, Phys. Rev. A, 89
  (2014), p.~043613.

\bibitem{zampetaki}
{\sc A.~V. Zampetaki, R.~Carretero-Gonz{\'a}lez, P.~G. Kevrekidis, F.~K.
  Diakonos, and D.~J. Frantzeskakis}, {\em Exploring rigidly rotating vortex
  configurations and their bifurcations in atomic {Bose}-{Einstein}
  condensates}, Phys. Rev. E, 88 (2013), p.~042914.

\end{thebibliography}
\end{document}